%% file: fl-with-privacy-from-compressor.tex
\documentclass[sigconf,manuscript,nonacm]{acmart}

\AtBeginDocument{%
  \providecommand\BibTeX{{%
    \normalfont B\kern-0.5em{\scshape i\kern-0.25em b}\kern-0.8em\TeX}}}

\copyrightyear{2023}
\acmYear{2023}
\setcopyright{rightsretained}
\acmConference[DistributedML '23]{Proceedings of the 4th International Workshop on Distributed Machine Learning}{December 8, 2023}{Paris, France}
\acmBooktitle{Proceedings of the 4th International Workshop on Distributed Machine Learning (DistributedML '23), December 8, 2023, Paris, France}
\acmDOI{10.1145/3630048.3630182}
\acmISBN{979-8-4007-0447-5/23/12}

\usepackage[utf8]{inputenc} 
\usepackage[T1]{fontenc}    
\usepackage[colorinlistoftodos,bordercolor=orange,backgroundcolor=orange!20,linecolor=orange,textsize=scriptsize]{todonotes}

\usepackage{amsmath, latexsym, amsthm}
\usepackage{mathtools}

\usepackage{hyperref}               
\usepackage{url}                    
\usepackage{booktabs}               
\usepackage{amsfonts}               
\usepackage{nicefrac}               
\usepackage{microtype}              
\usepackage{natbib}
\usepackage{comment}
\usepackage{tablefootnote}
\usepackage{graphicx}
\usepackage{xcolor}                 
\usepackage{hyperref}               
\usepackage{cmap}                   
\usepackage{mathtext}
\usepackage{icomma}                 
\usepackage{longtable}              
\usepackage{multirow}               
\usepackage{multicol}               
\usepackage{color}
\usepackage{url}                    
\usepackage{cleveref}               
\usepackage{algorithm}
\usepackage{algorithmic}
\usepackage{xspace}
\usepackage{color}
\usepackage{pifont}

\usepackage{float}
\usepackage{subcaption}
\usepackage{mdframed}
\usepackage{enumitem}
\usepackage{latexsym}
\usepackage{etoolbox}\AtBeginEnvironment{algorithmic}{\footnotesize}
\usepackage{tikz}
\usetikzlibrary{shapes.geometric, arrows}

\input{new_commands}         

\input{theorem_environments} 
\input{notes}                

\begin{document}

\title[Federated Learning is Better with Non-Homomorphic Encryption]{Federated Learning is Better with \\ Non-Homomorphic Encryption}

\author{Konstantin Burlachenko}
\affiliation{%
  \institution{AI Initiative, KAUST\thanks{King Abdullah University of Science and Technology}} 
  \city{Thuwal}
  \country{Saudi Arabia}
}

\author{Abdulmajeed Alrowithi}
\affiliation{%
  \institution{Saudi Data and AI Authority}
  \city{Riyadh}
  \country{Saudi Arabia}
}

\author{Fahad Ali Albalawi}
\affiliation{%
	\institution{Saudi Data and AI Authority}
	\city{Riyadh}
	\country{Saudi Arabia}
}

\author{Peter Richt\'{a}rik}
\affiliation{%
  \institution{AI Initiative, KAUST}
  \city{Thuwal}
  \country{Saudi Arabia}
}

\renewcommand{\shortauthors}{}

\begin{abstract}
Traditional AI methodologies necessitate centralized data collection, which becomes impractical when facing problems with network communication, data privacy, or storage capacity. Federated Learning (\abr{FL}) offers a paradigm that empowers distributed AI model training without collecting raw data. There are different choices for providing privacy during FL training. One of the popular methodologies is employing Homomorphic Encryption (\abr{HE}) - a breakthrough in privacy-preserving computation from Cryptography. However, these methods have a price in the form of extra computation and memory footprint. To resolve these issues, we propose an innovative framework that synergizes permutation-based compressors with Classical Cryptography, even though employing Classical Cryptography was assumed to be impossible in the past in the context of \abr{FL}. Our framework offers a way to replace \abr{HE} with cheaper Classical Cryptography primitives which provides security for the training process. It fosters asynchronous communication and provides flexible deployment options in various communication topologies.
\end{abstract}

%
%

\renewcommand\keywordsname{Keywords}

\keywords{Federated Learning, Privacy Preserving Machine Learning, Asynchronous Training, Security, Optimization, AES, CKKS.}



\maketitle

\section{Introduction}
Effective machine learning models necessitate vast training data from diverse sources \citep{zhang2020batchcrypt}. However, such data, often dispersed across various entities, faces sharing restrictions due to privacy concerns \citep{jiang2021flashe,liu2022privacy}. Federated learning (\abr{FL}) provides a viable solution, enabling collaborative global model training without exposing sensitive information \citep{mcmahan2017communication,kairouz2021advances}. \abr{FL} algorithms bifurcate into cross-device \abr{FL}, involving big amount of devices, and cross-silo \abr{FL}, typically engaging several distinct organizations \citep{liu2022privacy}. In both cases, preserving the privacy of clients' datasets is significant. \abr{FL} strives to ensure confidentiality by retaining private data on the client side, yet it falls short of providing substantial privacy assurance. Model parameters and derived quantities from it, transmitted from clients to a server, may embed sensitive data (see  Appendix \ref{app:reconstruction}). Detailed information about privacy methods for \abr{FL} is presented in Appendix \ref{app:overview_of_privacy_mechanisms}. We briefly discuss two of them.

Differential Privacy (\abr{DP}) is an approach to assess the privacy of a specific Algorithm provided by \citet{dwork2006our,dwork2006calibrating}. \abr{DP} algorithms ensure that, for any set of training examples, no attacker, no matter how powerful, can not learn much more information about a single training example than they could if this example had been excluded from the training data. Despite offering privacy protection for individual users, integrating \abr{DP} with \abr{FL} could amplify communication overhead, diminish accuracy. When striving for strong privacy guarantees in scenarios where the user count is small (which is possible in cross-silo settings) the noise impact from \abr{DP} technics makes training challenging. Understanding the convergence of \algname{SGD} with \abr{DP} is an active research topic \citep{chen2020understanding}. 

While \abr{DP} statistically hides the training data, Homomorphic Encryption (\abr{HE}) provides means to perform computation operations under encrypted numbers from $\mathbb{Z}^d$, $\mathbb{R}^d$, $\mathbb{B}^d$ without decryption, protecting input and output from execution part. Therefore \abr{HE} can be used for aggregating encrypted gradients \citep{liu2019secure,aono2017privacy} in a server that clients do not trust, given that master has a public key.

\paragraph{Classical Cryptography in FL} Classical Cryptography operates on the binary representation of data. It does not preserve even linear relations in a homomorphic way. The fact of exhibiting poor algebraic properties is an underlying reason why using Classical Cryptography was considered challenging by previous research papers \citep{kaissis2020secure,jain2023revisiting}. In these works, authors stated that Advanced Encryption Standard (\aesname{AES}) \citep{daemen2001reijndael} is not suitable for \abr{FL} or is challenging.

\paragraph{Communication Compression in FL}
Numerous compression methods such as quantization \citep{wen2017terngrad,safaryan2019stochastic}, sparsification \citep{wangni2018gradient,alistarh2018convergence}, and dithering \citep{alistarh2017qsgd,horvoth2022natural} have been explored to mitigate communication cost during \abr{FL} training. However, these techniques necessitate secure server-side aggregation, and there are no guarantees that communication reduction techniques and techniques aimed to preserve privacy or security are combinable.

\paragraph{Research Contributions}
We discovered that recently proposed permutated correlated compressors \citep{szlendak2021permutation} \compname{PermK} exhibit properties essential for using \textit{Classical Cryptography in FL} while preserving the ability of \textit{Communication Compression}. In our work, we introduce a framework that provides privacy and secure preserving training process to \abr{FL} applications in which previously \abr{HE} methods have been used. Summary of our contributions:

\begin{enumerate}
	\item We addressed the challenge previously mentioned in  \citep{kaissis2020secure,jain2023revisiting} that usage of \aesname{AES} is challenging in \abr{FL}.
	\item We demonstrated the operational advantages of the proposed privacy and secure aware optimization Alg.~\ref{alg:dcgd_permk_aes} over \abr{HE} in the setting in which \abr{HE} is typically applied in \abr{FL}. 
	\item We illustrated the framework's capability to train \modelname{ResNet-18} \citep{resnet18_56} \abr{DL} model in \dataname{CIFAR-10}, and carried the discussion of potential benefits for \abr{DL} in Appendix~\ref{app:flexibility_for_dl_training}. 
	\item We demonstrated a possibility of computation communication overlap and handling compute heterogeneity in Appendix~\ref{app:simulation_experiment}. Deployment flexibility in communication topologies is discussed in Appendix~\ref{app:comm_networks}.
\end{enumerate}

To support readers with various backgrounds we provide (i) glossary in Appendix~\ref{app:glossary}; (ii) details about \aesname{AES} in Appendix~\ref{app:aes_details}; (iii) overview of privacy mechanisms in \abr{FL} in Appendix~\ref{app:overview_of_privacy_mechanisms}, discussion about the difference between privacy and security in Appendix~\ref{app:privacy_vs_security}; (iv) overview of \ecryptname{CKKS} in Appendix~\ref{app:ckks_details} (see Table of Content at p.\pageref{app:toc_1}).


\section{Problem Formulation}

In this work, we develop a practical communication efficient privacy and secure aware framework for \abr{FL} training. 

\paragraph{Requirements for Optimization Objective}
From the perspective of machine learning (\abr{ML}), our objective is to select a function from a parameterized function class $\mathcal{F}$ indexed by $x \in \mathbb{R}^d$, by solving the following optimization problem:
\begin{equation}\label{eq:main}
	\squeeze \min \limits_{x\in \R^d} \left\{f(x)\eqdef \frac{1}{n}\sum \limits_{i=1}^n f_i(x) \right\}, 
\end{equation}

Here, $n\in \mathbb{N}$ represents the number of clients, and $x\in \mathbb{R}^d$ denotes the $d$ parameters or weights of a model $\hat{F}(\cdot;x) \in \mathcal{X}\to \mathcal{Y}$ that need to be learned across all $n$ clients. The function $f_i:\mathbb{R}^d \to \mathbb{R}$ provides the score criteria for using the model $\hat{F}(\cdot;x)$ on the client's $i$ data. In the context of \abr{FL} the functions $f_i$ typically represented as:

\begin{equation}\label{eq:fi_for_erm} 
	\squeeze  f_i(x) = \frac{w_i}{n_{i}} \sum \limits_{j=1}^{n_{i}} \left(\mathcal{L}_{ij}(b_{ij}, \hat{F}(a_{ij};x)) + R_{i}(x) \right),
\end{equation}

Here, $n_i \in \mathbb{R}$ denotes the number of data points at client $i \in [n]$, and $(a_{ij}, b_{ij}) \in \mathcal{X} \times \mathcal{Y}$ represent the input-output pairs at client $i$. The function $\mathcal{L}_{ij}(y_{\mathrm{real}}, y_{\mathrm{pred}} ): \mathcal{Y} \times \mathcal{Y} \to \mathbb{R}$ is a loss function that scores prediction,  $R_{i}: \mathbb{R}^d \to \mathbb{R}$ is the regularization function used for parameter $x$ at client $i$. The weight $w_i \in \mathbb{R}$ encodes knowledge about the role of client $i$. 
For example: (i) $w_i \eqdef {\left(n_i \cdot n\right)}/{\left( \sum_{i=1}^{n}n_i \right)}$ corresponds to a case when \textit{all data point}s are equally important; (ii) $w_i \eqdef 1$ corresponds to a case when \textit{all devices} are equally important. We aim to solve Problem \ref{eq:main} under the next requirements:



\paragraph{Security Requirements for the {FL} Training Procedure.}

\begin{enumerate}[label=(\Alph*)]
	\item Clients never transfer the training data to the master.
	\item Clients do not trust communication devices.
	\item Detect attempts of message tampering by adversaries.
	\item Clients distrust a server to store information that could compromise the training data.
	\item Preventing competitor worker interference.
	\item Limit memory traffic from clients to the master. 
	\item Overlapping communication and computing in clients because separate physical devices implement it. 
\end{enumerate}

\paragraph{Assumptions for applying our framework} (i) $f(x)$ is differentiable in training variable $x\in \mathbb{R}^d$; (ii) Clients trust each other through the established key; (iii) $\mathrm{dom}(f)=\RD$.


\paragraph{Real-world scenarios for applying our framework} (a) An individual using multiple \abr{IoT} devices and wants to train an \abr{ML} model via servers provided by a third-party company. In reality, engineers in the companies may have unlimited access to server software facilities and this social aspect should be considered while making a decision about sending plain messages to servers; (b) In cross-device settings several \abr{IoT} mutually trusted clients perform \abr{FL} training, but still there is a need to have a master: some client is temporarily unavailable and updates should be buffered; applied steps need to be stored to reconstruct the optimization trajectory for further statistical tests, but \abr{IoT} devices do not have the needed storage capacity; the training should be restarted in critical system failures; (c) We want to train \abr{FL} model without a physical master in a situation when communication topology effectively supports broadcasting. Our framework can be instantiated in this regime (see Appendix \ref{app:comm_networks}).


\section{Framework of Security Aware {FL} with Permuted Compressors}

The Distributed Compressed Gradient Descent 
(\algname{DCGD (Baseline)})  \citep{khirirat2018distributed}, presented as  (Algorithm \ref{alg:dcgd}, \myred{Option B}), enables the use of independent unbiased compressors $\mathcal{C}_i$ if $\Exp{\|\mathcal{C}_i(x) - x\|^2} \le w\|x\|^2$. If $w\ne0$ this algorithm does not induce variance for a $\mu$-strongly convex objective $f$ only in a specific (\textit{overparameterized}) mode: $\nabla f_i(x)=0, \forall i \in [n]$. If $\mathcal{C}_i (\nabla f_i(x)) \eqdef \nabla f_i(x)$, it reconstitutes distributed Gradient Descent (\algname{GD}). One line of research involves removing variance from the use of compressors. For instance, \algname{MARINA} \citep{gorbunov2021marina} and \algname{COFIG/FRECON} \citep{zhao2021faster} do not induce variance when using unbiased independent compressors, while \algname{EF21} \citep{richtarik2021ef21} avoids inducing variance from any independent contractive compressors. In all these algorithms, the logic in the master starts to include extra state updates based on obtaining messages from clients. But this is what we aim to avoid. Thus, the development of our framework is based on stateless \algname{DCGD (Baseline)}. For plain \algname{DCGD (Baseline)} we have two issues: (a) In Line 6, the algorithm sends sparsified information about $\nabla f_i (x)$ to the master, which is insecure; (b) In Line 8, the master computes the average of $g_i^k \in \mathbb{R}^d$, but the master should not obtain $g_i^k=\nabla f_i(x^k)$ or quantities that are subject to reconstruction attacks (see Appendix \ref{app:reconstruction}). One way to address these problems is to use the \abr{HE} schemes. The current practical state-of-the-art \abr{HE} scheme which works $\mathbb{R}^d$ is the \ecryptname{CKKS} \citep{cheon2017homomorphic}. The \ecryptname{CKKS} scheme is the most prevalent for \abr{ML} applications \citep{lauter2022protecting} and is a practical one. As we will demonstrate in experiments, the \ecryptname{CKKS} carries an approximation error that arises due to the scheme being lossy by design. See Appendix \ref{app:ckks_details}, \ref{app:overview_of_privacy_mechanisms} for details about \ecryptname{CKKS} and \abr{HE}.

\begin{algorithm}[H]
	\footnotesize
	\algsetup{linenosize=\footnotesize}
	\captionsetup{position=top}
	\begin{algorithmic}[1]
		\label{algo:1}
		\STATE  \textbf{Parameters:} step size $\gamma>0$, iterate $x^0\in\R^d$, clients' compressors $\cC_i$ \newline
		\myblue{Option A}: Clients negotiate a secret key ${\color{blue}sk}$ 
		\FOR {$k=0,1,2, \ldots$}
		\FOR {{\bf all workers $i \in \{1,2,\dots,n\}$ in parallel}}
		\STATE {Compute and compress local gradient $g_i^k = \cC_i^k(\nabla f_i(x^k))$} 
		\STATE \myblue{Option A}: {Send tuple $m_i^k = ({nonce}_i, {mac}_i, \tilde{g}_i^k) = {\color{blue}Enc}(g_i^k, {\color{blue}sk})$ to master, $\tilde{g}_i^k$ is encryption of $g_i^k$}
		\STATE \myred{Option B}: {Send $g_i^k$ to master as message $m_i^k$}
		\ENDFOR      
		\STATE {Master collects the messages from clients $G^k = (m_1^k,\dots,m_n^k)$}
		\STATE \myblue{Option A:} {Master broadcasts $G^k \in \mathbb{R}^{dn + |nonce|n + |mac|n}$ to workers}
		\STATE \myred{Option B:} {Master computes the aggregate $\hat{g}^k = \frac{1}{n}\sum_{i=1}^n  g_i^k$, $g_i^k \eqdef m_i^k$}
		\STATE \myred{Option B:} {Master broadcasts $\hat{g}^k \in \mathbb{R}^d$ to all $n$ workers}
		\FOR {{\bf all workers $i \in \{1,2,\dots,n\}$ in parallel}} 
		
		\STATE \myblue{Option A:} {From ${G}^k$ unpack ${mac}_j$ and $\tilde{g}_j^k \in \mathbb{R}^d, \forall j \in [n]$; \newline ${g}_j^k = {\color{blue}Decrypt}( \tilde{g}_j^k, {\color{blue}sk})$; $Verify({\color{blue}sk}, {g}_j^k, mac_j)$}
		\STATE \myblue{Option A:} {Workers computes the aggregate $\hat{g}^k = \frac{1}{n} \sum_{i=1}^n  g_i^k$}
		\STATE \myred{Option B:} {Workers obtain $\hat{g}^k$ from master during round $k$.}
		\STATE Compute the next iterate $x^{k+1} = {x^k - \gamma \hat{g}^k}$
		\ENDFOR
		\ENDFOR
	\end{algorithmic}
	\caption{\footnotesize{\algname{DCGD} with \myblue{Naive usage of AES  (A)} and \myred{Baseline (B)}}}
	\label{alg:dcgd}    
\end{algorithm}

\vspace{-0.4cm}

\begin{algorithm}[H]
	\footnotesize
	\algsetup{linenosize=\footnotesize}
	\begin{algorithmic}[1]
		\label{algo:3}
		\STATE  \textbf{Parameters:}  Dimension of opt. problem $d>0$, number of clients $n>0$.
		\STATE U.a.r. generate permutation $z$ of sequence $[d]=\{1,2,\dots,d\}$
		\STATE Split z into $n$ buckets, where each bucket has a size at least $B=\left\lfloor \frac{d}{n}\right\rfloor$
		\STATE Each bucket $b_i$ is initialized with $\{z_{ (i\cdot B) - B + 1}, \dots, z_{ (i\cdot B)}\}, 1 \le i \le n$
		\STATE Compute the residual $t = d - n \cdot \left\lfloor \frac{d}{n}\right\rfloor$
		\STATE Sample without replacement $t$ clients from $n$ as a set $S$, $|S| = t$
		\STATE Scan the set $S=\{s_1,\dots,s_k, \dots, s_t\}$ and update $b_{s_k} = b_{s_k} \cup z_{d - t + k}$
		\STATE {Setup compression $\cC=(\cC_1,\dots, \cC_n)$: $[\cC_i(x)]_j = n \cdot x_j \cdot I( j \in b_i)$}
	\end{algorithmic}
	\caption{\footnotesize{Sampling of Correlated Permutation Compressors (\compname{PermK}) configuration ($d>n$)}}
	\label{alg:perm_k_gen}    
\end{algorithm}

\vspace{-0.4cm}

\begin{algorithm}[H]
	\footnotesize
	\algsetup{linenosize=\footnotesize}
	\begin{algorithmic}[1]
		\label{algo:4}
		\STATE  \textbf{Parameters:} learning rate $\gamma>0$, start iterate $x^0\in\R^d$ \newline
		All clients negotiate a secret key ${\color{blue}sk}$.	All clients and master negotiate a ${seed}$ for pseudo-random number generator (\algname{PRG}).
		\FOR {$k=0,1,2, \ldots$}
		\FOR {{\bf all workers $i \in \{1,2,\dots,n\}$ in parallel}}
		\STATE {Generate permutation compressor $\cC_i^k(\cdot; {seed})$ for round $k$ at worker $i$ with Algorithm~\ref{alg:perm_k_gen} using the known $seed$.}
		\STATE {Compute and compress local gradient $g_i^k = \cC_i^k(\nabla f_i(x^k))$}
		\STATE {Represent $g_i^k \in \mathbb{R}^d$ in sparse form.}
		\STATE {Send $m_i^k = ({nonce}_i, {mac}_i, \hat{g}_i^k)={\color{blue}Enc}(g_i^k,{\color{blue}sk})$ to the master.}
		\ENDFOR      
		\STATE {Master concatenates the message ${G}^k = concat(m_1^k, \dots, m_n^k)$}
		\STATE {Master broadcasts the compressed aggregate ${G}^k$ to all workers}
		\FOR {{\bf all workers $i \in \{1,2,\dots,n\}$ in parallel}}      		
		\STATE {Reconstruct indices from (\algname{PRG}) for all compressors $\mathcal{C}_1^k, \dots, \mathcal{C}_n^k$ with Algorithm~\ref{alg:perm_k_gen} using the known ${seed}$.}
		\FOR {{\bf all block of coordinates $b \in \{1,2,\dots,n\}$ in parallel}}
		\STATE Obtain part of $G^k$ corresponds to $m_b^k$
		\STATE Unpack ${nonce}_b$, ${mac}_b$, $\hat{g}_b^k$, and ${g}_b^k = {\color{blue}Decrypt}(\tilde{g}_b^k, {\color{blue}sk})$
		\STATE $Verify({\color{blue}sk}, \hat{g}_b^k, {mac}_b)$. If no - halt the training process.
		\STATE Compute the next iterate $x_b^{k+1} = {x_b^k - \frac{\gamma}{n} \cdot \hat{g}_b^k}$
		\ENDFOR
		\ENDFOR
		\ENDFOR
	\end{algorithmic}
	\caption{\footnotesize{\algnamewithaes{DCGD/PermK/AES}, $d \ge n$}}
	\label{alg:dcgd_permk_aes}    
\end{algorithm}

The clients need to send in parallel $n$ \textit{encrypted} messages to the master. Since clients trust each other, symmetric key encryption is a natural option to use. In the world of symmetric ciphers, we have decided to use a block cipher, specifically the industry-standard \aesname{AES} \citep{daemen1999aes}, instead of stream ciphers like \algname{SALSA} \citep{bernstein2008salsa20}. The choice was made due to the advantages of block ciphers, as: (i) hardware support within CPU \footnote{AES support: Intel x86 Westmere, AMD x86 Bulldozer, ARM Cortex-A53}; (ii) the provision of multiple security levels ($128,192,256$ bits); (iii) the necessary flexibility to work with available parts of the vector $G^k$ (In coming Algorithm \ref{alg:dcgd_permk_aes}). The \aesname{AES} block-cipher provides security for a single 16-byte block. The \textit{Modes of Operation} provide a way to use \aesname{AES} for more than one block. \textit{Message Authentication Code} (\abr{MAC}) provides guarantees from tampering. In our work, we used \aesname{AES/EAX} mode of operation. For details about \aesname{AES} see Appendix  \ref{app:aes_details}.

The naive way to use \aesname{AES} is to use Algorithm \ref{alg:dcgd}, \myblue{(Option A)}. It solves problems with an untrusted server and channels. Without knowing $sk$, the untrusted party cannot join the training procedure. Due to the use of \abr{MAC} (Lines 5, 13), the training process verifies the integrity and protects against malicious attacks. To provide semantic security against a chosen-plaintext attack (\attackname{CPA}) and have the ability to use $sk$ in a distributed way, we employ random $nonce$ based selection. It eliminates the need for client coordination in a $nonce$ selection if $nonce$ is big enough such as $128$ bits. Details about the role of $nonce$ in block ciphers are presented in Appendix~\ref{app:aes_details}.

With this Algorithm~\ref{alg:dcgd} \myblue{Option-A}, solves (A)--(E). However, this strategy has two big downsides: (1) Computation at Lines 13 and 14 is repeated by all clients with complexity per client $\mathcal{O}(dn)$; (2) Amount of information from the master in Line 9 is $\mathcal{O}(dn)$, not $\mathcal{O}(d)$. We pay this price due to the use of \aesname{AES}, which does not allow to perform aggregation in the server. From one point of view, the natural way is to perform averaging for vectors in $\mathbb{R}^d$ in the master (Line 10), but on another side, we perform bit-wise operations (Line 5) inside \aesname{AES} cipher. There is no way to connect it from the first point of view, but we have discovered how to do it!

The \compname{PermK} correlated compressor from work \citep{szlendak2021permutation} possesses a compelling property that has yet to be recognized. The correlated compressors operate interdependently for clients and the schema was outlined in Algorithm \ref{alg:perm_k_gen}. Let $\cC_1,\dots,\cC_n:\R^d\to \R^d$ be randomized \compname{PermK} compressors, then $$\Exp{\frac{1}{n}\sum_{i=1}^n \cC_i(v_i)} = \frac{1}{n}\sum_{i=1}^n v_i, \forall v_i \in \R^d.$$

If $d \ge n$, then $\frac{1}{n}\sum_{i=1}^n \cC_i(v_i)$ fulfills the variance bound: \newline $$\Exp{\norm{ \frac{1}{n}\sum_{i=1}^n \cC_i(v_i) - \frac{1}{n}\sum_{i=1}^n v_i}^2} \leq \frac{1}{n}\sum_{i=1}^n \norm{v_i}^2 - \norm{\frac{1}{n}\sum_{i=1}^n v_i}^2$$. 

This compressor is used in our \algname{DCGD/PermK} Algorithm \ref{alg:dcgd_permk_aes}. Its unbiasedness property ensures the absence of systematic errors and the variance bound guarantees that if we are in an overparameterized setting and $x^k \to x^*$, then the estimator's variance will decay to zero, which should remove any oscillation behavior near the solution. The crucial property of \algname{DCGD/PermK} is that the aggregation of compressed gradients can be replaced by concatenation, which is algebraic monoid for $(\mathbb{R}^*, \mathrm{concat})$, and not the algebraic group. Scaling an encrypted vector by $\frac{1}{n}$ is unfeasible in the master, but this can be delegated to the iterate update Algorithm \ref{alg:dcgd_permk_aes}, Line 17. We can use non-secure pseudo-random generators as \citep{matsumoto1998mersenne} for sampling indices because they are independent of the input. 
\section{The Resilience for Attacks}
\label{sed:resilience}

In chosen-plaintext-attack (\attackname{CPA}), an attacker may adaptively ask for the encryption $(e_1, \dots)$ of arbitrary messages $(m_1,\dots)$ to obtain the ability to correctly guess from the set of two encrypted cipher-texts $\{E_1, E_2\}$ which encryption belongs to which message in the set $\{M_1, M_2\}$, given that attacker knows the set $\{M_1, M_2\}$. \ecryptname{CKKS} and \aesname{AES} in \aesname{EAX} mode of operation \citep{bellare2004eax} are secure against \attackname{CPA}. 

In a chosen-ciphertext-attack (\attackname{CCA}), the attacker has unlimited access to a decryption oracle, with the same goal as in \attackname{CPA}. All \abr{HE} schemas cannot achieve \attackname{CCA} security \citep{fauzi2022ind}. It has been proved that \aesname{AES}/\aesname{EAX} is secure against \attackname{CCA} \citep{bellare2004eax}. 

The algorithms executed in general-purpose microprocessors from an execution point of view are a stream of instructions. However, the energy consumption for different operations is different  \citep{horowitz20141}. It leads to the possibility of side-channel attacks, where the attacker can use information about physical leakages when the number (or type) of operations is some function of secret key or plaintext. The \aesname{AES} block cipher and \textit{Operation Modes} perform the same stream of operation types (see Appendix~\ref{app:aes_details}). In \ecryptname{CKKS} polynomial multiplication, encoding, decoding, and randomization can lead to a trace to plain message \citep{aydin2022reveal} using side-channel attacks. During the past time from standardization of \aesname{AES} the works targeted to protect \aesname{AES} from side-channel attacks have been carried \citep{rahaman2008side}, \citep{gross2017efficient}. 

Protecting \abr{HE} libraries from side-channel attacks is an open question. For example, according to \citep{aydin2022reveal} the \abr{HE} implementation in a reference \libname{SEAL} library \citep{chen2017simple} and any derived libraries such as \libname{TenSEAL} \citep{benaissa2021tenseal} requires protection against side-channel attacks. The work describes how to exploit an asymmetry in ciphertext generation and decrease the security level from $2^{128}$ to $2^{4.4}$ for Brakerski/Fan-Vercauteren(\abr{BFV}) \citep{fan2012somewhat} Ring Learning With Error (\abr{RLWE}) (see Appendix~\ref{app:lwe}). Therefore, protection from side-channel attacks for \aesname{AES} can be considered to be more well-developed.


\section{Experiments}

We made an experimental comparison of several optimization algorithms with different compression and security methods. In Table~\ref{tbl:list_of_optimization_algos} presented in Appendix~\ref{app:list_of_optimization_algos} we summarized their qualitative aspects.

\subsection{Synthetic Experiments}
\label{sec:syntetic_exp}

We're going to illustrate the advantages of employing block cipher \aesname{AES} (see Appendix \ref{app:aes_details}), as opposed to the \ecryptname{CKKS} (see Appendix \ref{app:ckks_details}). The experiments were conducted on a simulated environment using \libname{FL\_PyTorch} \citep{burlachenko2021fl_pytorch}. Details on the computing environment are in Appendix~\ref{app:reconstruction}. We configured the \ecryptname{CKKS} to offer security guarantees as \aesname{AES} with $128$ bit ($16$ bytes) key. In this case, the size of public and private keys for \ecryptname{CKKS} has a lower bound $420\,000$ bytes (see Appendix~\ref{app:ckks_details} for underlying reason). During usage of \ecryptname{CKKS}, the public key should be reported somehow to the master once. As we will see while using \algnamewithaes{DCGD/PermK/AES} no key at all should be reported to master for master to operate. Therefore the key size can be a problem for \ecryptname{CKKS} already when the volume of communicated information is far smaller than $0.42 \cdot 10^6$ Bytes. In our experiments, we ignore the overhead from key negotiations between parties.

\paragraph{Optimization Problem and Experimental Setup}
In our synthetically controlled experiments, we consider a specific smooth convex optimization problem which is obtained from Equation \ref{eq:main} via $f_i(x) = \frac{1}{n_i} \|A_i x - b_i\|^2$, $A_i \in \mathbb{R}^{n_i \times d}, b_i \in \mathbb{R}^{d}$.

\begin{figure*}[t]
	\centering             
	
	\captionsetup[sub]{font=footnotesize,labelfont={},labelformat=empty}		
	\captionsetup[subfigure]{font=footnotesize,labelfont={},labelformat=empty}
	\captionsetup[figure]{font=footnotesize,labelfont={},labelformat=empty}
	
	\begin{subfigure}[ht]{0.99\textwidth}
		\includegraphics[width=\textwidth]{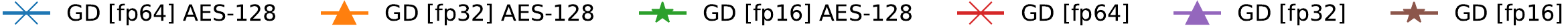} 
	\end{subfigure}
	
	\begin{subfigure}[ht]{0.475\textwidth}
		\includegraphics[width=\textwidth]{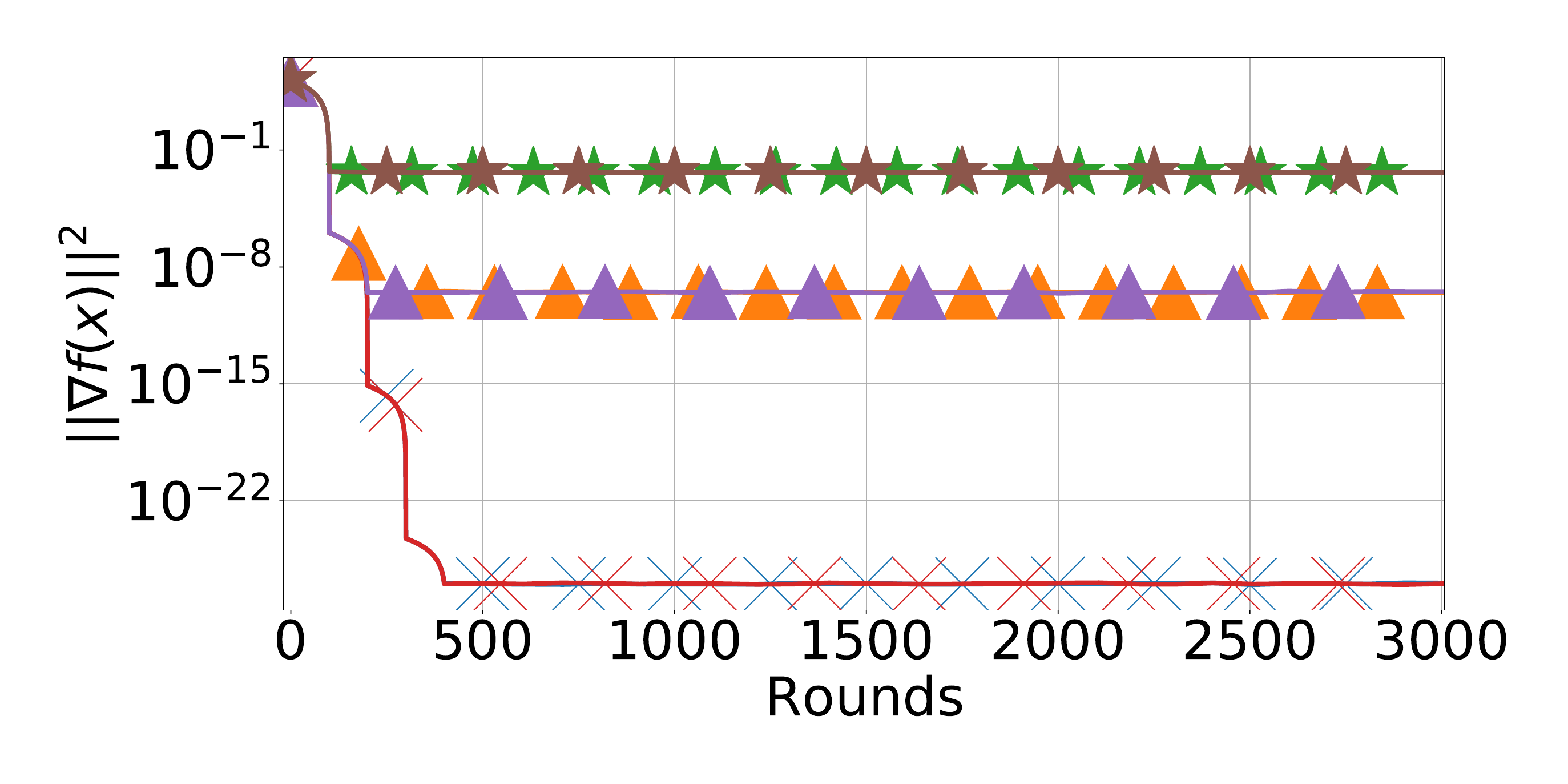} 
		\vspace{-1.5\baselineskip}
		\caption{{ (a) }}
	\end{subfigure}
	\begin{subfigure}[ht]{0.475\textwidth}
		\includegraphics[width=\textwidth]{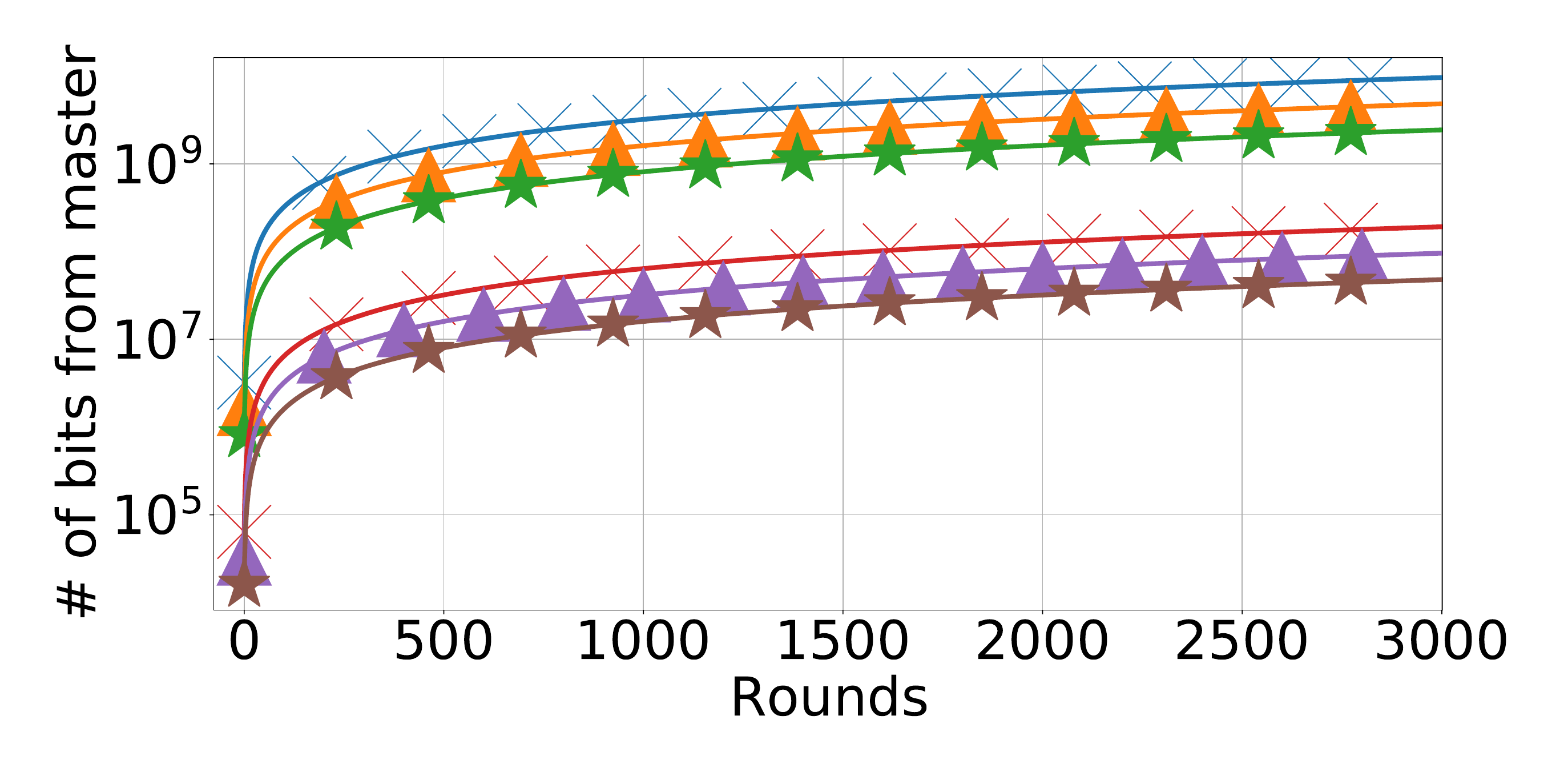} 
		\vspace{-1.5\baselineskip}
		\caption{{ (b) }}
	\end{subfigure}
	\begin{subfigure}[ht]{0.475\textwidth}
		\includegraphics[width=\textwidth]{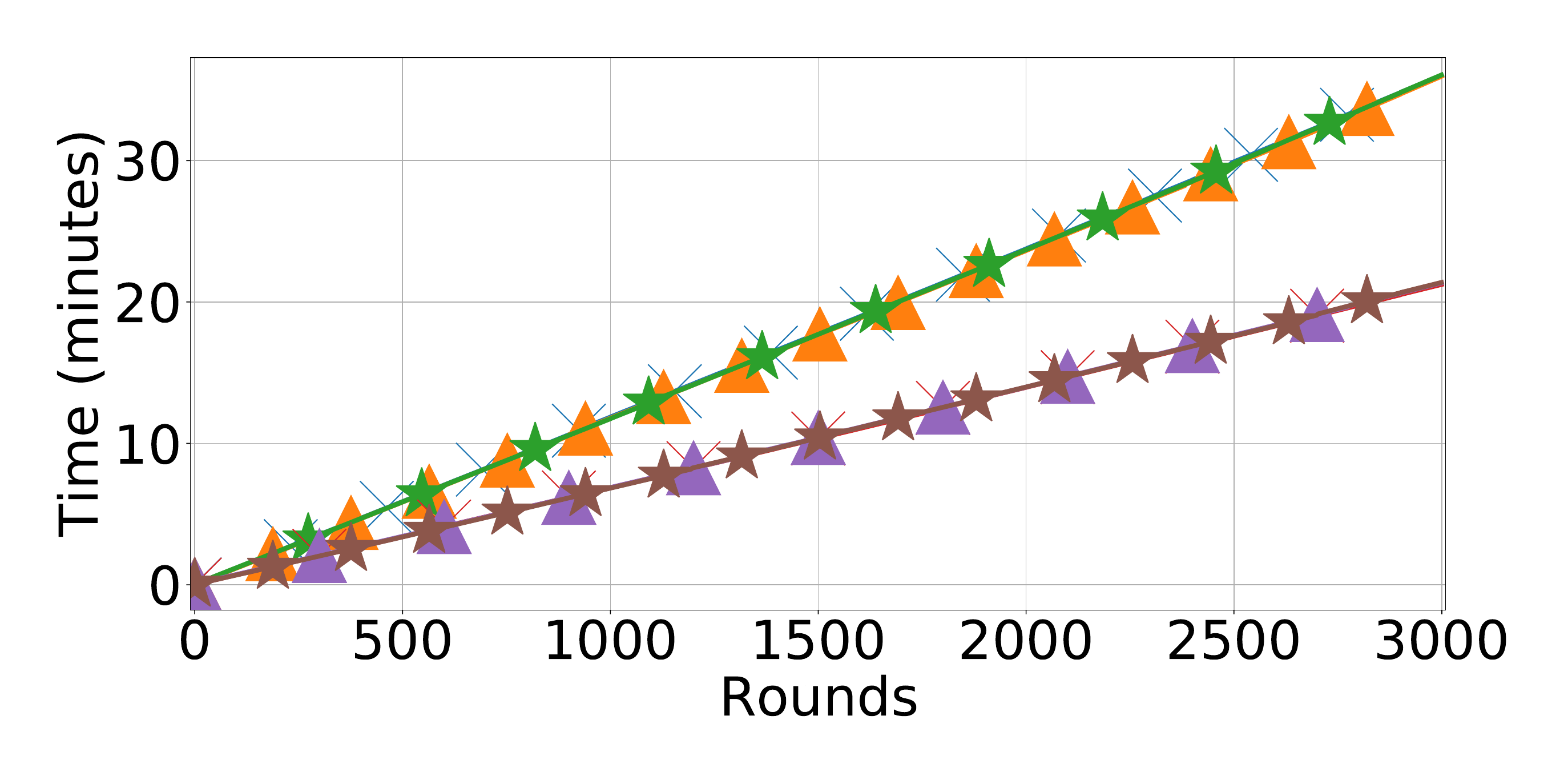} 
		\vspace{-1.5\baselineskip}
		\caption{{ (c) }}
	\end{subfigure}
	
	\caption{{Synthesized \modelname{Linear Regression} in interpolation mode, $n_i=12$, $n=50$, $d=1000$. No compression. Th. step sizes.}}
	\label{fig:exp_syn_1}
\end{figure*}

\begin{figure*}[t]
	\centering
	\captionsetup[sub]{font=footnotesize,labelfont={},labelformat=empty}		
	\captionsetup[subfigure]{font=footnotesize,labelfont={},labelformat=empty}
	\captionsetup[figure]{font=footnotesize,labelfont={},labelformat=empty}
	
	\begin{subfigure}[ht]{0.99\textwidth}
		\includegraphics[width=\textwidth]{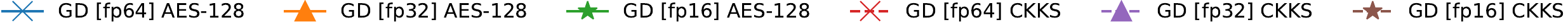} 
	\end{subfigure}
	
	\begin{subfigure}[ht]{0.475\textwidth}
		\includegraphics[width=\textwidth]{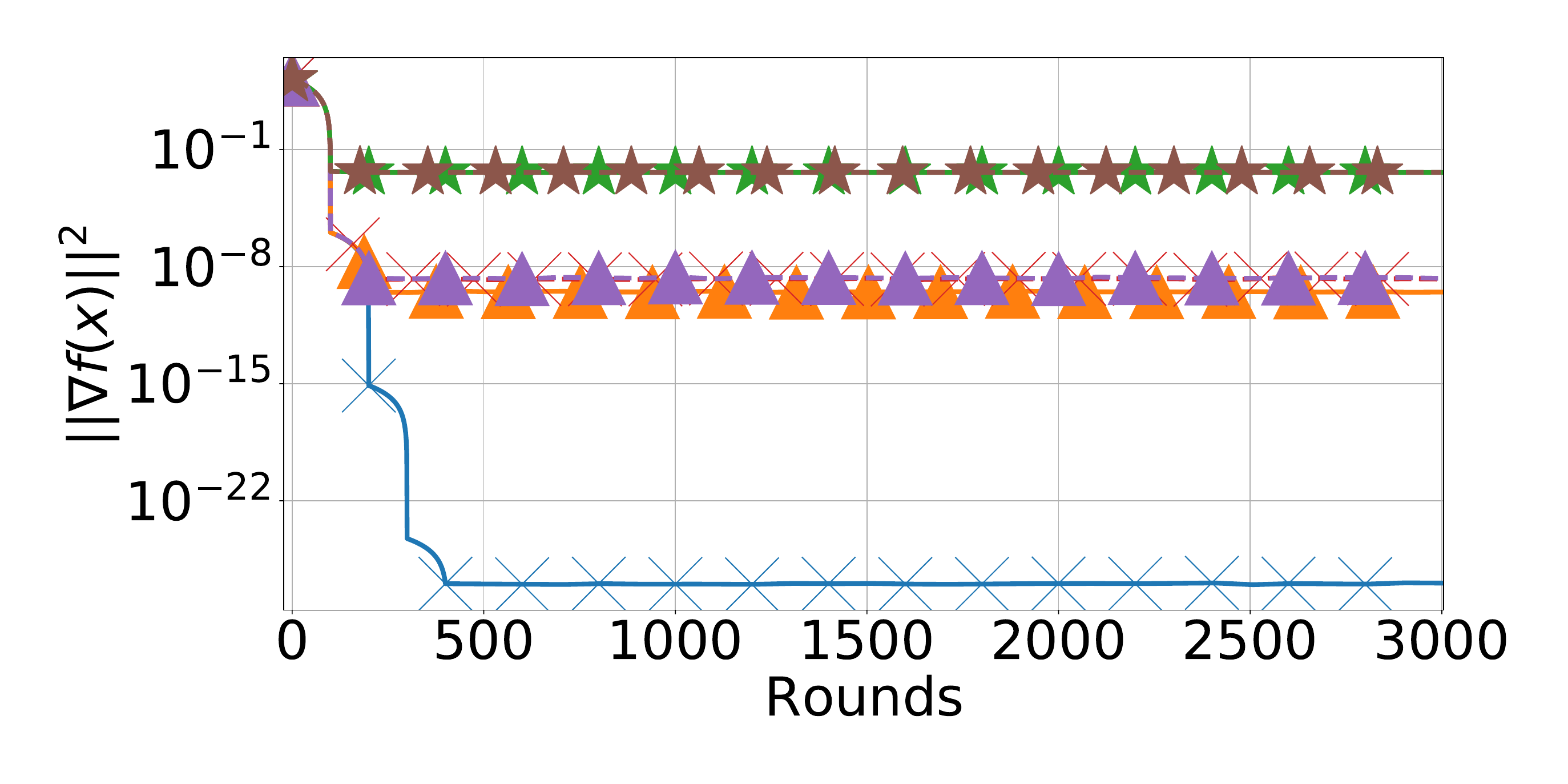} 
		\vspace{-1.5\baselineskip}
		\caption{{ (a) }}
	\end{subfigure}
	\begin{subfigure}[ht]{0.475\textwidth}
		\includegraphics[width=\textwidth]{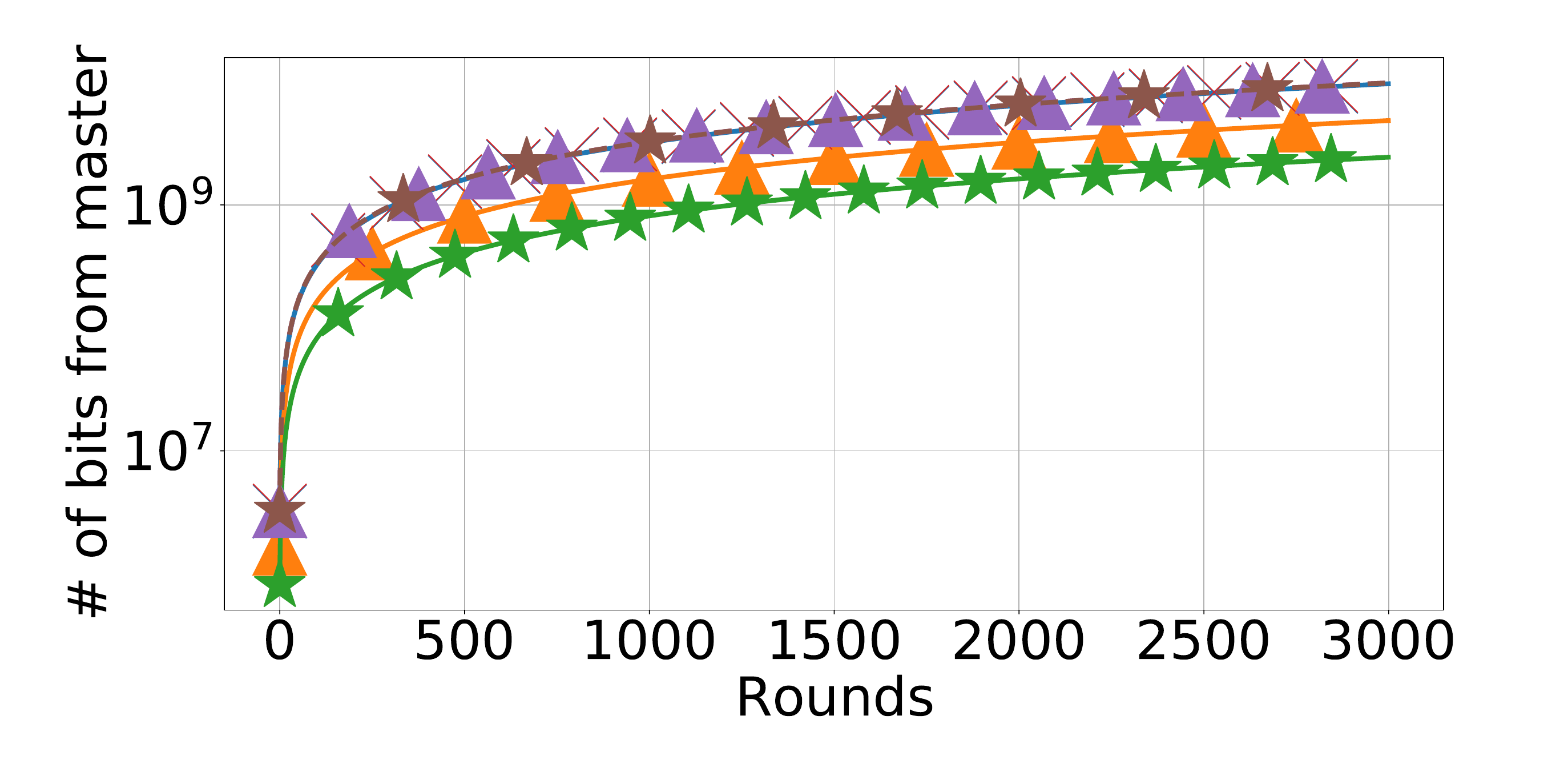} 
		\vspace{-1.5\baselineskip}
		\caption{{ (b) }}
	\end{subfigure}
	\begin{subfigure}[ht]{0.475\textwidth}
		\includegraphics[width=\textwidth]{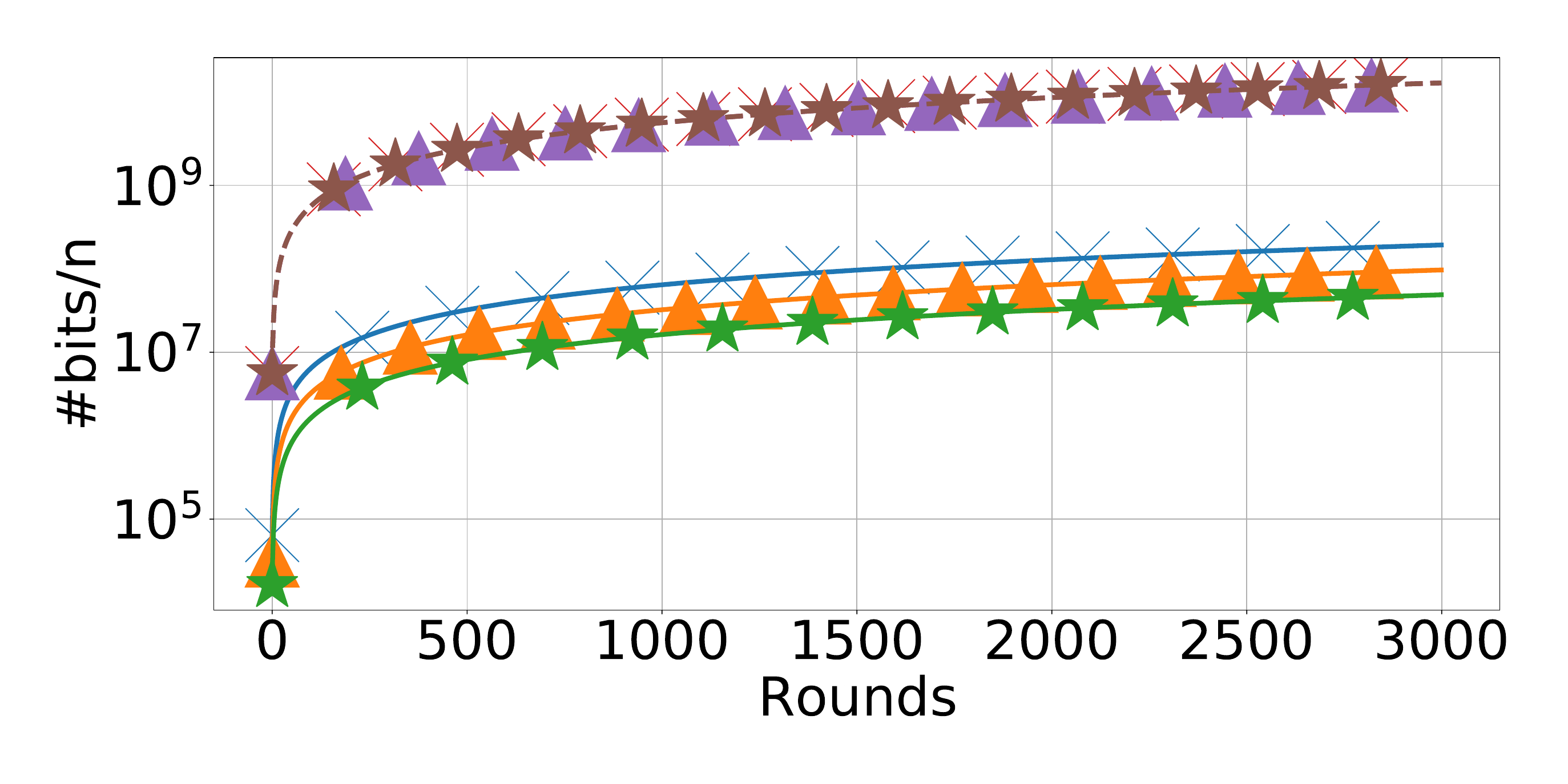} 
		\vspace{-1.5\baselineskip}
		\caption{{ (c) }}
	\end{subfigure}
	\begin{subfigure}[ht]{0.475\textwidth}
		\includegraphics[width=\textwidth]{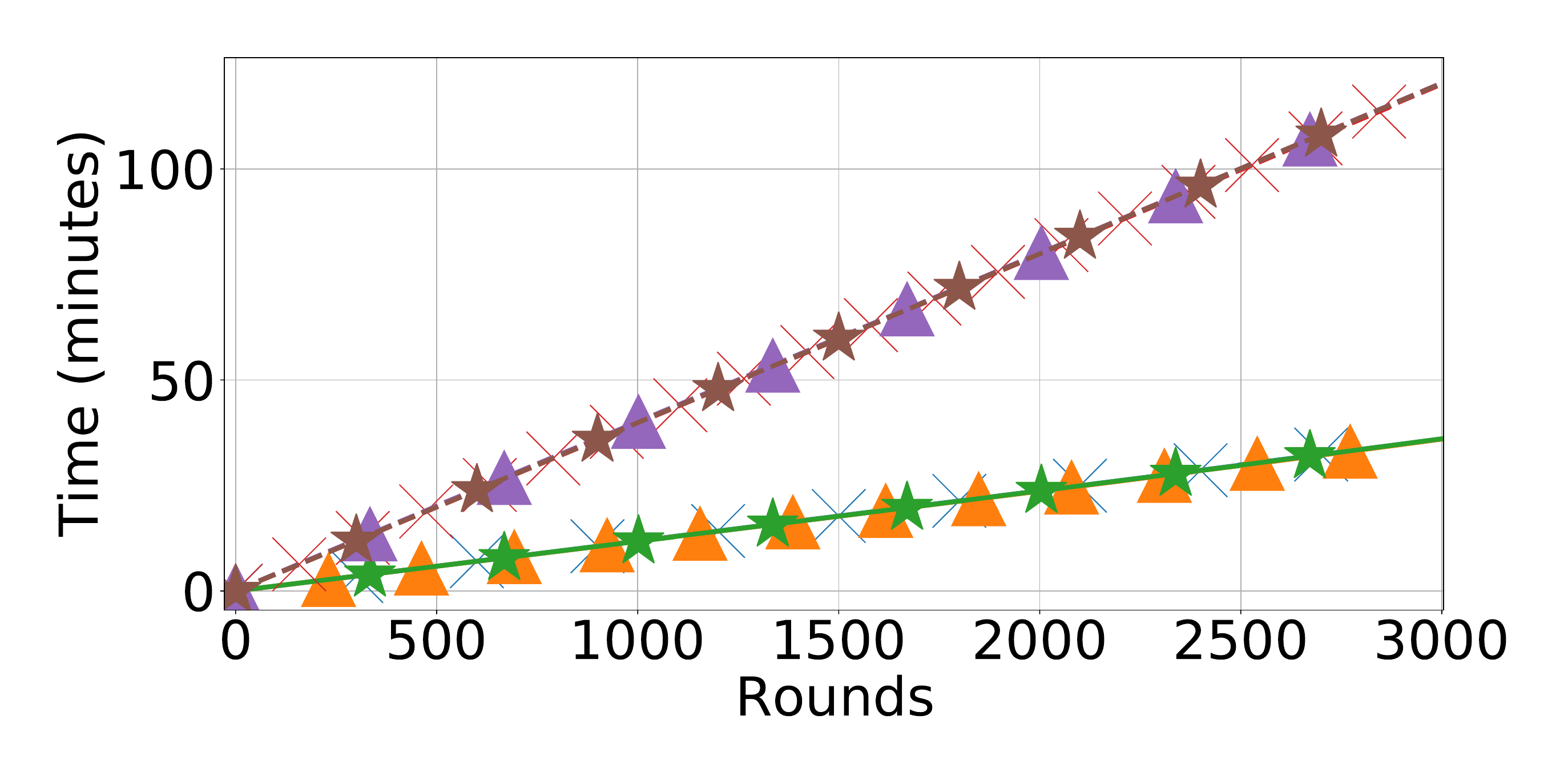} 
		\vspace{-1.5\baselineskip}
		\caption{{ (d) }}
	\end{subfigure}
	
	\caption{{Synthesized \modelname{Linear Regression} in interpolation mode, $n_i=12$, $n=50$, $d=1000$. No compression. Th. step sizes.}}
	\label{fig:exp_syn_2}
\end{figure*}

\paragraph{Case 1: Distributed GD with/without AES/CKKS} We conducted experiments with $d=1000$, $n=50$, and $n_i=12$. In our designed experimental setup we filled the Hessian $\nabla ^2 f(x)$ such that its nonzero eigenvalues lie uniformly in $[1.0, 10.0]$, therefore $L_f=10$. We used the maximum theoretical constant step size $\gamma=1/L_f$ for \algname{GD}. Fig.~\ref{fig:exp_syn_1} (a) shows the impact of \texttt{IEEE-754 FP16}, \texttt{FP32}, and \texttt{FP64} formats. It shows that \algnamewithaes{GD/AES} does not hurt float arithmetic. However, \algnamewithaes{GD/AES} increases traffic between the master and clients by a factor of $n$ as we see from  Fig.~\ref{fig:exp_syn_1} (b) and increases wall clock time by a factor of 1.7 as in Fig.~\ref{fig:exp_syn_1} (c). Next, Fig.~\ref{fig:exp_syn_2} compares \algname{GD} with \aesname{AES} and \ecryptname{CKKS}. According to Fig.~\ref{fig:exp_syn_2} (a), \algname{GD/CKKS} and FP64 arithmetic, it's possible to obtain $| \nabla f(x^k)|^2 \approx 10^{-9}$, when \algnamewithaes{GD/AES} attains $| \nabla f(x^k)|^2 \approx 10^{-23}$. From Fig.~\ref{fig:exp_syn_2} (b), (c) we see that \algname{GD/CKKS} increases load from master  slightly compared to \algnamewithaes{GD/AES}, but increases the load from clients by $\times 10^4$. As per Fig.~\ref{fig:exp_syn_2} (d), \ecryptname{CKKS} is approximately $\times 3$ slower compared to \aesname{AES}. Using \algnamewithaes{GD/AES} is reasonable only if  $d$ and $n$ are small.

\begin{figure*}[t]
	\centering
	\captionsetup[sub]{font=footnotesize,labelfont={},labelformat=empty}		
	\captionsetup[subfigure]{font=footnotesize,labelfont={},labelformat=empty}
	\captionsetup[figure]{font=footnotesize,labelfont={},labelformat=empty}
	
	\begin{subfigure}[ht]{0.85\textwidth}
		\includegraphics[width=\textwidth]{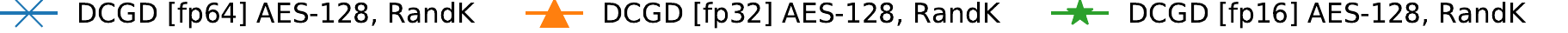}
	\end{subfigure}
	\begin{subfigure}[ht]{0.65\textwidth}
		\includegraphics[width=\textwidth]{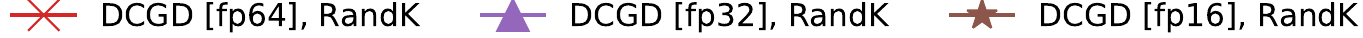}
	\end{subfigure}
	
	\begin{subfigure}[ht]{0.475\textwidth}
		\includegraphics[width=\textwidth]{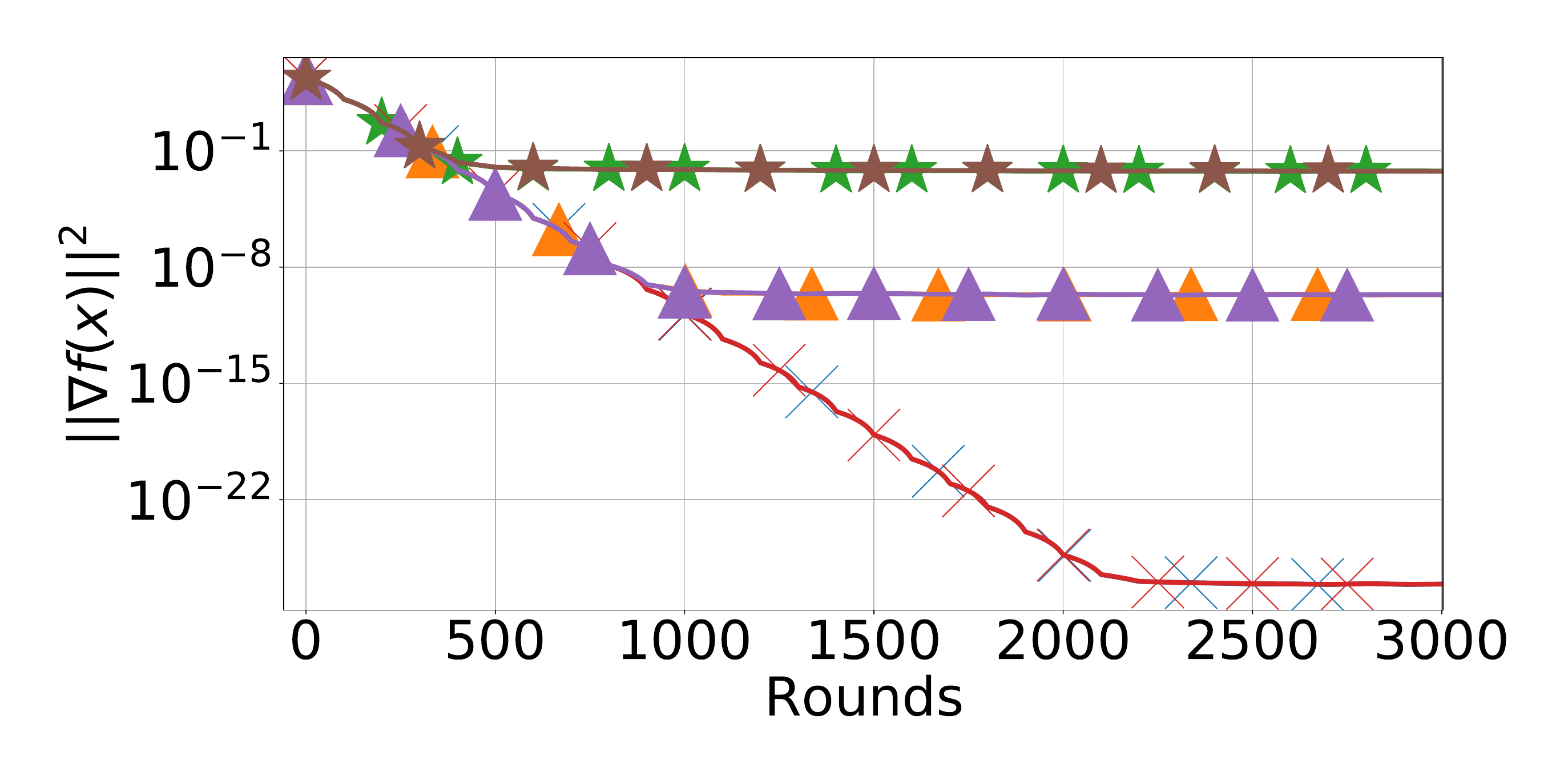} 
		\vspace{-1.5\baselineskip}
		\caption{{ (a) }}
	\end{subfigure}
	\begin{subfigure}[ht]{0.475\textwidth}
		\includegraphics[width=\textwidth]{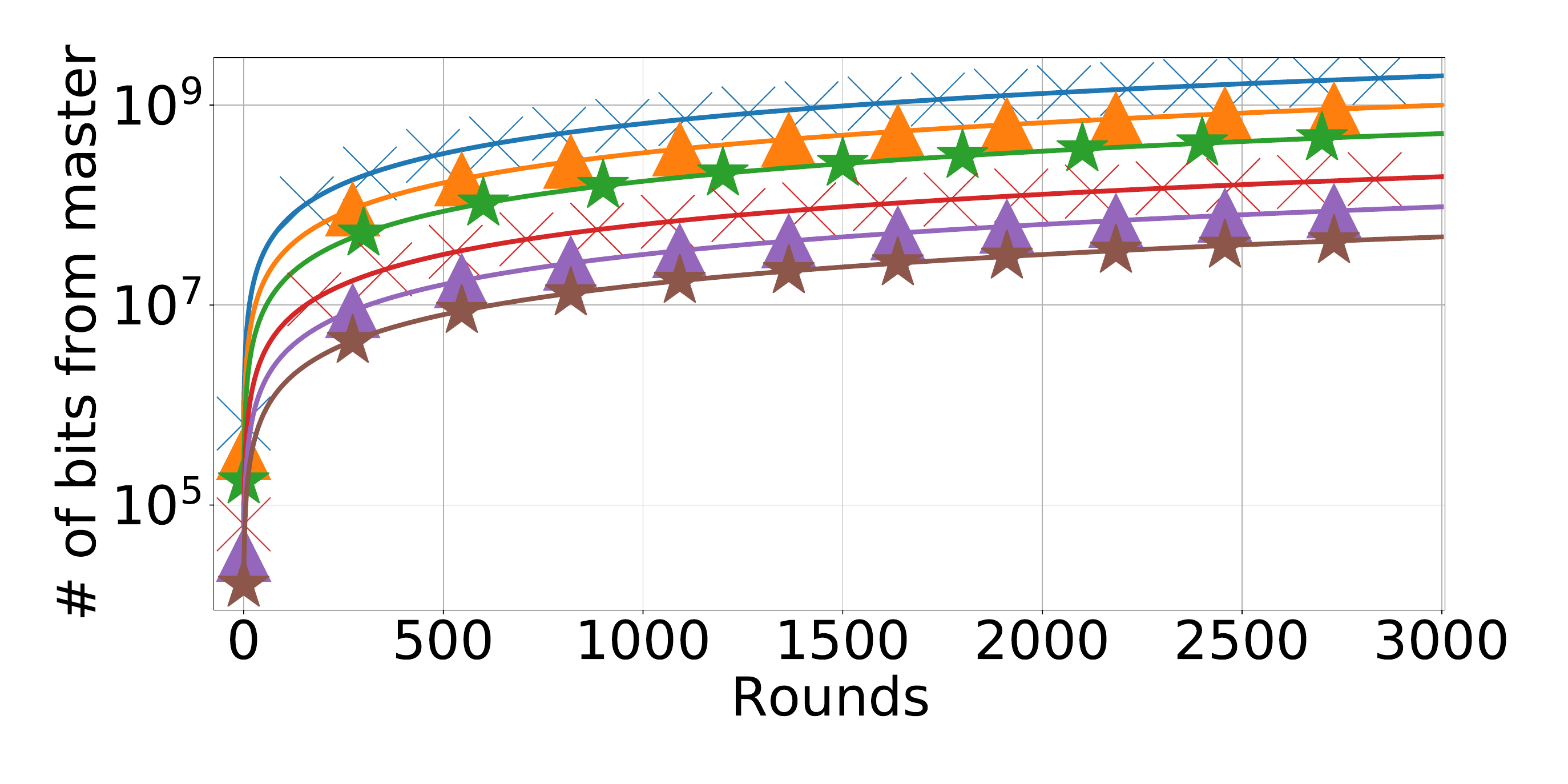} 
		\vspace{-1.5\baselineskip}
		\caption{{ (b) }}
	\end{subfigure}
	\begin{subfigure}[ht]{0.475\textwidth}
		\includegraphics[width=\textwidth]{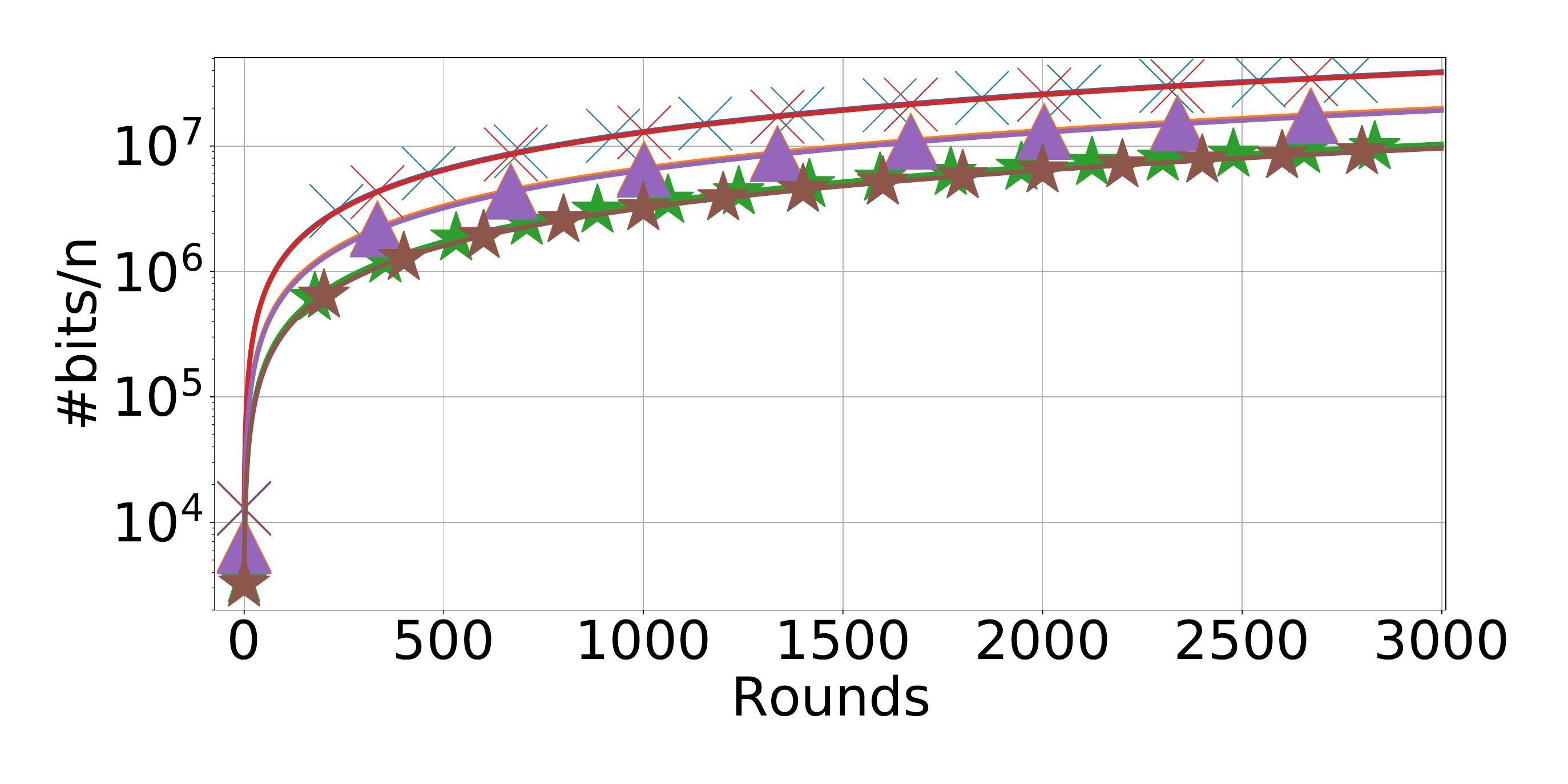} 
		\vspace{-1.5\baselineskip}
		\caption{{ (c) }}
	\end{subfigure}
	
	\caption{{Synthesized \modelname{Linear Regression} in interpolation, $n_i=12$, $n=50$, $d=1000$. Compressors: \compname{RandK}[$K=0.2d$]. Th. step sizes.}}
	\label{fig:exp_syn_3}
\end{figure*}

\begin{figure*}[t]
	\centering
	\captionsetup[sub]{font=footnotesize,labelfont={},labelformat=empty}		
	\captionsetup[subfigure]{font=footnotesize,labelfont={},labelformat=empty}
	\captionsetup[figure]{font=footnotesize,labelfont={},labelformat=empty}
	
	\begin{subfigure}[ht]{0.85\textwidth}
		\includegraphics[width=\textwidth]{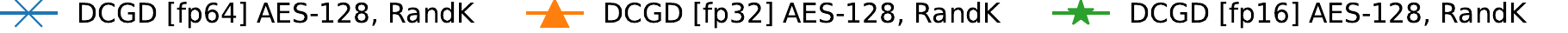}
	\end{subfigure}
	\begin{subfigure}[ht]{0.75\textwidth}
		\includegraphics[width=\textwidth]{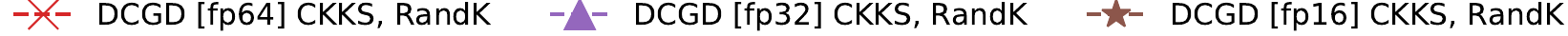}
	\end{subfigure}

	\begin{subfigure}[ht]{0.475\textwidth}
		\includegraphics[width=\textwidth]{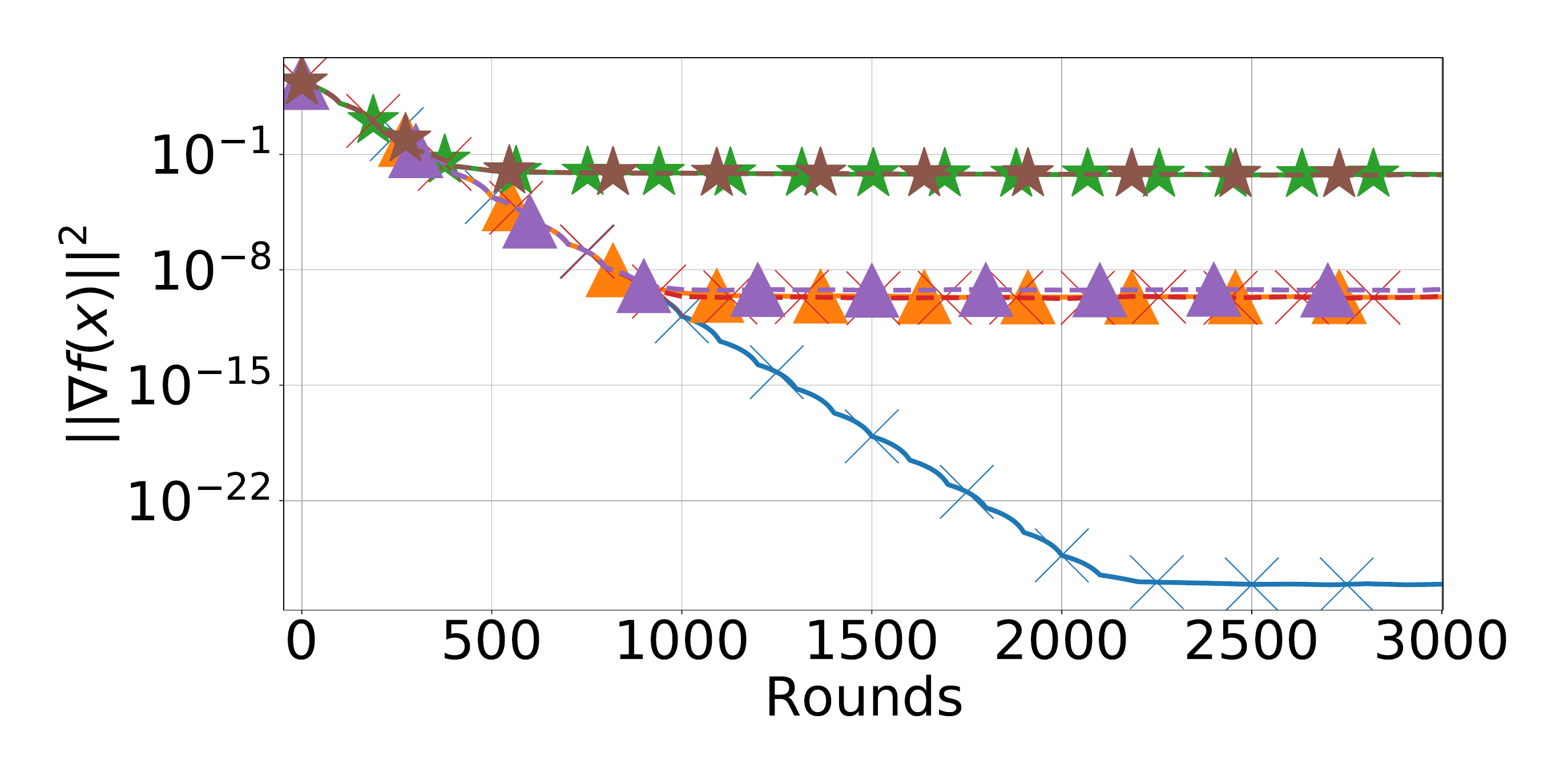} 
		\vspace{-1.5\baselineskip}
		\caption{{ (a) }}
	\end{subfigure}
	\begin{subfigure}[ht]{0.475\textwidth}
		\includegraphics[width=\textwidth]{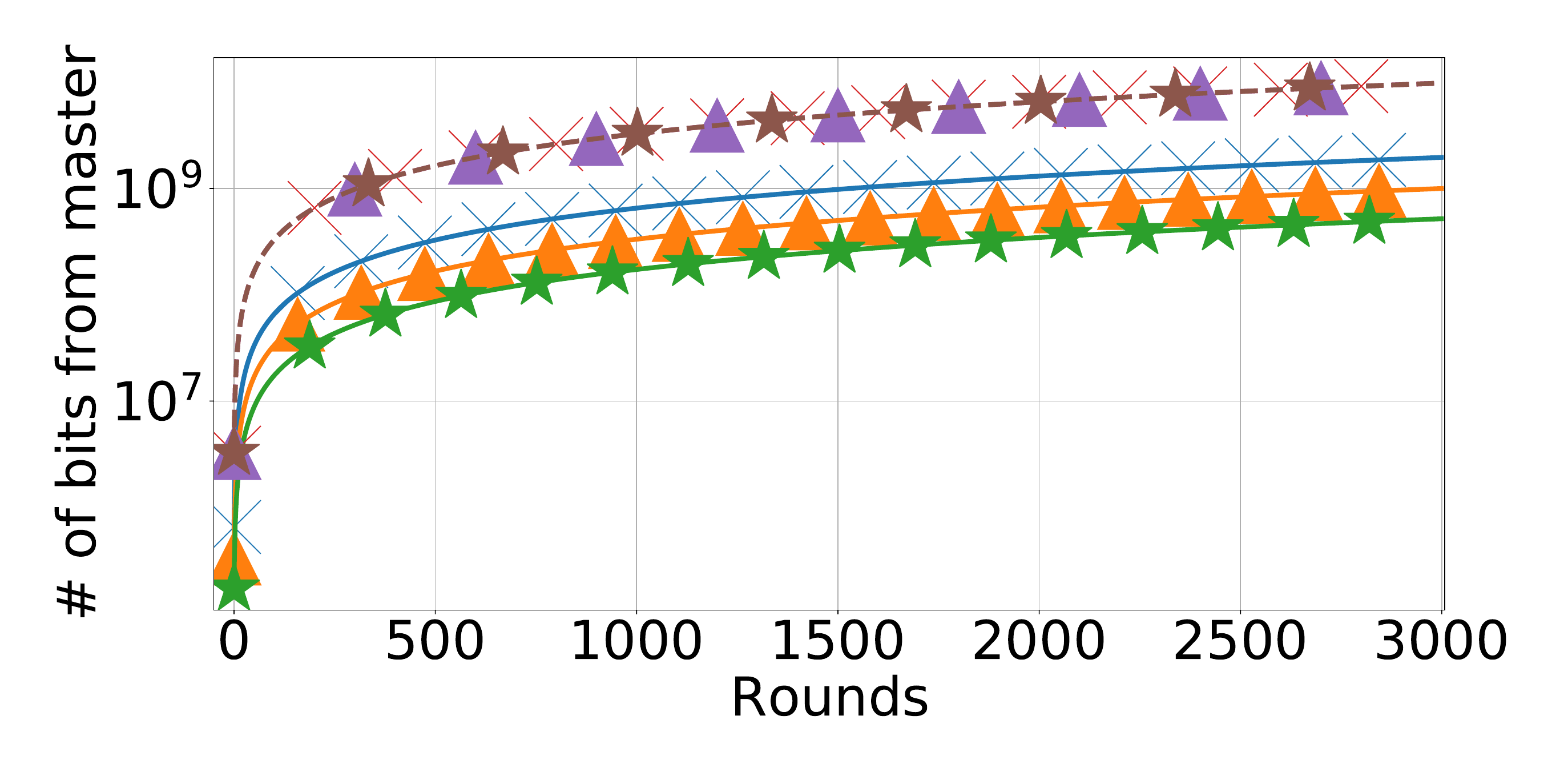} 
		\vspace{-1.5\baselineskip}
		\caption{{ (b) }}
	\end{subfigure}
	\begin{subfigure}[ht]{0.475\textwidth}
		\includegraphics[width=\textwidth]{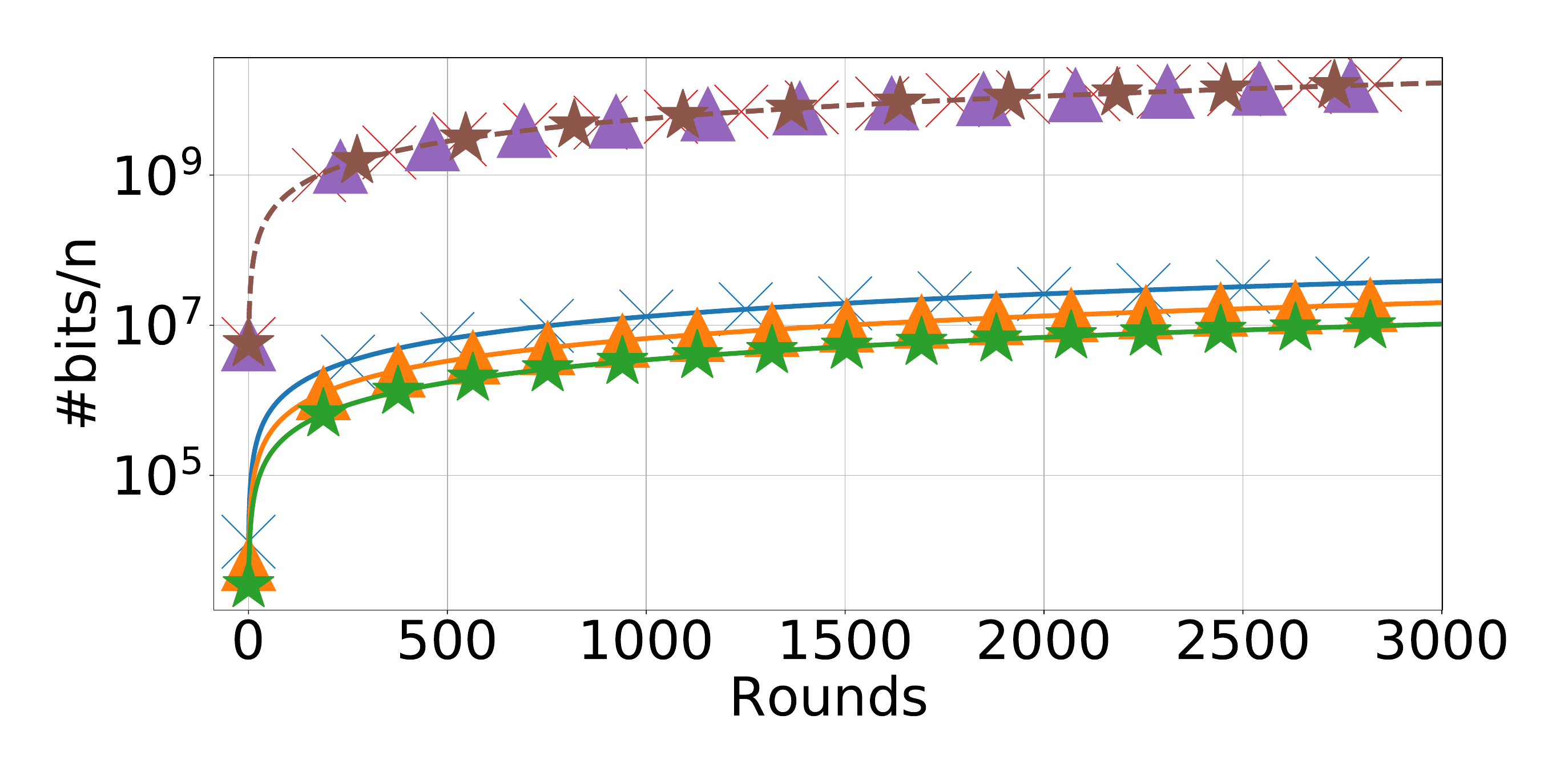} 
		\vspace{-1.5\baselineskip}
		\caption{{ (c) }}
	\end{subfigure}
	\begin{subfigure}[ht]{0.475\textwidth}
		\includegraphics[width=\textwidth]{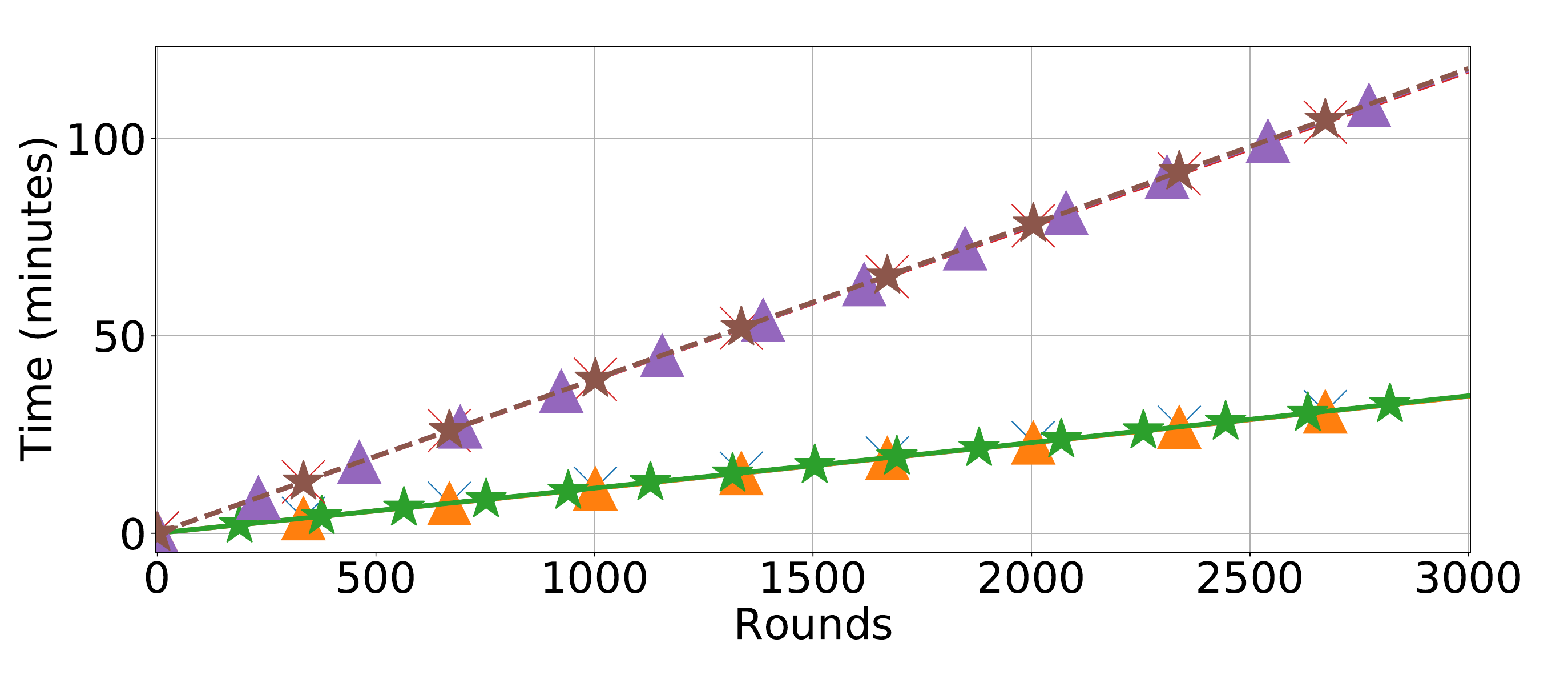} 
		\vspace{-1.5\baselineskip}
		\caption{{ (d) }}
	\end{subfigure}
	
	\caption{{Synthesized \modelname{Linear Regression} in interpolation, $n_i=12$, $n=50$, $d=1000$. Compressors: \compname{RandK} [$K=0.2d$]. Th. step sizes.}}
	\label{fig:exp_syn_4}
\end{figure*}

\paragraph{Case 2: DCGD with AES/CKKS} In this experiment, we employ \algname{DCGD} with \compname{RandK} sparsification compressor to analyze the possibility of gradient sparsification while preserving client's privacy from the master. To compress $\nabla f_i(x)$ each client creates a set $S_i\subset\{1,2,\dots,d\}$ of size $K$ chosen uniformly at random, and compute $C(\nabla f_i(x)) \eqdef \frac{d}{K} \sum_{j \in S_i} [\nabla f_i(x)]_j \cdot e_j$, where $e_j$ are unit vectors of standard basis of $\mathbb{R}^d$. Results are presented in Fig.~\ref{fig:exp_syn_3}, and Fig.~\ref{fig:exp_syn_4}. Fig.~\ref{fig:exp_syn_3} (a) and Fig.~\ref{fig:exp_syn_4} (a) demonstrates that using \aesname{AES} does not lead to numerical issues, whereas \ecryptname{CKKS} for FP64 does. In \algname{GD}, the master broadcasts a vector from $\mathbb{R}^d$ in each round. For \algnamewithaes{DCGD/RandK/AES}, if clients can reconstruct sparsified indices, then the master has to broadcast $d = K \cdot n = \frac{1}{5} \cdot n$ encrypted scalars and $32n$ bytes from employing \textit{nonce}, and \textit{mac} for privacy and integrity. This process reduces the number of bits transmitted from the master to the clients by a factor $\times 5$, compared to standard \algnamewithaes{GD/AES}, as depicted in Fig.~\ref{fig:exp_syn_3} (b) and Fig.~\ref{fig:exp_syn_1} (b). Fig.~\ref{fig:exp_syn_4} (c) shows that \ecryptname{CKKS} does not leverage the sparsity. For \ecryptname{HE} schemas, encoding of any two vectors (e.g. sparse and dense) should be indistinguishable due to semantic security requirements. However, from a computational perspective, ignoring sparsity is sometimes highly impractical, and this gap presents an open research question for \ecryptname{HE}. Fig.~\ref{fig:exp_syn_4} (b) highlights that \aesname{AES} adapts to any bit representation of scalars. In contrast, \libname{TenSEAL} \citep{benaissa2021tenseal} implementation of \ecryptname{CKKS} does not exhibit this property.

\begin{figure*}[t]
	\centering
	\captionsetup[sub]{font=footnotesize,labelfont={},labelformat=empty}		
	\captionsetup[subfigure]{font=footnotesize,labelfont={},labelformat=empty}
	\captionsetup[figure]{font=footnotesize,labelfont={},labelformat=empty}
	
	\begin{subfigure}[ht]{0.90\textwidth}
		\includegraphics[width=\textwidth]{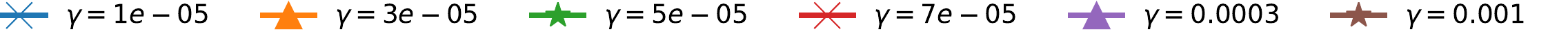} 
	\end{subfigure}

	\begin{subfigure}[ht]{0.65\textwidth}
		\includegraphics[width=\textwidth]{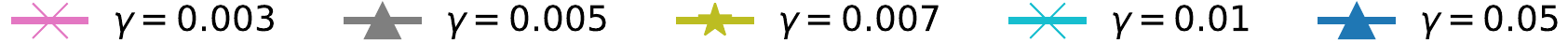} 
	\end{subfigure}
	
	\begin{subfigure}[ht]{0.475\textwidth}
		\includegraphics[width=\textwidth]{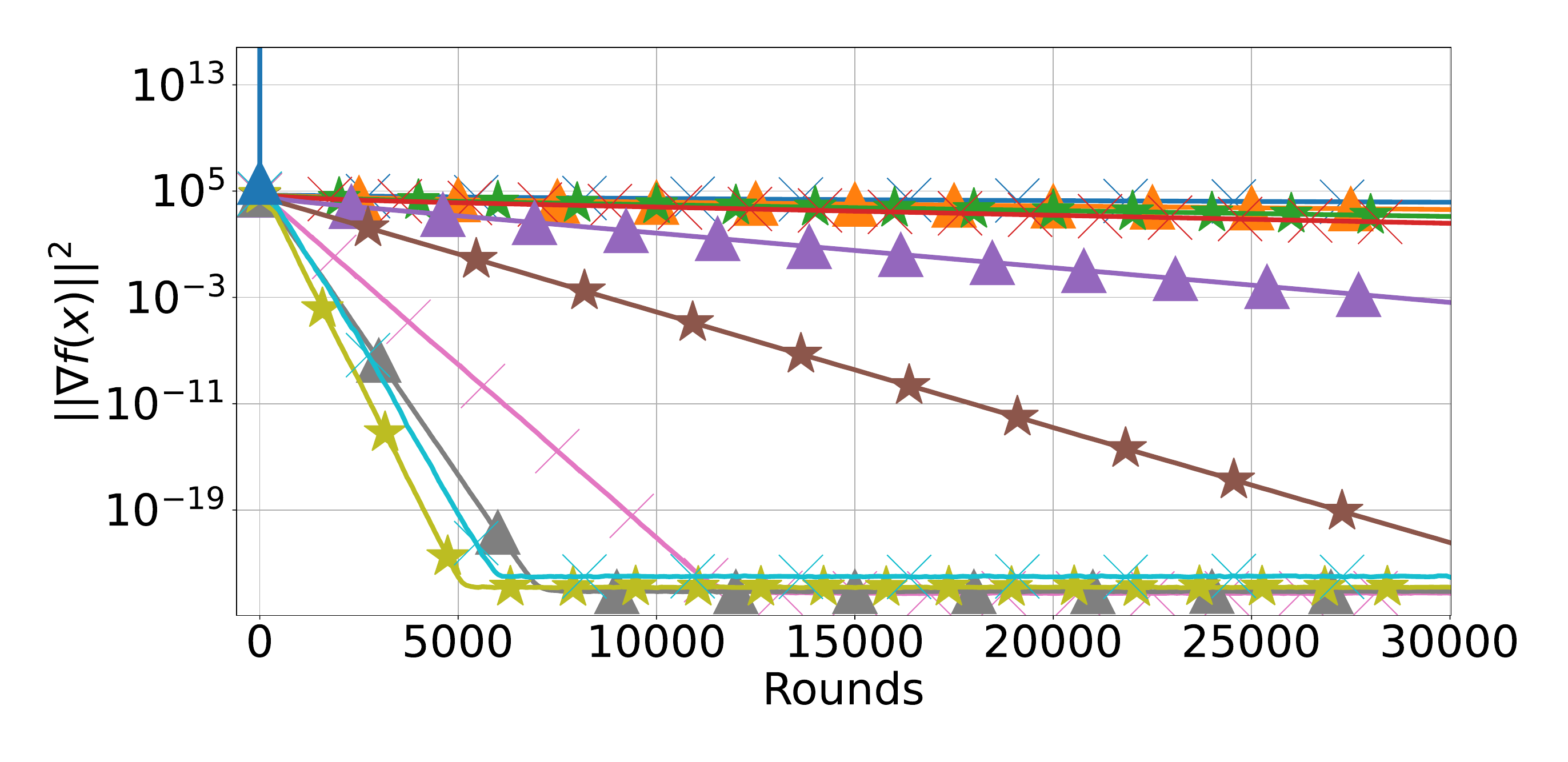}
		\vspace{-1.5\baselineskip}
		\caption{{ (a) }}
	\end{subfigure}
	\begin{subfigure}[ht]{0.475\textwidth}
		\includegraphics[width=\textwidth]{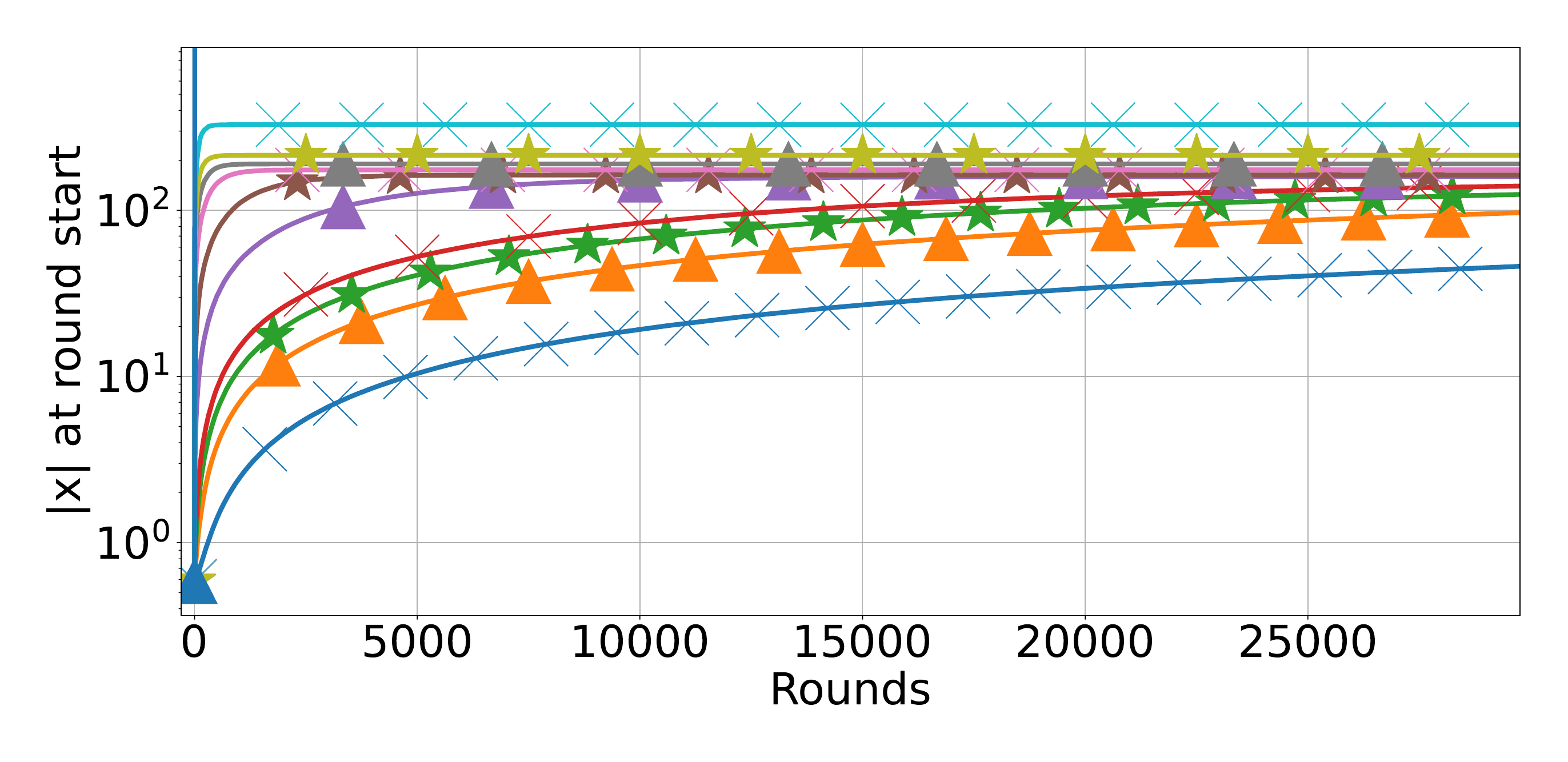}
		\vspace{-1.5\baselineskip}
		\caption{{ (b) }}
	\end{subfigure}
	\begin{subfigure}[ht]{0.475\textwidth}
		\includegraphics[width=\textwidth]{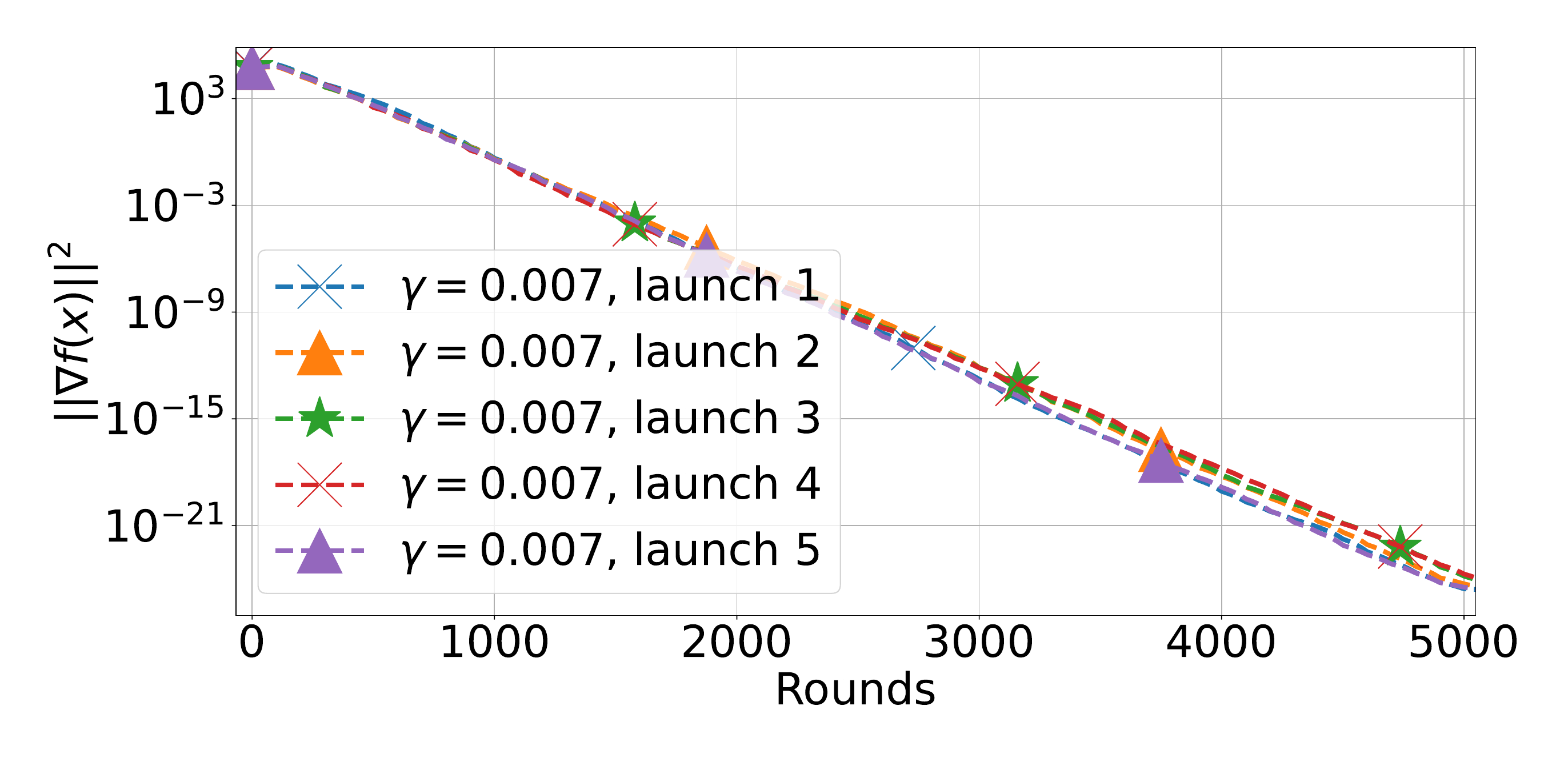}
		\vspace{-1.5\baselineskip}
		\caption{{ (c) }}
	\end{subfigure}
	
	\caption{{Tuning step size $\gamma$ without decay for \algname{DCGD/PermK}. Synthesized \modelname{Linear Regression} in interpolation, $5$ launches, FP64.}}
	\label{fig:exp_syn_5}
\end{figure*}

\paragraph{Case 3: DCGD with PermK} Both \algnamewithaes{GD/AES} and \algnamewithaes{DCGD/AES} require a significant amount of data to be sent from the master to the clients. In the case of using correlated compressors as \compname{PermK}, all clients do not intersect in supports of sparsified gradients by design. The encryption of the global direction can be obtained by concatenating encrypted messages from clients. When using \compname{PermK}, clients do not need to perform any aggregation on their side, and decryption of the whole global direction obtained from master can be done in $\mathcal{O}(d)$ independent on $n$. We aim to find an approximate $\gamma$ for \algname{DCGD/PermK}. We generated $5$ problems with matrices $A_i \sim U[0,1)^{n_i \times d}$, projected $A=[A_1, \dots, A_n]^\top$ to have $L_{f} = 10$, and computed $b_i \eqdef A_i x_{\mathrm{fixed}}$. We tested various step sizes demonstrated in Fig.~\ref{fig:exp_syn_5}. We found that using a step size $\gamma \ge \frac{1}{2L_{f}}=0.05$ led to divergence, as shown in Fig.~\ref{fig:exp_syn_5} (a), (b). From Fig.~\ref{fig:exp_syn_5} (a), we see that the method exhibits linear convergence without oscillation near the solution, similar to \algname{DCGD} with \compname{RandK} (Here $\nabla f_i (x^*)=0, \forall i \in [n]$, because $d > n_i \cdot n$). From Fig.~\ref{fig:exp_syn_5} (c), we see that the variance of the optimization path using fixed step size $\gamma=0.007$ and fixed $d, n_i, n, L_f$ is negligible.

\begin{figure*}[t]
	\centering
	\captionsetup[sub]{font=footnotesize,labelfont={},labelformat=empty}		
	\captionsetup[subfigure]{font=footnotesize,labelfont={},labelformat=empty}
	\captionsetup[figure]{font=footnotesize,labelfont={},labelformat=empty}
	
	\begin{subfigure}[ht]{0.85\textwidth}
		\includegraphics[width=\textwidth]{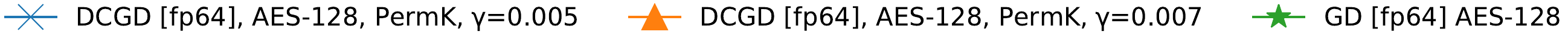}
	\end{subfigure}
	\begin{subfigure}[ht]{0.85\textwidth}
		\includegraphics[width=\textwidth]{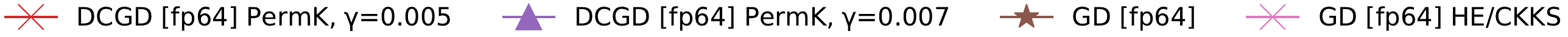} 
	\end{subfigure}
	
	\begin{subfigure}[ht]{0.475\textwidth}
		\includegraphics[width=\textwidth]{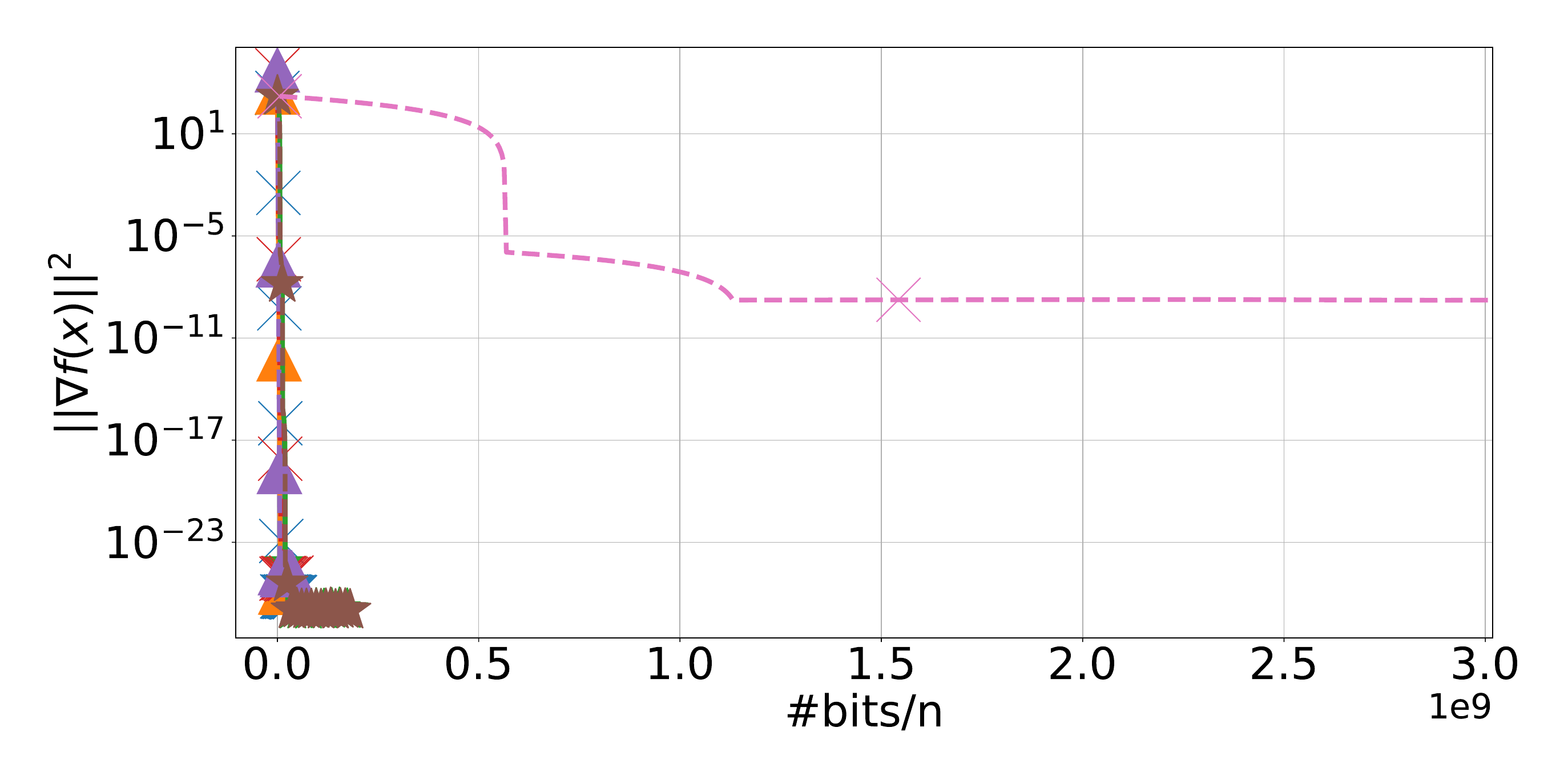} 
		\caption{{ (a) }}
	\end{subfigure}
	\begin{subfigure}[ht]{0.475\textwidth}
		\includegraphics[width=\textwidth]{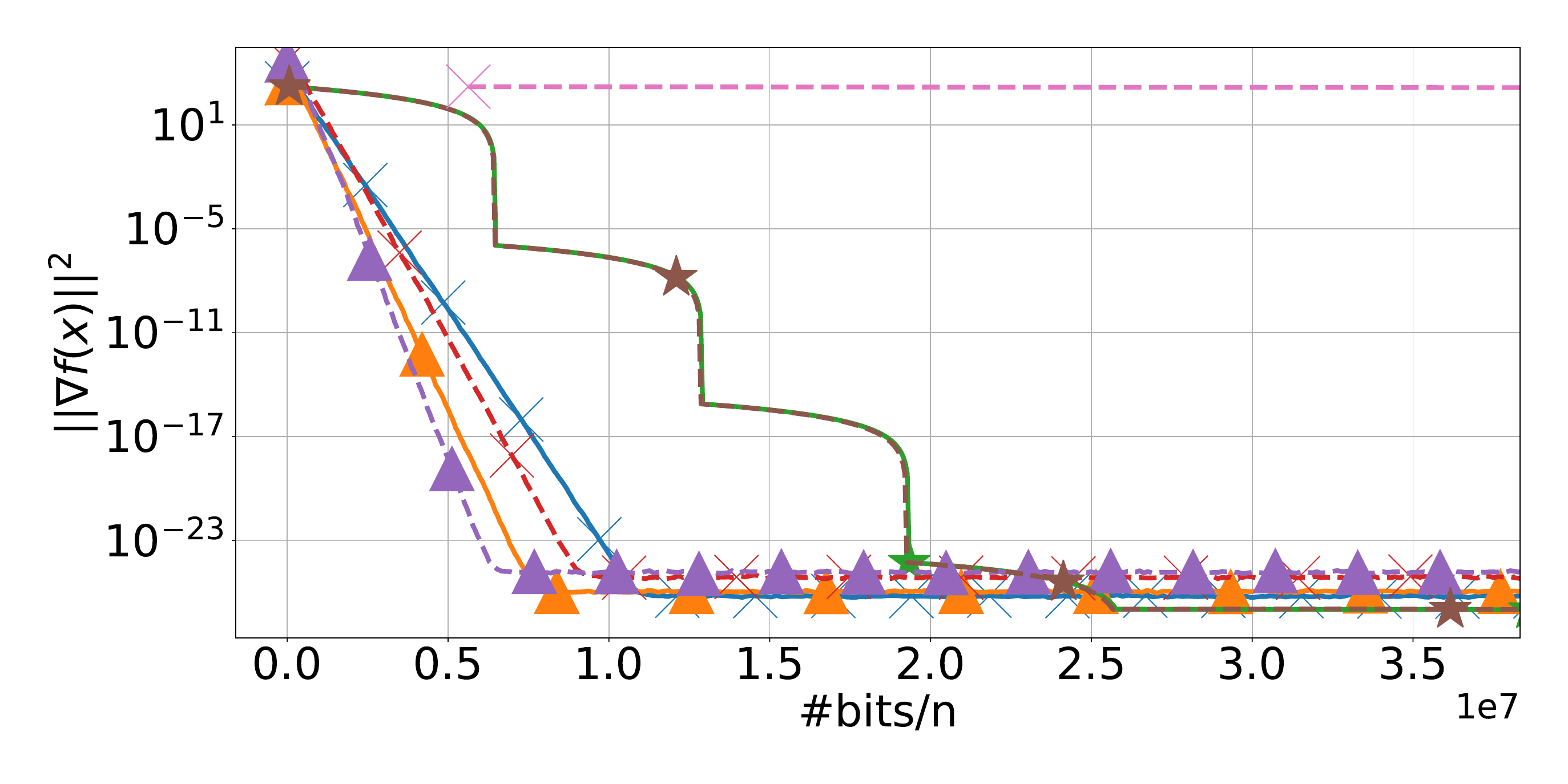} 
		\caption{{ (b) }}
	\end{subfigure}
	\begin{subfigure}[ht]{0.475\textwidth}
		\includegraphics[width=\textwidth]{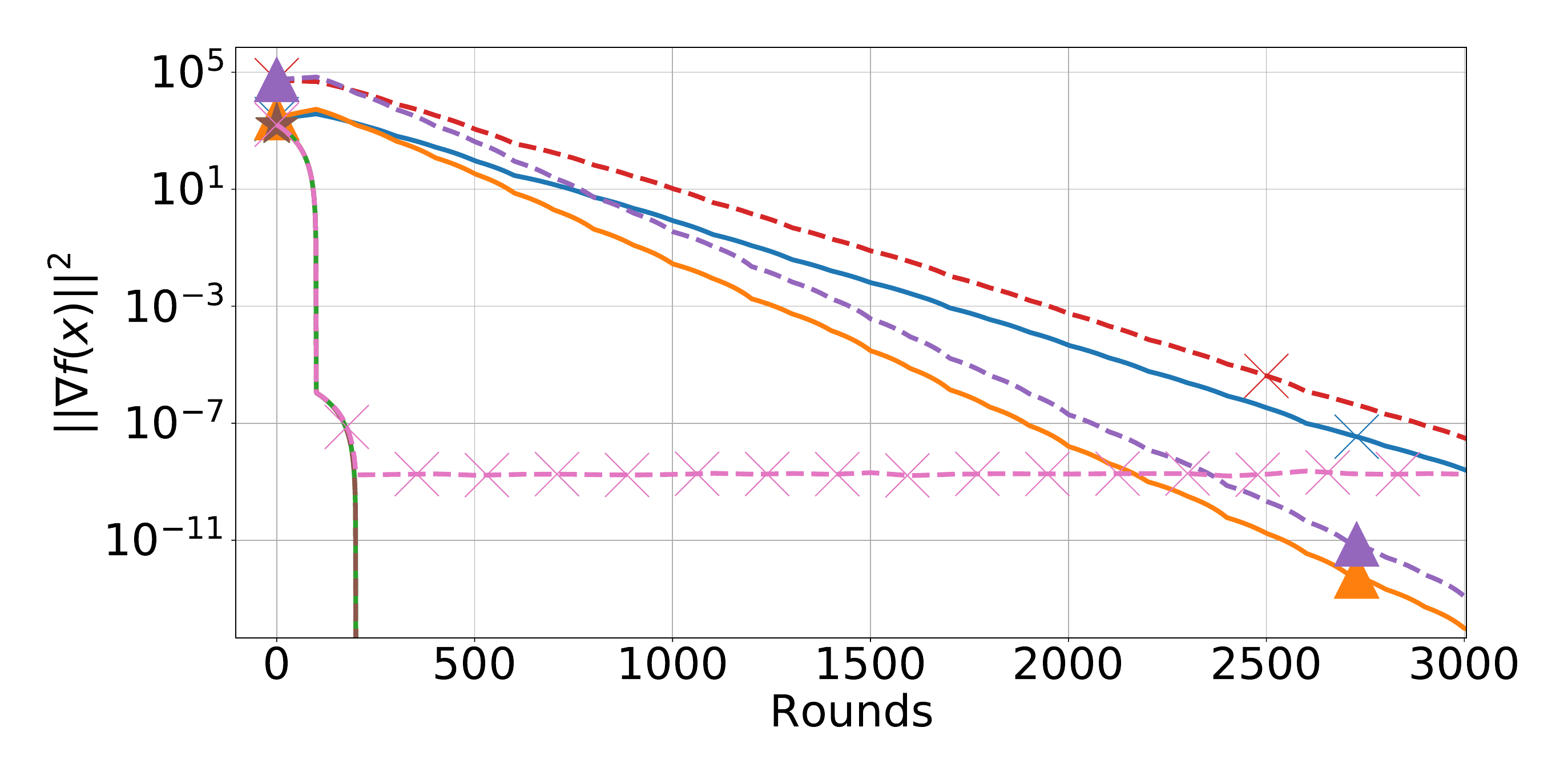} 
		\caption{{ (c) }}
	\end{subfigure}
	\begin{subfigure}[ht]{0.475\textwidth}
		\includegraphics[width=\textwidth]{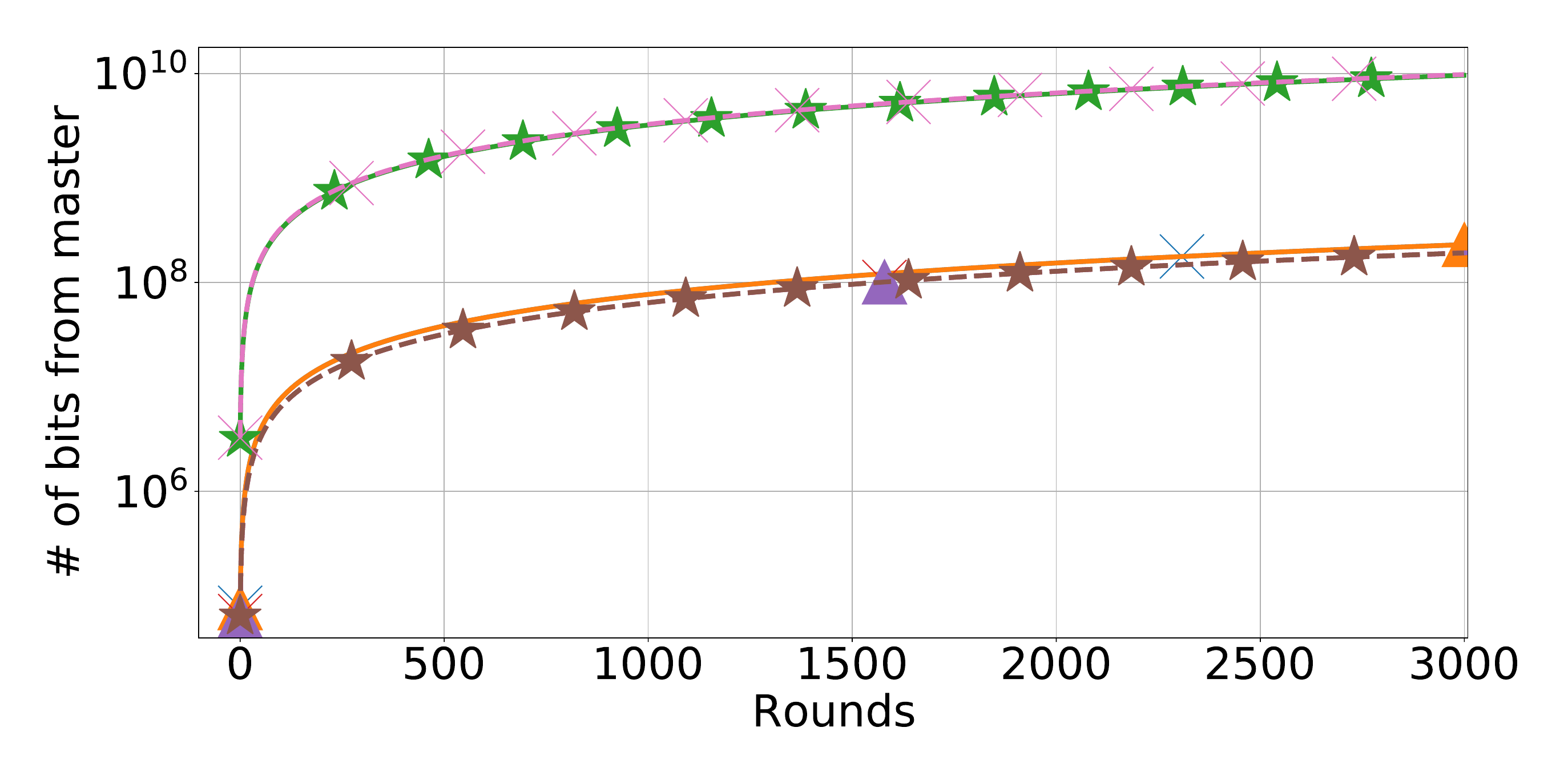} 
		\caption{{ (d) }}
	\end{subfigure}	
	\caption{{\modelname{Linear Regression} in an interpolation. \algname{DCGD} use tuned step size. \algname{GD}, \algname{GD/CKKS}, \algnamewithaes{GD/AES} use theoretical.}}
	\label{fig:exp_syn_6}
\end{figure*}

\paragraph{Comparison of {DCGD/PermK/AES} and {GD/CKKS}} Fig.~\ref{fig:exp_syn_6} compares \algname{GD}, \algname{GD/CKKS}, \algname{GD/PermK} with \algnamewithaes{GD/AES}, \algnamewithaes{GD/PermK/AES}. We see that the \ecryptname{CKKS} schema does not leverage the sparsity of vectors, making sparsification-based compression ineffective in reducing communication during privacy-preserving training. Therefore, there is no benefit from using \compname{RandK} for \algname{DCGD/CKKS}, and it's better to use vanilla \algname{GD}. For \algname{GD}, we used theoretical step size, which in practice is extremely tight. If communication isn't free, Fig.~\ref{fig:exp_syn_6} (a) and Fig.~\ref{fig:exp_syn_6} (d) suggest that \ecryptname{CKKS} is impractical in settings where client-master communication is a bottleneck. Fig.~\ref{fig:exp_syn_6} (b) shows that \algnamewithaes{GD/AES} does not increase client-to-master traffic but does significantly increase master-to-client traffic, as seen in Fig.~\ref{fig:exp_syn_6} (d). If communication is free, Fig.~\ref{fig:exp_syn_6} (c) shows that the best convergence in terms of rounds is attained for \algname{GD} or \algnamewithaes{GD/AES} with preserving security. If communication is free, \algname{GD/CKKS} remains suboptimal due to the approximate nature of floating-point operations in \ecryptname{CKKS}. Suppose the key metric is convergence in $\|{\nabla f(x^k)}\|$ relative to the number of bits from client to master. Fig.~\ref{fig:exp_syn_6} (b) shows that \algname{GD/CKKS} uses approximately $0.6 \cdot 10^7$ bits per client after the first round of optimization. \algnamewithaes{DCGD/PermK/AES} with this transfers can attain $\|{\nabla f(x)}\|^2 \approx 10^{-20}$.

\begin{figure*}[t]
	\centering
	\captionsetup[sub]{font=scriptsize,labelfont={}}	
	\captionsetup[subfigure]{labelformat=empty}
	\begin{subfigure}[ht]{0.45\textwidth}
		\includegraphics[width=\textwidth]{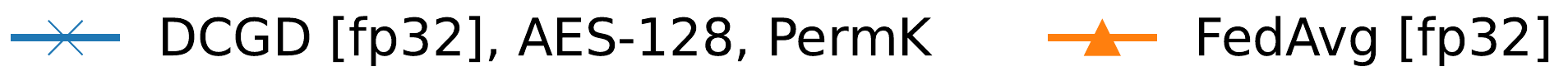} \caption{{  }}
	\end{subfigure}
	
	\vspace{-1.2\baselineskip}
	
	\begin{subfigure}[ht]{0.495\textwidth}
		\includegraphics[width=\textwidth]{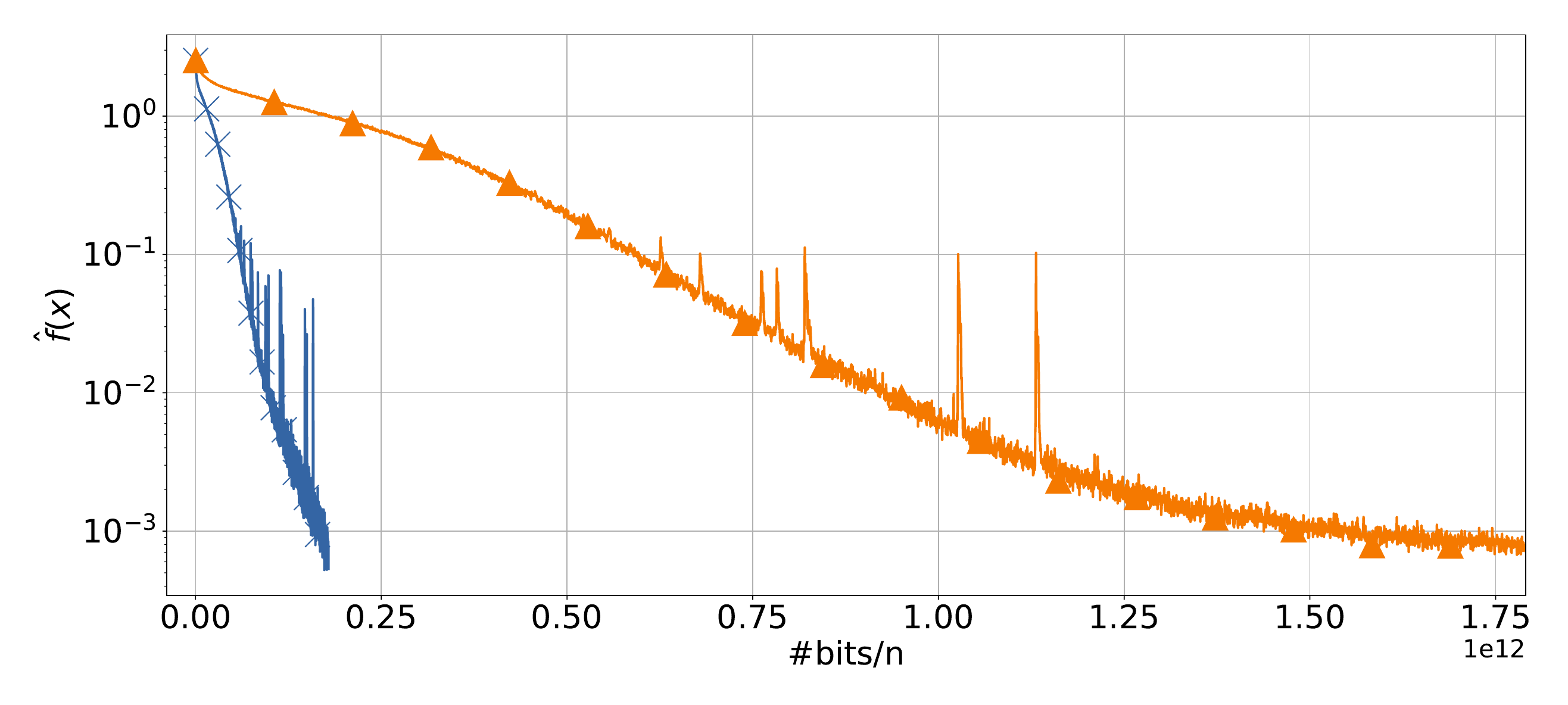} \caption{{  }}
	\end{subfigure}
	\begin{subfigure}[ht]{0.495\textwidth}
		\includegraphics[width=\textwidth]{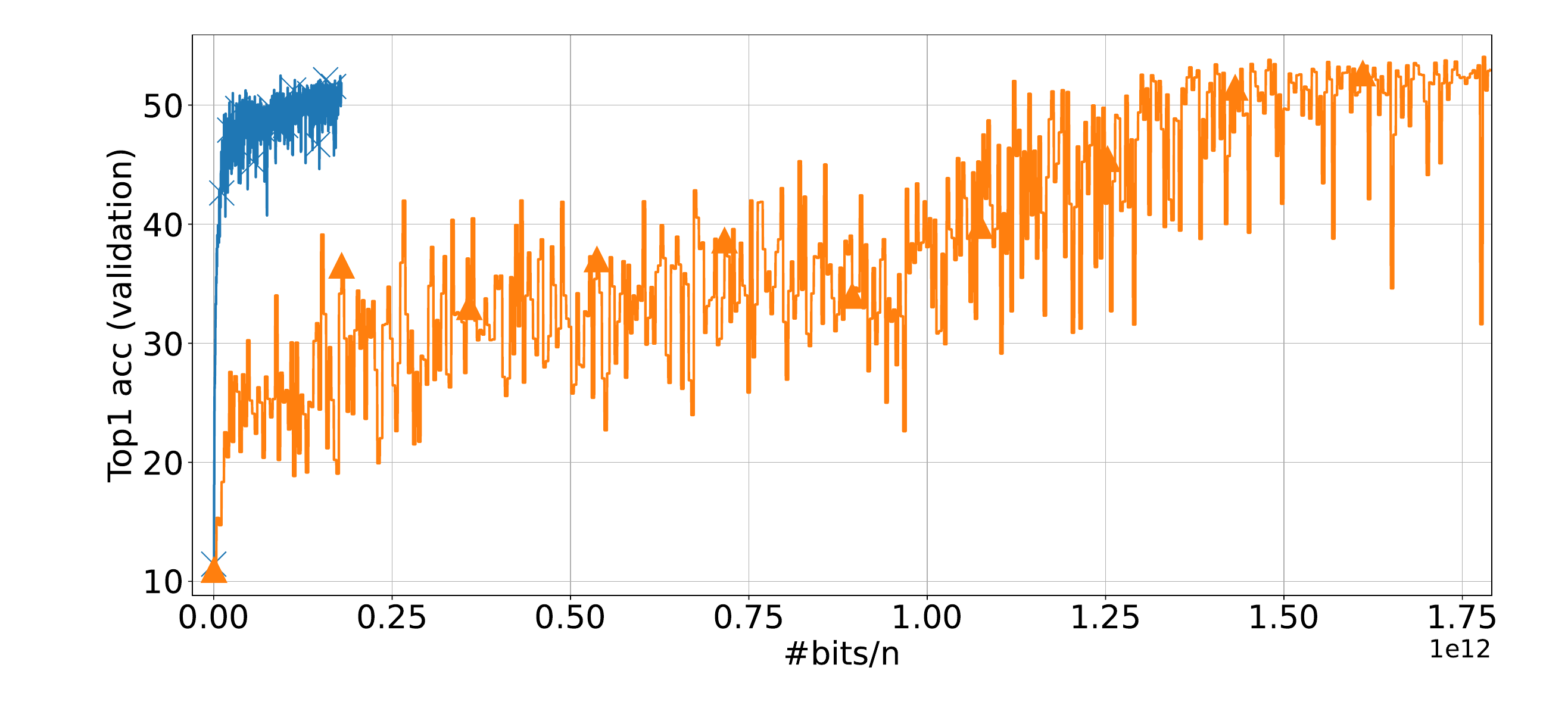} \caption{{  }}
	\end{subfigure}
	\begin{subfigure}[ht]{0.495\textwidth}
		\includegraphics[width=\textwidth]{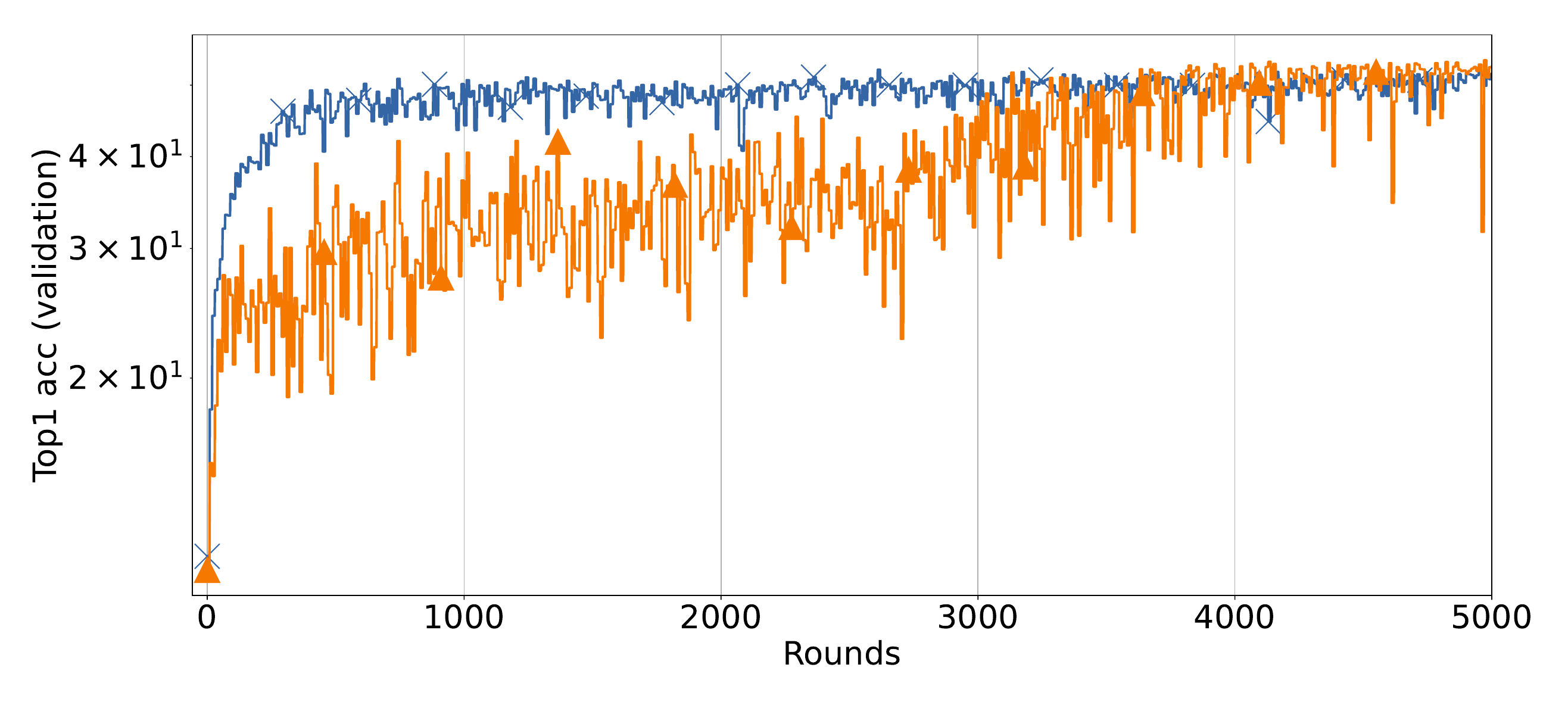} \caption{{  }}
	\end{subfigure}
	
	\vspace{-1.5\baselineskip}
	
	\caption{{\modelname{ResNet-18} in \dataname{CIFAR-10}, $n=10$, $d = 11\,181\,642$. \algnamewithaes{DCGD/PermK/AES} and \algname{FedAvg}.}}
	\label{fig:resnet_exp}
\end{figure*}

\subsection{Image Classification Application}
\label{sec:image_classification}

We evaluated the applicability of using \algnamewithaes{DCGD/PermK/AES} on Deep Neural Networks training. We used the \modelname{ResNet-18} architecture \citep{he2016deep} and trained it on the \dataname{CIFAR-10} dataset \citep{krizhevsky2009learning}, which consists of $60\,000$ images across $10$ classes with a resolution of $32\times32$ pixels. We used \modelname{ResNet-18} implementation from TorchVision library, part of the \libname{PyTorch} \citep{paszke2019pytorch}. The model size $d = 11,181,642$. For Optimization Problem \ref{eq:main}, we used a standard cross-entropy loss for $\mathcal{L}_{ij}$ terms in Equation \ref{eq:fi_for_erm}. Next, we distributed the dataset uniformly across $n = 10$ clients, with each client participating in every round. Global learning rate $0.1$, local learning rate $0.1$, local weight decay $5\cdot 10^{-4}$, number of rounds $5\,000$. During each round, clients evaluate $\nabla f_i(x)$ using a fixed 10\% subset of each client's local data points. In \abr{FL}, there are cases where clients cannot store a large number of data points (e.g., high-resolution images, recorded voice in \abr{IoT} devices) due to storage limitations. Our experiment simulates this scenario. 
We compared our implementation with \algname{FedAvg} \citep{mcmahan2017communication}, where no compression or encryption is performed. We tracked during training communicated message size,  accuracy, and convergence. Our reported metrics in Fig. \ref{fig:resnet_exp} are computation framework and communication topology independent. In these experiments, we did not carry out an analysis of the spent time for communication, compression, and \aesname{AES} encryption/decryption, because the actual numbers are highly implementation dependent quantities.

In \algnamewithaes{DCGD/PermK/AES}, the message size from the client to the master contains $\left\lceil \frac{d}{10} \right\rceil = 1,118,165$ parameters with $4$ bytes each one. The overhead from \aesname{AES-128/EAX} Operation Mode is $32$ bytes per single message, which is negligible. Fig.~\ref{fig:resnet_exp} shows that the message size was reduced from $42.65$ (for \algname{FedAvg}) {MBytes} to $4.26$ {MBytes} (for \algname{DCGD/PermK}). We cannot launch \ecryptname{CKKS} with security guarantees as \aesname{AES-128} in our environment due to memory overhead. In Appendix \ref{sec:syntetic_exp}, we demonstrated that problems start appearing in \ecryptname{CKKS} with \aesname{AES-128} security guarantees already for $d=10^6$. For a discussion about extra flexibility in Training DL models see Appendix~\ref{app:flexibility_for_dl_training}. 

\section{Deployment Flexibility}

The \textit{Physical Network Topologies} describes the  arrangement of the computation and routing devices. In a Mesh topology, every pair of nodes is connected with a dedicated link. It has high bandwidth and fault tolerance but requires a lot of cables. In this setting, the \algname{DCGD/PermK} is the natural choice. The  Algorithm~\ref{alg:dcgd_permk_aes} can be observed as a series of broadcast operations to reconstitute the optimization step. The benefits to other topologies are discussed in Appendix~\ref{app:comm_networks}. 

In Appendix~\ref{app:simulation_experiment}, we present the refined scheduled communication and computation plan that utilizes the possibility of computation communication overlap and handling computation heterogeneity during training \modelname{Linear Regression}. The modeled situation that we studied contained $5$ clients (one of them is a straggler) which all are connected to the master with a shared channel. We demonstrated that the actual execution plan can be refined. The possible gained speedup for \algname{GD} is $\times 1.31$, for \algname{DCGD/PermK} is $\times 3.33$. The \algname{DCGD/PermK} has flexibility to start several compute operations during waiting the straggler, which are limited for \algname{GD}.

\section{Conclusions}
We proposed a novel secure \abr{FL} framework that uses symmetric-key encryption with permutation compressors to protect the gradients during communication and simultaneously compress them.  We conducted experiments on real and synthetic data. Additional studies about the effect of problem dimension, overlapping communication and computation, and deployment options in various network topologies are presented in Appendices~\ref{app:extra_experiment}, ~\ref{app:comm_networks}. Our work opens a new possibility for applying Classical Cryptography to \abr{FL} and challenges some existing claims about its limitations. Possible future research and current limitations are described in Appendix \ref{app:limitation_and_future_research}.

\clearpage

\bibliographystyle{ACM-Reference-Format}
\bibliography{bibliography.bib}

\newpage
\appendix
\onecolumn

\part*{Appendix}

\label{app:toc_1}
\tableofcontents

\newpage

\clearpage

\section{Glossary}
\label{app:glossary}

Our work is multidiscipline. To make our paper more readable for researchers with different backgrounds, we have constructed this Glossary.

\begin{longtable}{|c|p{0.87\textwidth}|}
	\caption{General Terminology} \\
	\hline
	\textbf{Term} & \textbf{Meaning} 
	\endfirsthead
	\hline
	Term & Meaning 
	\endhead 
	\hline
	\hline
	\endfoot
	\hline
	FL & Federated Learning. \\
	GD & Gradient Descent. \\
	DCGD & Distributed Compressed Gradient Descent.\\
	TEE & Trusted Execution Environments. \\
	DP & Differential Privacy. \\
	MPC & The term is overloaded. In the context of Federated Learning literature typically means Multi-Party Computation. \\
	HE & Homomorphic Encryption. \\
	IoT & The collective network of connected devices.
\end{longtable}

\begin{longtable}{|c|p{0.87\textwidth}|}
	\caption{Optimization Terminology} \\
	\hline
	\textbf{Term} & \textbf{Meaning} 
	\endfirsthead
	\hline
	Term & Meaning 
	\endhead 
	\hline
	\hline
	\endfoot
	\hline
	$d$ & Dimension of optimization variable. \\
	$n$ & Number of clients/agents/devices. \\
	$f_i$ & Local Loss function on client number $i$. \\
	$f$ & Objective function with we want to minimize with image $\mathbb{R}$ and domain $\mathbb{R}^d$. \\
	$\gamma$ & Step size of learning rate. \\
	$x$ & Trainable or Optimization variable from $\mathbb{R}^d$. \\
	$x^k$ & Trainable or Optimization variable at most outer loop of optimization algorithm number $k$. \\
	round & The iteration in the outermost loop of the optimization algorithm. \\
	$e_i$ & The vector from $\mathbb{R}^d$ whose i-th component is equal $1$ while all the others are zeros. \\
	u.a.r. & uniformly at random. \\
	r.v. & random variable. \\
	p.d.f. & probability distribution function. \\
	$\mathbb{E}[.]$ & Expectation of some Random Variable. \\
	PermK & Permutated correlated compressors. \\
	Master & Entity in \abr{FL} which performs aggregation and other forms of reductions.\\
\end{longtable}

\begin{longtable}{|c|p{0.80\textwidth}|}
	\caption{Discrete Math Terminology} \\
	\hline
	\textbf{Term} & \textbf{Meaning} 
	\endfirsthead
	\hline
	Term & Meaning 
	\endhead 
	\hline
	\hline
	\endfoot
	\hline
	Algebra & Set of elements and defined operations on set which lead to elements of the same set.\\
	Monoid & Any algebra with the binary, associative operation, which also has a neutral
	element. Example - concatenation.\;\\
	Group & Any algebra with the binary, associative operation, with a neutral
	element and each element has an inverse.\\
	Ring & Algebra with summation, and multiplication. A summation is a commutative group,  multiplication is a monoid. Multiplication is distributive with respect to summation. In Ring without extra assumptions can have a situation such that $a\ne0$, $b\ne0$, but $ab=0$.
	\\
	Polynomial & $p(x) = p_0 + p_1 \cdot x + p_2 \cdot x^2 + \dots$. \\
	$K[x]$ & Let K be some field. The set of all polynomials with coefficients in some field $K$. It is called commutative called the polynomial ring over $K$. In the Ring of polynomials $K[X]$ there is no division in general similar to Algebra $N$. However, it's possible to perform division with residual. One basis for such set is $\{1, x, x^2, \dots\}$.\\
	$Z[X]$ & Integer polynomial rings over commutative ring $Z$ is denoted as $Z[X]$. This is the set of polynomials whose coefficients are integers and polynomials depend only on one variable.\\
	$P[x]/(x^2 + 1)$ & This notation means polynomials which are obtained in the following way. We take the polynomial ring $P[x]$ and perform modulus division by the polynomial $x^2 + 1$. This modulus arithmetic restricts obtained polynomials to have a power less than $2$. Such a ring is an example of a quotient ring.\\
	Unity Roots & Roots of unity are roots of the following equation $Z^n=1$. If $Z\in \mathbb{C}$ the roots are: $z_k=\exp(2 \pi k i / n )$ for $k=0,1,2,\dots,n-1$.\\
	GCD & The greatest common divisor (GCD) of two or more integers, that are not zero, is the largest positive integer that divides each of the integers. For example $\gcd(8, 12) = 4$. \\
	\hspace{0.4cm} Homomorphism \hspace{0.4cm} & Homomorphism of two groups $G_1$ and $G_2$ is a mapping $f:G_1 \to G_2$ between two groups, such that $\forall x,y \in G_1$ the following holds: $f(x*y)=f(x)*f(y)$\\	
	Isomomorphism & Isomomorphism of groups $G_1$ and $G_2$ is a mapping $f:G_1 \to G_2$ between two groups, such $\forall x,y \in G_1$ the following holds: $f(x*y)=f(x)*f(y)$ and $f$ is bijection mapping. From the point of view of Algebraic structures, two isomorphic groups are the same, even if they have different natures of their elements.\\	
\end{longtable}

\begin{longtable}{|c|p{0.75\textwidth}|}
	\caption{Cryptography Terminology} \\
	\hline
	\textbf{Term} & \textbf{Meaning} 
	\endfirsthead
	\hline
	\textbf{Term} & \textbf{Meaning}
	\endhead 
	\hline
	\hline
	\endfoot
	\hline
	FHE & Fully Homomorphic Encryption. \\
	SWHE & Somewhat Homomorphic Encryption. \\
	LFHE & Leveled Fully Homomorphic Encryption. \\
	CKKS & Cheon-Kim-Kim-Song Homomorphic Encryption schema, which is formally only SWHE. \\
	AES & Advanced Encryption Standard.\\
	sk & Secret Key.\\
	pk & Public Key.\\
	Symmetric Encryption & Encryption schema in which secret and public key is the same.\\
	PRP & Pseudo Random Permutation.\\
	PRF & Pseudo Random Function. \\
	MAC & Message Authentication Code. \\
	CRC & Cyclic Redundancy Check.\\
	CPA & Chosen plaintext attack.\\
	CCA & Chosen ciphertext attack.\\
	CTR & Counter Mode Randomized.\\
	CBC & Cipher Block Chaining.\\
	EAX & Encrypt then Authenticate then Translate Mode.\\
	LWE & The Learning With Errors search problem to find a solution to a noisy system of linear equations in a finite field. \\
	RLWE & The Ring Learning With Errors search problem. Find a solution to a noisy system of the linear equation when an underlying algebra is a polynomial over a ring. \\
	Integer Lattice& It is a set of points which correspond to all possible linear combinations with integer coefficients of the $n$ vectors $b_i \in \mathbb{R}^d$, i.e. it the set $\{x:x=\sum_{i=1}^{n} {\alpha}_i b_i,{\alpha}_i \in \ZS\}$.\\
\end{longtable}

\begin{longtable}{|c|p{0.80\textwidth}|}
	\caption{Systems and Communication  Terminology} \\
	\hline
	\textbf{Term} & \textbf{Meaning} 
	\endfirsthead
	\hline
	\textbf{Term} & \textbf{Meaning} 
	\endhead 
	\hline
	\hline
	\endfoot
	\hline
	DRAM & Dynamic Random Access Memory.\\
	GPU & Graphics Processing Unit.\\
	FP & Float point format.\\
	\hspace{0.38cm} FP16, FP32, FP64 \hspace{0.38cm} & Float point format with a specified number of bits per single scalar.\\
	CUDA & Compute Unified Device Architecture.\\
	CPU & Central Processing Unit.\\
	AVX & Advanced Vector Extensions.\\
	NIC & Network Interface Controller.\\
\end{longtable}

\clearpage

\section{Compared Optimization Algorithms}
\label{app:list_of_optimization_algos}

In our work, we made a comparison of several optimization algorithms with different compression and security methods. In Table~\ref{tbl:list_of_optimization_algos} below we summarize their qualitative and qualitative features.

\begin{longtable}{|l|p{0.65\textwidth}|}
	\caption{Compared Optimization Algorithms with Compression and Privacy Mechanisms} \\
	\hline
	\textbf{Name} & \textbf{Description} 
	\endfirsthead
	\hline
	\textbf{Name} & \textbf{Description} 
	\endhead 
	\hline
	\hline
	\endfoot
	\hline
	1. \algname{GD [FP16|FP32|FP64]}	
	& Algorithm~\ref{alg:dcgd}, Baseline (\myred{B}), $\mathcal{C}_i(x) \eqdef x$. 
	\newline
	\newline
	Gradient Descend. In this algorithm, clients in a distributed way compute the gradient for the function $f_i$ defined in Equation \ref{eq:fi_for_erm} in the current iterate. Master obtains gradients from the clients and averages them. The master broadcasts the result directly to all clients. The specified Floating-Point format is used to represent both data and trainable variables in all clients and during aggregation in the master. 
	\newline
	\newline
	\textbf{Quantitative and qualitative characteristics:}
	\begin{enumerate}
		\item It's completely insecure (see Appendix~\ref{app:reconstruction}).
		\item Volume information from clients to master is $\mathcal{O}(dn)$ bytes per round.
		\item Broadcasted information from master to clients is $\mathcal{O}(d)$ bytes per round.
	\end{enumerate}
	For details on why information about $\nabla f_i(x; D_i)$ can potentially reveal information about dataset $D_i$ see discussion in  Apeendix~\ref{app:reconstruction}.	
	\\
	\hline
	2. \algnamewithaes{GD [FP16|FP32|FP64]/AES-128} & 
	Algorithm~\ref{alg:dcgd}, Naive usage of \aesname{AES} (\myblue{A}), $\mathcal{C}_i(x) \eqdef x$. 
	\newline
	\newline
	Gradient Descend with symmetric \aesname{AES} encryption which uses a key size of \myblue{$128$} bits ($16$ bytes). In this algorithm, clients in a distributed way compute the gradient for the function $f_i$ defined in Equation \ref{eq:fi_for_erm} in the current iterate. Then clients encrypt computed $\nabla f_i(x)$ in parallel. After that Master collects the encoded gradients as a communication hub. Unfortunately master can not perform any algebraic operations, because $\nabla f_i(x)$are encrypted, and the master needs to broadcast all directions to the clients. The specified Floating-Point format is used to represent both data and trainable variables in all clients and during aggregation in the master. 
	
	During using \aesname{AES/EAX} there exists a fixed negligible overhead from using EAX Mode Operation Mode of $32$ bytes in each client to master communication information transfer. Specifically, this Mode Operation includes the following overheads: \textit{Nonce} - a random $16$-byte value, and the \textit{Tag} - $16$-byte message authentication code.	
	\newline
	\newline
	\textbf{Quantitative and qualitative characteristics:}
	\begin{enumerate}
		\item Master cannot average obtained encrypted clients' local gradient.
		\item Volume information from clients to master is $\mathcal{O}(dn + 32n)$ bytes per round.
		\item Broadcasted information from master to clients is $\mathcal{O}(dn + 32n)$ bytes per round.
	\end{enumerate}
	For details about \aesname{AES} see Apeendix~\ref{app:aes_details}.
	\\
	\hline
	3. \algnamewithaes{DCGD [FP16|FP32|FP64]/AES-128/RandK} &
	
	Algorithm~\ref{alg:dcgd}, Naive usage of \aesname{AES} (\myblue{A}), $\mathcal{C}_i(x) \eqdef \frac{d}{k} \sum_{i \in S} x_i \cdot e_i, S \sim_{u.a.r} \{s: s\in 2^{[d]}, |s|=k \}$.
	\newline	
	\newline
	Distributed Compressed Gradient Descend with symmetric \aesname{AES} encryption which uses a key size of \myblue{$128$} bits ($16$ bytes), and with using \compname{RandK} sparsifier.
	
	In this algorithm, clients in a distributed way compute the gradient for the function $f_i$ defined in Equation \ref{eq:fi_for_erm} in the current iterate. After this clients compress them by selecting $k$ components from $d$ u.a.r. Then clients use a sparse bitwise representation of the sparsified gradient and encode non-zero values with \aesname{AES/EAX} mode of operation. Then clients encrypt computed $\mathcal{C}(\nabla f_i(x))$ in parallel. After that Master collects the encoded gradients as a communication hub. Unfortunately master can not perform any algebraic operations, because $\nabla f_i(x)$are encrypted, and the master needs to broadcast all directions to the clients. The specified Floating-Point format is used to represent both data and trainable variables in all clients and during aggregation in the master. 
	
	During using \aesname{AES/EAX} there exists a fixed negligible overhead from using EAX Mode Operation Mode of $32$ bytes in each client to master communication information transfer. Specifically, this Mode Operation includes the following overheads: \textit{Nonce} - a random $16$-byte value, and the \textit{Tag} - $16$-byte message authentication code.	
	The implementation of \compname{RandK} sparsifier in the case of using a pseudo-random generator can be implemented in a way that no indices should be transferred during training because they can be reconstituted. 
	\newline
	\newline
	\textbf{Quantitative and qualitative characteristics:}
	\begin{enumerate}
		\item Master cannot average obtained encrypted and compressed clients' local gradient.
		\item Volume information from clients to master is $\mathcal{O}(kn + 32n)$ bytes per round. 
		\item Broadcasted information from master to clients is $\mathcal{O}(kn + 32n)$ bytes per round.
		\item Fundamentally there is some redundancy induced by the inability to perform aggregation in the master.
	\end{enumerate}
	\\
	\hline
	4. \algname{GD [FP16|FP32|FP64]/CKKS} & 
	Algorithm~\ref{alg:dcgd}, Baseline (\myred{B}), $\mathcal{C}_i(x) \eqdef x$.
	Extra encryption is carried via employing  \ecryptname{CKKS}.
	\newline
	\newline
	In this algorithm, clients in a distributed way compute the gradient for the function $f_i$ defined in Equation \ref{eq:fi_for_erm} in the current iterate. After this each client using the public key in \ecryptname{CKKS} schema performs encryption of gradient vectors. Master obtains the encoded gradients from the clients. Then master using a public key performs aggregation and relinearization to reduce the size of a ciphertext after arithmetic operations. After this master broadcast aggregated gradient estimator. Clients using private (secret) keys perform for decryption of the encrypted direction from the master and then make a step.
	\newline
	\newline
	\textbf{Quantitative and qualitative characteristics:} 
	\begin{enumerate}
		\item \ecryptname{CKKS} encryption considerably increases ciphertext space. The volume of information from clients to master $\mathcal{O}(C \cdot dn)$ bytes, where $C$ is a constant which depends on $d$, security level, degree of the used polynomial, and cardinality of space $\mathbb{Z}_q$ to represent integer coefficient of two transferred polynomials after encoding.
		\item Size of public and private key to guarantee \aesname{AES-128} for \ecryptname{CKKS} is approximately $420\,000$ bytes, while for \aesname{AES-128} the key size is $16$ bytes. In some tasks this key size is negligible, in some tasks, it is not. For details see Appendix~\ref{app:ckks_details}.
		\item \ecryptname{CKKS} does not operate on the level of bits only and can operate only with fp64 float point format in \libname{TenSeal} implementation. Therefore there is no possibility of using less precision in combination with \ecryptname{CKKS}. 		
	\end{enumerate}
	For details about \ecryptname{CKKS} see Appendix~\ref{app:ckks_details}.
	\\
	\hline
	5. \algname{DCGD [FP16|FP32|FP64]/RandK/CKKS} & 
	Algorithm~\ref{alg:dcgd}, Baseline (\myred{B}), $\mathcal{C}_i(x) \eqdef \frac{d}{k} \sum_{i \in S} x_i \cdot e_i, S \sim_{u.a.r} \{s: s\in 2^{[d]}, |s|=k \}$. Extra encryption is carried via employing  \ecryptname{CKKS}.
	\newline
	\newline
	Distributed Compressed Gradient Descend with \ecryptname{CKKS}. In this algorithm, clients in a distributed way compute the gradient for the function $f_i$ defined in Equation \ref{eq:fi_for_erm} in the current iterate. After this clients compress them by selecting $k$ components from $d$ u.a.r. After this each client using the public key in \ecryptname{CKKS} schema performs encryption of sparsified gradient vectors. Master obtains the encoded gradients from the clients. Then master using a public key performs aggregation and relinearization to reduce the size of a ciphertext after arithmetic operations. After this master broadcast aggregated gradient estimator. Clients using private (secret) keys perform for decryption of the encrypted direction from the master and then make a step.
	\newline
	\newline
	\textbf{Quantitative and qualitative characteristics:} 
	\begin{enumerate}
		\item \ecryptname{CKKS} encryption considerably increases ciphertext space. The volume of information from clients to master $\mathcal{O}(C \cdot dn)$, where $C$ is a big constant which depends on $d$, security level, degree of the used polynomial, and cardinality of space $\mathbb{Z}_q$ to represent integer coefficient of two transferred polynomials after encoding. For details see Appendix~\ref{app:ckks_details}.
		\item Size of public and private key to guarantee \aesname{AES-128} for \ecryptname{CKKS} is approximately $420\,000$ bytes, while for \aesname{AES-128} the key size is $16$ bytes. In some tasks this key size is negligible, in some tasks, it is not. For details see Appendix~\ref{app:ckks_details}.
		\item \ecryptname{CKKS} does not operate on the level of bits only and can operate only with fp64 float point format in \libname{TenSeal} implementation. Therefore there is no possibility of using less precision in combination with \ecryptname{CKKS}.
		\item The \ecryptname{CKKS} does not support linear operations over sparse vectors from $\RD$. For \ecryptname{HE} schemas, encoding of any two vectors (e.g. sparse and dense) should be indistinguishable due to semantic security requirements. However, from a computational perspective, ignoring sparsity is sometimes highly impractical. This gap presents an open research question for \ecryptname{HE}.
	\end{enumerate}
	\\
	\hline
	6. \algnamewithaes{DCGD [FP16|FP32|FP64]/PermK/AES} & 
	Algorithm~\ref{alg:dcgd_permk_aes}.
	\newline
	\newline
	Distributed Compressed Gradient Descend with using \compname{PermK} sparsifier. In this algorithm, clients in a distributed way compute gradients in the current iterate. After this clients compress them by selecting $k$ components from $d$ jointly	via using Algorithm~\ref{alg:perm_k_gen}. 
	
	Then clients send a sparse bitwise representation of the sparsified gradient. Master collects the encoded gradients as a communication hub. 
	
	The \compname{PermK} compressor garantees that what is needed to perform at master is only concatenation. Therefore the algorithm can be implemented in practice in situations when the master can store, but cannot compute. 
	
	The specified floating-point format is used to represent both data and trainable variables. The implementation of \compname{PermK} in the case of using a pseudo-random generator can be implemented in a way, that no indices should be transferred during training because they can be reconstituted and negotiation between clients should be carried in runtime if clients have negotiated initial seed. 
	\newline
	\newline
	\textbf{Quantitative and qualitative characteristics:}
	\begin{enumerate}
		\item There is $32$ byte overhead during message transfers compared to \algname{DCGD/RandK}.
		\item The current implementation works only in case $d \ge n$, when \algname{DCGD/RandK} and \algname{GD/CKKS} do not require this
		\item Volume information from clients to master is $\mathcal{O}(d + 32n)$ bytes per round. 
		\item Broadcasted information from master to clients is $\mathcal{O}(d + 32n)$ bytes per round.
	\end{enumerate}
	\\
	\hline
	7. \algnamewithaes{DCGD [FP16|FP32|FP64]/PermK} & 
	Algorithm~\ref{alg:dcgd_permk_aes} in which $\myblue{Encrypt}(x) \eqdef x$, $\myblue{Decrypt}(x) \eqdef x$.
	\newline
	\newline
	Distributed Compressed Gradient Descend with using \compname{PermK} sparsifier. In this algorithm, clients in a distributed way compute gradients in the current iterate. After this clients compress them by selecting $k$ components from $d$ jointly	via using Algorithm~\ref{alg:perm_k_gen}. Then clients send a sparse bitwise representation of the sparsified gradient. Master collects the encoded gradients as a communication hub. The \compname{PermK} compressor garantees that what is needed to perform at master is only concatenation. Therefore the algorithm can be implemented in practice in situations when the master can store, but cannot compute. The selected floating point format is used to represent both data and trainable variables. The implementation of \compname{PermK} in case of using a pseudo-random generator can be implemented in a way, that no indices should be transferred during training because they can be reconstituted and negotiation between clients should be carried in runtime if clients have negotiated initial seed. 
	\newline	
	\newline
	\textbf{Quantitative and qualitative characteristics:}
	\begin{enumerate}
		\item The current implementation works only in case $d \ge n$ only, when \algname{DCGD/RandK} and \algname{GD/CKKS} does not require this.
		\item This algorithm ignores privacy and security aspects.
		\item Volume information from clients to master is $\mathcal{O}(d + 32n)$ bytes per round. 
		\item Broadcasted information from master to clients is $\mathcal{O}(d + 32n)$ bytes per round.
	\end{enumerate}
	\label{tbl:list_of_optimization_algos}
\end{longtable}

\clearpage

\section{Computing and Software Environment}
\label{app:environment}

We conducted numerical experiments using the Python software suite \libname{FL\_PyTorch} \citep{burlachenko2021fl_pytorch}, running on Python 3.9. The target machine is a server-grade system operating on Ubuntu 18.04 and Linux Kernel v5.4.0-148. It is equipped with a 48-core Intel(R) Xeon(R) Gold 6246 CPU (2 sockets with 24 cores per socket) running at 3.3 GHz. The machine is equipped with 256 GBytes of DDR4 DRAM system memory operating at 2.9GHz. The installed CPU does not support both the AVX512FP16 Instruction Set Architecture
\footnote{\href{https://www.intel.com/content/www/us/en/content-details/669773/intel-avx-512-fp16-instruction-set-for-intel-xeon-processor-based-products-technology-guide.html}{https://www.intel.com/content/www/us/en/content-details/669773/intel-avx-512-fp16-instruction-set-for-intel-xeon-processor-based-products-technology-guide.html} - {Intel AVX-512 FP16 Instruction Set}} and it does not support \texttt{FP16} arithmetic. The machine also has an NVIDIA GeForce RTX 3090 GPU built on Ampere microarchitecture with 24 GBytes of DRAM@9.7GMHz GPU memory. This GPU supports the CUDA Compute Capability 8.6. and as a consequence, supports the \texttt{FP16} arithmetic, which is supported for devices with Compute Capability 5.3. and higher. \footnote{\href{https://docs.nvidia.com/cuda/cuda-c-programming-guide/index.html\#features-and-technical-specifications}{https://docs.nvidia.com/cuda/cuda-c-programming-guide/index.html\#features-and-technical-specifications} - {NVIDIA  Specification for Compute Capabilities}}. Numerical experiments with FP16 arithmetic were carried out in this GPU.

We have patched \libname{FL\_PyTorch} to include necessary functionality from the \libname{TenSeal} \citep{benaissa2021tenseal} library version 0.3.14 and PyCryptodome \citep{pycryptodome} version 3.17. The experiment's scheduler for computation and communication from Appendix \ref{app:simulation_experiment} generates text descriptions in dot graphviz format. Finally, we used Graphviz 8.0.5 \citep{ellson2002graphviz} to generate a directed graph from this description.

\clearpage

\section{Overview of Existing Privacy Mechanisms in Context of {FL}}
\label{app:overview_of_privacy_mechanisms}

\paragraph{Trusted Execution Environments (\abr{TEE})} The \abr{TEE} brings the idea of handling self-isolation of processes more seriously than it has been done in the Operation Systems before. An example of the particular implementation of this idea is ARM TrustZone \citep{pinto2019demystifying} and Intel SGX \citep{costan2016intel}, where both technologies have been implemented at the hardware level. Such technologies allow the protection of memory from reading and writing operations not only from another process in the Operation System (OS) but also from the kernel of the OS by itself. The goal of \abr{TEE}: 
\begin{center}
	\textit{TEE protects the execution environment from illegal intervention, which will defect training}.
\end{center}

\paragraph{Differential Privacy ({DP})} The goal of \abr{DP} is to ensure statistical analysis does not hurt the privacy aspect. The intuitive definition can be as follows: "An algorithm A is differentially private if an observer seeing its output cannot tell if a particular individual's data was used in the computation." \abr{DP} criteria assess the private property of the algorithm, but it does not dictate how this functionality should be implemented. In general, as more noise is appending the data or derived quantities from the data that are released publicly, the algorithm starts to be more \abr{DP}, but from another side, the statistical properties of the data for the purpose of solving original task start to be loosed.

\begin{definition} [Approximate Differential Privacy] \label{def:approx_dp}
	An algorithm $M:\mathcal{X}^n \to \mathcal{Y}$ satisfies approximate $(\varepsilon, \delta)$~-~DP if $\forall X, X' \in \mathcal{X}^n$ such that datasets are different in one point (neighboring datasets denoted as $X \sim X'$) and $\forall T \subseteq \mathcal{Y}$ the following holds:
	\[
	P[M(X) \in T] \le \exp(\varepsilon) P[M(X') \in T] + \delta
	\]
\end{definition}

The definition \ref{def:approx_dp} is a relaxation of pure-DP, first proposed by Dwork et al. in \citep{dwork2006our} in 2006. The pure-DP was given by in \citep{dwork2006calibrating}, requiring $\delta=0$. The Approximate DP possesses weaker privacy guarantees but allows the addition of less noise.

DP-ERM aims to output $\hat{\theta}$, which is $(\varepsilon, \delta)$ - DP with respect to the training dataset $D$. To do that, there are three types of approaches: (1) Output perturbation; (2) Objective perturbation; (3) Gradient perturbation. Algorithms that solves that DP-ERM problem are quantified by \textit{expected excess empirical risk} $\E[\mathcal{L}_{erm}(\hat{\theta}, D) - \mathcal{L}_{erm}(\theta_{erm}^*, D)]$, and for DP-RM problem by  \textit{expected population risk} $\E[\mathcal{L}_{rm}(\hat{\theta}, D) - \mathcal{L}_{rm}(\theta_{rm}^*, D)]$. These quantities are sometimes named as \textit{utility}. \\
The goal of \abr{DP}:
\begin{center}
	\textit{DP protects output of algorithms so that users' data are not leaking from Algorithm execution}.
\end{center}

One important classification in \abr{DP} algorithms targeted to distributed environments is their separation into two classes: \textit{Centralized} and \textit{Local} Differential Private settings. In the centralized model, a client trusts curator (or master) and DP protection mechanism are applied in master. In the local model, each individual applies a differential private mechanism to their own data before sending it to an untrusted curator (or master). Centralized models have lower privacy loss, since the noise is added only once at the end of the process such as aggregation at master. However, the centralized model also requires more trust in the aggregator.

\paragraph{Aggregation with Multi-Party Computation ({MPC})} The MPC is a sub-field of Cryptography concerned with the problem of having a set of parties that compute an agreed function of their private inputs. The goal of secure multi-party computation (\abr{MPC}) is to enable independent data owners who do not trust each other or any common third party to jointly compute a function that depends on all of their private inputs. \abr{MPC} protocols are typically implemented with (a) \textit{Secret sharing} - in this case, it requires a lot of total communication rounds to compute the average across $n$ clients; (b) \textit{Garbled circuits} - in this case there both communication and computation overhead is added to the training \cite{zhao2019secure}. The goal of \abr{MPC}: 

\begin{center} 
	\textit{MPC allows for protecting inputs for the algorithm at the cost of communication.}
\end{center}

\paragraph{Homomorphic Encryption (\ecryptname{HE})}

Homomorphic Encryption (\ecryptname{HE}) enables numerical computation on encrypted data, e.g. aggregating vectors from $\mathbb{R}^d$.  However, the result can only be decrypted with the private key. The concept dates back to 1978 when R. L. Rivest et al. \citep{rivest1978data} proposed this but without a complete solution. \ecryptname{HE} allows any device party to compute functions on encrypted data using only a public key and encrypt the result. \ecryptname{HE} aims to hide the plaintext input and output from the executor, who only sees the encrypted versions of it. This is possible with  \ecryptname{HE}. On the other hand, obfuscation is the process of encrypting the program (function $f$), not the input or output from the caller. Obfuscation is impossible under weak technical conditions \citep{barak2001possibility}.

Fully Homomorphic Encryption (\ecryptname{FHE}) allows any computable algorithm to be executed on encrypted data without any restrictions on the binary operations. The first \ecryptname{FHE} scheme was proposed by Craig Gentry \citep{gentry2009fully}, based on lattices and a novel bootstrapping technique. Let $c_i$ be the encryption of message $m_i$, and $c$ be the result of evaluating a function $\hat{f}(c_1,\dots c_n)$, which should decrypt to $m=f(m_1,\dots m_n)$. In his Ph.D. thesis, C. Gentry listed some requirements that any \ecryptname{FHE} scheme should satisfy, which we summarize below:

\begin{enumerate}
	\item \textit{Correctness:} Decryption should always recover the correct evaluation of the function, i.e., $P( {Decrypt(c) = f(m_1, m_2,\dots)} = 1)$.
	\item \textit{Semantic security:} The encryption of any two messages should be computationally indistinguishable.
	\item \textit{Efficiency:} Decryption should not be more expensive than evaluating the function itself.
	\item \textit{Compactness:} Ciphertexts should have a polynomial size in the security parameter, independent of the size of the function evaluated.
	\item \textit{Security:} The best-known attack should have exponential complexity in the security parameter, i.e., $\Omega({2^k})$ ($k$ is a security parameter).
	\item \textit{Feasibility:} Key generation, encryption, and decryption should have polynomial complexity in the security parameter.	
\end{enumerate}

By C.Gentry, only algorithms that allow executing any computable algorithm on encrypted data that satisfies properties (1) - (6) can be called \ecryptname{FHE}. The \ecryptname{FHE} with this requirement can firstly be hard to construct, and secondary in practice, they may not be computationally efficient. To mitigate these issues, two main strategies are the following:

\begin{itemize}
	\item \textit{Somewhat Homomorphic Encryption} (\ecryptname{SWHE}). One way to make \ecryptname{FHE} more efficient in practice is to restrict the class of functions that can be evaluated. This leads to Somewhat Homomorphic Encryption, which can handle functions from restricted classes (e.g. they are low-degree polynomials). 
	
	\item \textit{Leveled FHE} (\ecryptname{LFHE}). Another way is to limit the depth of the binary circuit that represents the function. This leads to a notion of Leveled FHE, which can handle arbitrary functions represented by boolean circuits but with a fixed bound on the circuit depth.
\end{itemize}

Challenges of applying \ecryptname{HE} for training Machine Learning models and scientific computation:

\begin{enumerate}
	\item The \ecryptname{HE} adds noise to the plaintext to ensure security. The challenge is to manage it with error-correction techniques, as it typically grows after each arithmetic operation.
	\item The \ecryptname{HE} cannot perform random access on encrypted data without revealing information.
	\item The \ecryptname{HE} cannot exploit the advantages of Random Access Machines, which can compute some algorithms faster than binary circuits, e.g., Binary Search.
	\item The \ecryptname{HE} does not support multiple keys natively. Originally, it was designed as a single-private-key system.
	\item The \ecryptname{HE} does not obfuscate the function itself, only the input and output. As mentioned, obfuscation is impossible under weak conditions \citep{barak2001possibility}.
	\item The \ecryptname{HE} works on binary circuits or functional schemas. This class of representation is Turing Complete, assuming we can create circuits of different levels for different inputs. The pure \ecryptname{FHE} operations are computationally intensive and currently for practical purposes the \ecryptname{SWHE} and \ecryptname{LFHE} schemas should be considered instead.	
	\item The \ecryptname{HE} methods require large ciphertext sizes.
	\item Choosing the right \ecryptname{HE} scheme for a given machine learning task is not trivial.
\end{enumerate}

The goal of \abr{HE}: 
\begin{center}
	\textit{HE allows meaningful manipulation under encrypted data without revealing it.}
\end{center}

\clearpage

\section{Discussions}
\label{app:discussions}

\subsection{The Imperative of Safeguarding Against Eavesdropping}
\label{app:reconstruction}

Assume that during training with first-order optimization method, each client $i$ discloses the following information at each iteration $k \in {1, \dots, K}$: $$\dfrac{\partial f_i(x^k)}{\partial x_j} \eqdef \lim_{dx_j \to 0} \dfrac{f_i(x^k + e_k \cdot {dx}_j; {\color{red}D_i}) - f(x^k; {\color{red}D_i})}{{dx}_j}.$$ If $f_i$ is twice differentiable and has bounded Hessians near point $x^k$, the knowledge of partial derivative approximates the following equality: $\dfrac{\partial f_i (x^k)}{\partial x_j} \cdot dx_j \approx {f_i(x^k + e_j \cdot {dx}_j; {\color{red}D_i}) - f(x^j; {\color{red}D_i})}$. In the last approximate equality, $e_j$ is a unit norm vector of the standard basis of $\mathbb{R}^d$. Let's assume that:
\begin{enumerate}
	\item The $f_i(x)$ is linear with respect to ${\color{red}D_i}$ in original form, or there exists a bijective change of variable ${\color{red}D_i \to {D_i}'}$ such that $f_i(x)$ is linear in ${\color{red}D_i'}$.	
	\item Iterates $x^k$ are uniformly distributed in $\mathbb{R}^d$. 
\end{enumerate}

Under these assumptions, a training process might reveal sensitive information. Indeed, if an adversary obtains this information, then information about partial derivative provides noisy response linear function in ${\color{red}D_i}$. This general setting has been studied by \citep{dinur2003revealing}. Authors demonstrated that there exist linear attacks that could, with high probability, expose ${\color{red}D_i}$.

Specifically, let's denote the length of ${\color{red}D_i}$ is equal to $N$ bits. Let's assume adversarial obtains $K$ noisy answers (for a single partial derivative) with an additive error of $E=o(\sqrt{N})$ to each answer. Having $K \ge \theta(N^2/E^2)$, such adversarial queries can reveal information about ${\color{red}D_i}$ with high probability. The hidden constant is around $256$ but can be improved. A single full gradient $\nabla f_i(x; {\color{red}D_i})$ provides $d$ equations instead of $1$ for each response. This demonstrates that in specific modes the pretty big amount of partial derivative already reveals enough information to reconstruct the client's dataset. Such attacks are not only in theoretical interest but can be practically mounted \citep{cohen2018linear}, \citep{kasiviswanathan2013power}.

\subsection{Privacy and Security}
\label{app:privacy_vs_security}

As we have described in Appendix~\ref{app:overview_of_privacy_mechanisms}, different mechanisms can be used for \abr{FL} to provide privacy for different aspects training process. Each mechanism has a specific target of protection (such as input, output, execution, or communication channel) and a specific adversary model. The precise meaning of protection is defined by rigorous formalization and the details are crucial. For example, \abr{DP} protects the output of an algorithm so that it does not reveal sensitive information about the input data, while \abr{MPC} protects the input data from being exposed to other parties during computation. Recent research papers on \abr{FL} deployments have mostly used Local DP \citep{bhowmick2018protection}, or a combination of \abr{MPC} and centralized \abr{DP}.

In contrast, the proposed design in our work can be understood as a mechanism for providing security for the training process rather than privacy. The response obtained from the server by the client is computed securely and accurately with protection against attacks on the server or the communication channels to the server. The notion of privacy covers the guarantee that the target of protection can participate in an algorithm without being observed by unauthorized parties. Security, on the other hand, can be considered as a more holistic concept that specifies: the algorithm or protocol, how personal information and derived quantities are protected, the type and strength of attacks that are resisted, and the requirement of having a secret key to access the data. Sometimes, a complex security protocol may not be feasible or desirable for certain situations, and a privacy mechanism that is less comprehensive but more flexible may be a better choice.

\clearpage

\section{Usage of {AES} cipher during Distributed Training}
\label{app:aes_details}

This section overviews the current state-of-the-art block cipher \aesname{AES} and explains how it can be used for symmetric key encryption in scenarios where multiple client messages must be encrypted and decrypted securely.

\subsection{{AES} Block Cipher} 
A block cipher is a fundamental cryptographic primitive that transforms a fixed-length block of bits into another block of the same length using a secret key. The \aesname{AES} is today's most widely used secure block cipher. Some CPUs have hardware support for it. Examples of CPUs with x86 instruction set architecture that support \aesname{AES} in the hardware level are Intel Westmere, AMD Bulldozer, and ARM AARCH64 instruction set architecture example is ARM Cortex-A53. 

\aesname{AES} block cipher supports three key sizes: $128$, $192$, or $256$. The key size determines the level of security and the computational cost of the encryption and decryption operations. The key space $K$ for \aesname{AES-128} has size of $|K|=2^{128}$. The input message space $M$ and the output cipher text space $C$ of the block cipher \aesname{AES} for all types of keys have the same size equal to $128$ bits. Therefore cardinality of $M$ and $C$ is equal to $|M|=|C|=2^{128}$.

For each key $k\in K$, the \aesname{AES} block cipher maps $M \to C$ with a bijective function, with $M = C$, and $|M|=|C| \le \infty$. Essentially, the key $k$ is a selector of bijective mapping or permutation. Each key realizes the permutation of the input message. If we assume that each of two distinct keys $k_1, k_2 \in K$ implements different bijective mappings $M \to C$, then the number of permutations that \aesname{AES} can realize is $|K|=2^{128}$. Even though it's a big number, this is much smaller than $2^{128}!$, all possible permutations if input and output are $128$ bits in length. In the Cryptography community, the block cipher \aesname{AES} is sometimes observed as a primitive that implements a Pseudo Random Permutation (\abr{PRP}). In other words, this means that for a fixed key, it defines a permutation. The discrepancy between the number of all possible permutations and the number of \abr{PRPs}  that \aesname{AES} can realize is elegantly resolved in the Cryptography community. It is resolved by introducing the notion of a \abr{Secure PRP}. The algorithm which implements \abr{PRP} implements a \abr{Secure PRP}  if an adversary from observing the realization of permutation $f$ cannot distinguish by using an arbitrarily tractable algorithm between two events:
\begin{enumerate}
	\item The permutation function $f:M \to M$ is chosen uniformly at random from all possible $M!$ permutations.
	\item The permutation function $f:M \to M$ is chosen as one of the permutations that block cipher $\mathrm{BlockCipher}(\cdot,k):M \to M$ with $k \sim_{u.a.r} K$ can realize.
\end{enumerate}

By the current status in Cryptography, the \aesname{AES} is believed to be a secure \abr{PRP}. Having two input and output pairs, adversarial may wish to derive the secret key. The brute force search on \aesname{AES-128} is computationally infeasible because $2^{128}$ is too large to enumerate. The best-known attack on the full version of \aesname{AES-128} that can recover the complete key has a complexity of $2^{126}$ \citep{bogdanov2011biclique}. The \aesname{AES-128} is secure against brute force search and also against linear and quantum attacks \citep{jang2022quantum}.

\subsection{Internals of {AES} Block Cipher} The \aesname{AES} block cipher has $10$ rounds for \aesname{AES-128} and $14$ rounds for \aesname{AES-256}. The key is expanded into $11$ or $15$ subkeys of $128$ bits in length each. Subkeys are used in each round. Each round consists of four invertible steps: key addition (in the sense of exclusive boolean \textit{or}, which we will denote as XOR), byte substitution, row shift, and column mix. These steps transform a $4 \times 4$ matrix of bytes representing each round's input. Byte substitution adds non-linearity, row shift rotates each row cyclically, and column mix applies a linear transformation to each column. The last subkey is used for a final key addition (XOR) to mask the output. The \aesname{AES} decryption reverses the encryption steps. The \aesname{AES} has different implementations for different devices and code size requirements. All occurred transformations are stateless; consequently, the \aesname{AES} block cipher is stateless.

\subsection{Apply {AES} Block Cipher for more than one input block}

The \aesname{AES} is a secure block cipher, but it is a bad idea to use the same key $k$ to encrypt multiple blocks deterministically. This breaks the notion of Semantic Security, which essentially means that adversarial can deduce some information from analyzing sequences of ciphertexts from the block cipher (e.g. adversarial can, from observing $c_1=c_2$  conclude that $m_1 = m_2$, even adversarial does not know exactly values $m_1,m_2$). To preserve Semantic Security, the encryption should have protection from chosen-plaintext-attack, which informally denotes this kind of attack.

Formally, in chosen-plaintext-attack (\attackname{CPA}) an attacker may adaptively ask for the encryption $(e_1, e_2, \dots)$ of arbitrary messages $(m_1,m_2,\dots)$ of his choice. The attacker's goal is to obtain the ability to correctly guess from two obtained encryption $\{e_a,e_b\}$ from experiment $A$ and experiment $B$ which encryption belongs to which message from the set  $\{m_a, m_b\}$. The attacker does  not know encryption of $\{m_a, m_b\}$ in advance, and the attacker does not know which plain message has been used in which experiment. In the context of this attack, the advantage of adversarial is defined as: $${ADV}_{cpa} = |P(\{ \mathrm{ExperimentA\, uses\,} m_a \}) - P(\{ \mathrm{ExperimentB\, uses\,} m_b\})|.$$ Fundamentally, there are two ways to provide security against \attackname{CPA} attacks:

\begin{enumerate}
	\item Increase ciphertext space and allow encryption to work randomly. This is the underlying reason why \ecryptname{CKKS} is \attackname{CPA} secure. The downside of this is that ciphertext space is increasing.
	\item Augment key space $K$ with extra counter, named as a \textit{nonce} from a nonce space $N$. The clients who perform encryption guarantee that the pair $(key, nonce)$ is unique during the life of the $k$. The $nonce$ is open to everybody.
\end{enumerate}

Next, there are two different ways to select \textit{nonce} in nonce-based protection against \attackname{CPA} attacks:
\begin{enumerate}
	\item \textit{Deterministic counter.} In this mode, a single deterministic integer counter is used. Sometimes there is no need to send nonce itself with ciphertext during communication if the receiver of ciphertext can recover counter from other information.
	\item \textit{Randomized counter.} If encryption happens on several devices, the coordination of \textit{nonce} complicates the process. One way around this is to select $nonce \sim_{u.a.r.} N$, where $N$ is sufficiently large, e.g. $|N|>2^{128}$. Such nonce space size will guarantee that we expect to obtain during the sampling at least one collision after sampling $\sqrt{2|N|} + 1 = 2^{64}$ due to the famous Birthday Paradox.
\end{enumerate}

The two popular ways to use nonce-based encryption for \aesname{AES}, also known as \textit{Operation Modes}, are the following:
\begin{enumerate}
	\item \textbf{Nonce-based Cipher Block Chaining Mode} (\ecryptname{CBC}). In this schema, the input message is split into buckets. These inputs are substituted into \aesname{AES} block cipher, but the inputs themselves are masked before encryption. The first plaintext bucket is masked (via XOR) with \textit{nonce}. Other plaintext buckets are masked (via XOR) with the output from \aesname{AES} block cipher operated in the previous block. The output of this mechanism is the public nonce and sequence of encrypted input buckets. It can be proved that \aesname{AES/CPA} secure with advantage ${ADV}_{cpa} \le 1/2^{32}$ if use \aesname{CBC} for $2^{48}$ \aesname{AES} ($128$ bits) blocks. If \textit{nonce} is generated with non-secure PRG, then it should be firstly encrypted by itself with secure \aesname{AES} Block to generate secure \textit{nonce}.
	
	\item {\textbf{Randomized Counter Mode} \ecryptname{(CTR)}}. The input message is split into buckets. But in this mode, the mask for each bucket $i$ is generated with \aesname{AES} block cipher   $\aesname{AES}(k, nonce+i)$. The input bucket is masked with this value $\aesname{AES}(k, nonce+i)$ via XOR operation. It can be shown to guarantee \attackname{CPA} security with the advantage $1/2^{32}$ for it the number of transferred messages with \aesname{AES} should be at most $2^{64}$ \aesname{AES} Blocks. In addition, this operation mode is more flexible in contrast to \ecryptname{CBC} because this mode does not require any padding and can be implemented by discarding not needed bits in the last bucket.
\end{enumerate}

The \ecryptname{CKKS} and \aesname{AES} in \aesname{EAX} mode of operation \citep{bellare2004eax} are secure against \attackname{CPA} attacks.

\subsection{Apply {AES} with Integrity Guarantees}

The \attackname{CPA} security and operations mode solve security against eavesdropping, but they don't provide integrity of the delivered messages. Message Authentication Code(\ecryptname{MAC}) is Cryptography protected analogous to (non-secure) Cyclic Redundancy Check (\abr{CRC}), which is used in Networking communication to protect from random (not adversarial) noise in communication channels. The goal of \abr{MAC} is that the value of \abr{MAC} coupled with a ciphertext guarantees that the message has not been modified during the transfer by the adversary. The \textit{signing algorithm} for generating \abr{MAC} tag takes a plain message $m$ and secret key $k$ as input. The signature is added to the transferred message with the public nonce. Adversarial without a secret key $k$ cannot produce valid $(Message, Mac)$ pairs in such a way that it will pass the \textit{verification algorithm}. The \textit{verification algorithm} takes as input the secret key $k$, decrypted message $m$, and \abr{MAC} value $tag$. In practice, nonce and \abr{MAC} are typically $16$ bytes long. If the optimization algorithm already requires sending far more than $2$ elements of $\mathbb{R}$ encoded in \texttt{IEEE-754 FP64} format then this overhead is negligible during the training.

Two popular constructions to provide integrity in the context of \aesname{AES} are the following:

\begin{enumerate}
	\item \textbf{Encrypted CBC-MAC}. The input message is split into $16$ bytes buckets. After this, the first bucket is plugged into \aesname{AES} as input. Output from this block is masked with XOR  with the next input $16$ byte bucket. After the XOR operation is finished, the output of XOR is plugged into the next \aesname{AES} block as input, and the process repeats in a chaining fashion. It is important to notice that all \aesname{AES} blocks use the same key $k$. Finally, the last output is plugged into one more \aesname{AES} block as input, which uses another key $k'$ and schema releases \abr{MAC} tag.
	
	\item \textbf{Nested MAC}. In this approach, the message is split into $16$ bytes buckets. After this, the first \aesname{AES} block obtains as input this first bucket and $k$ is used as a secret key. The output of this \aesname{AES} block is used as a key for the next block, and data for this \aesname{AES} block is the next input bucket. After processing all messages, the last output is plugged into the last \aesname{AES} block as input, but for the last \aesname{AES} block, another secret key $k'$ should be used similarly as in \textbf{CBC-MAC}.
\end{enumerate}

For using \textbf{CBC-MAC} and \textbf{Nested MAC} securely, the number of signed messages (of arbitrary length) should be on the order of $2^{64}$ \aesname{AES} $128$-bits blocks. For details, please check the analysis of these schemes.

\subsection{Chosen Ciphertext Attack {CCA} and Authentication}

In a chosen ciphertext attack (\attackname{CCA}), the attacker has access to a decryption oracle and can decrypt any ciphertext of his choice. Also, the attacker can access encryption oracles and obtain encryption of arbitrary messages of his choice. Adversarial aims to break semantic security in the \attackname{CPA} sense. To provide the \attackname{CCA} security, the two processes \texttt{Encrypt} and {\texttt{MAC}} should be performed sequentially, and typically it's preferable to perform them in this order. In our paper, we have used \aesname{EAX}, which is \texttt{CTR} mode for encryption combined with \texttt{CMAC} for integrity purposes.

\subsection{Using {AES} with {EAX} Operation Mode in our paper} 
We use \aesname{AES} encryption with \ecryptname{EAX} mode, which produces a triple as output: 
\[\langle m=C_i(\nabla f(x^t) \in \mathbb{R}^d (16/32/64 \cdot d\,\mathrm{bits}), nonce \in N (128\, \mathrm{bits}), Tag (128\,\mathrm{bits}, ciphertext)\rangle\].

The ciphertext has the same bit length as the original message. Because \ecryptname{EAX} mode employs \ecryptname{CTR} for stream encryption, it does not require any padding. The $nonce$ is a random 16-byte value, and the $Tag$ is a $16$ byte message authentication code. As it has been mentioned in Section \ref{sed:resilience}, the \ecryptname{CKKS} and all \abr{HE} schemas cannot achieve \attackname{CCA} security \citep{fauzi2022ind}. For \aesname{AES}/\aesname{EAX}, it has been proved in \citep{bellare2004eax} that it's secure against \attackname{CCA} attacks. Cryptography is an area of science by itself. Readers for can gain more information about Classical Cryptography from \citep{boneh2020graduate}.

\clearpage

\section{Homomorphic Encryption with {CKKS}}
\label{app:ckks_details}

\subsection{Learning With Errors Problem and  REGEV09 Algorithm as a Concrete Example of SWHE}
\label{app:lwe}

The security of the \ecryptname{CKKS} scheme depends mainly on the hardness of the Learning With Errors(\abr{LWE}) problem which we will overview next. The \textit{"Learning with errors"} search problem represents was analyzed in \cite{regev2009lattices}. For this work Oded Regev has obtained a G{\"o}del prize in 2018:

\begin{equation*}
	\begin{aligned}
		\mathrm{find}\,s\in \ZS_q^n \\
		\mathrm{such\,that:}\, s^T \bA + e^T &= b \,\mathrm{where}\,e\,\mathrm{is\,r.v.} \\
	\end{aligned}
\end{equation*}

Parameters have the following properties: 

\begin{itemize}
	\item The $q$ is a prime number.
	\item $m > n$.
	\item $\bA\in \ZS_q^{n \times m}$ is a random matrix drawn u.a.r. from $\ZS_q^{n \times m}$.	
	\item $\beta \in \RS$.
	\item $e\in \ZS_q^m$ is small (in terms of $L_2$ norm) r.v. For r.v. $e$ there is no assumption about its distribution. The only one constraint is that $e \ne 0$.
\end{itemize}

As can be observed from the description LWE problem operates on a system of over-determined sets of equations. When $e=0$ the system can be solved via \textit{Gaussian elimination}, but when $e \ne 0$ we believe it's a hard search problem. Specifically, Theorem 1 from \cite{regev2009lattices} makes a connection between LWE and the search problem in integer lattices via the following theorem:

\begin{center}
	\textit{Instance of LWE problems with parameter $n$ is as hard as the Short Integer Solution Problem}.
\end{center}

The best well-known solution for solving the LWE problem works in the following time: ${q}^{\mathcal{O}\left({n}/{\log(n)}\right)}$ \cite{blum2003noise}.

There exists a variation of the LWE problem named as Decision Learning With Errors problem. In theory, it has been shown that Decisional LWE is as hard as LWE. In usual LWE we can not find a solution for a noisy set of linear equations effectively based on knowledge of $A$ and $b$. In \textit{Decisional LWE} the adversaries can not from observing tuple $(\bA,b)$ distinguish the following two scenarios:
\begin{enumerate}
	\item The tuple $(\bA,b)$ has been generated completely uniformly at random. In some sense, there are no real hidden linear dependence structures between columns of $A$ and $b$.
	\item The tuple $(\bA,b)$ in fact has a specific structure $(\bA=\bA,b=\bA^T s + e)$.
\end{enumerate}

The LWE problem provides a way to publish a lot of perturbated linear equations in variable $s$ in the form of $s^T \bA + e^T = b$, and essentially hide $s\in\ZS^n$ from parties who do not know exactly $e$.

Next, we will describe concrete examples of using this idea. The method described next represents a symmetric Somewhat Homomorphic Encryption (\abr{SWHE}) Learning With Errors (LWE) based scheme known as \textbf{{REGEV09}}. The method was proposed in 2009 at work \cite{regev2009lattices}. Its security properties are based on hardness to solve the decisional LWE.

\paragraph{KeyGen.} $s\in \ZS_q^n \sim U$ is a secret key, $q$ is a prime, $n$ is a secure parameter.

\paragraph{Encrypt.} Encryption is working bit by bit. The message that we encrypt without loss of generality can be considered as  $m\in\{0,1\}$. The algorithm for encryption produces ciphertext as $c=(a,b) \in \ZS_q^n \times \ZS_q$ via following rules:

\begin{enumerate}
	\item $e \in \ZS$  is a "short" noise generated from some distribution, satisfied constraint $|e|<q/4$).
	\item $a \in \ZS_q^n$  is a random vector sampled uniformly from its domain.
	\item $b \eqdef \langle a,s \rangle + e + (m \cdot \lfk q/2 \rfk)|\mod  q$.
	\item The released ciphertext $c=(a,b) \in \ZS_q^n \times \ZS_q$
\end{enumerate}

\paragraph{Decrypt.} Decryption happens via using the following formula which involves function ${round}_a(x)$ which round to $0$ or $a$ depends on what is more close to $x$.
\begin{eqnarray*}
	\hat{m} &=& \dfrac{round_{q/2}\left(b- (\langle a,s \rangle \pmod q)
		\right) }{q/2} \\
	&=& \dfrac{round_{q/2}({\color{red}{(\langle a,s \rangle + e + m \cdot \lfk q/2 \rfk)}} - \langle a, s \rangle)}{q/2} = \dfrac{{round_{q/2}(e + {\color{red}{m}} \cdot \lfk q/2 \rfk))}}{q/2} .
\end{eqnarray*}

For decryption to work correctly, we need to have $e \in (-\lfk q/4 \rfk, \lfk q/4 \rfk )$. Now let's verify that this formula works correctly:
If ${\color{red}{m=0}}$, and because $|e| < \lfk q/4 \rfk$ $\implies \hat{m} = 0$. If ${\color{red}{m=1}}$, and because $|e| < \lfk q/4 \rfk \implies e > -\lfk q/4 \rfk$ $\implies \hat{m} = 1$.\\

\paragraph{HE Add.} We add two ciphertext messages $c_{i}=(a_{i},b_{i}),i \in \{0,1\}$ by components:
\begin{equation*}
	(a_1 + a_2, b_1 + b_2) = (a_1+a_2, \langle a_1 + a_2, s \rangle + (e_1 + e_2) + (m_1 + m_2) \lfk q/2 \rfk \pmod q .
\end{equation*}

This schema is SWHE and homomorphic additive. In the worst case error is doubled in each addition operation of each bit. If the adversary listens to the channel, he can collect $(a,b)$ for each transferred bit and construct $\bA=[a_1, a_2,\dots]$ and the right-hand side $b$. But encoding of ciphertext directly follows LWE problem description with additive error correction code $e + m \cdot \lfk q/2 \rfk$. By properties of the \textit{Decision LWE} Problem, adversaries can not get any information from it because data distribution $(\bA,b)$ is indistinguishable from a uniform.

\subsection{Introduction to CKKS}
There are different variants of \abr{FHE}, \abr{LFHE}, and \abr{SWHE}, but most of them operate on boolean or integer arithmetic. In our paper, we compare \algnamewithaes{DCGD/PermK/AES} against \algname{GD} with Cheon-Kim-Kim-Song (\ecryptname{CKKS}) \citep{cheon2017homomorphic} schema. This schema allows approximate arithmetic on encrypted real and complex numbers and dense linear vectors. 

The \ecryptname{CKKS} schema violates property (1) of \abr{FHE} from Appendix \ref{app:overview_of_privacy_mechanisms}, and it is specialized to work with these two linear spaces $\mathbb{R}^d$ and $\mathbb{C}^d$. Therefore, \ecryptname{CKKS} is an \abr{SWHE} scheme. Still, it is sufficient for most Machine Learning problems that operate on real spaces during the training phase, and it is the most popular schema in Machine Learning applications that require \abr{HE}. Supporting multiple keys (MK) with privacy guarantees is an active research topic \citep{kluczniak2023circuit}. None of \ecryptname{MK-CKKS} possible schemas are currently implemented 
in \libname{TenSEAL} \citep{benaissa2021tenseal} or \libname{SEAL} \citep{seal} libraries. In our work, we used the classical single-key \ecryptname{CKKS} scheme for comparison because it's suitable for our setting. 

As we mentioned, the \ecryptname{CKKS} scheme allows us to perform \abr{HE} computations on vectors of complex and real values. \ecryptname{CKKS} provides ways to perform element-wise addition, multiplication, and rotation of elements in encrypted form. The \ecryptname{CKKS} uses three different keys:

\begin{itemize}
	\item \textit{Public key.} The public key in \ecryptname{CKKS} schema is used for the encryption of plain messages.
	
	\item \textit{Relinearization key}. Relinearization key reduces the size of a ciphertext after arithmetic operations on ciphertexts.
	
	\item \textit{Private(secret) key}. The private (secret) key is used for decryption of encrypted messages.
	
\end{itemize}

To perform arithmetic operations on encrypted vectors on the master device, the master should know the public key and the master should know the relinearization keys to reduce the size of a ciphertext after arithmetic operations. The public key in \ecryptname{CKKS} schema is used for encryption and can be shared with the master. However, a private(secret) key is used for decryption and must be kept confidential from the master if clients do not trust the master. The security of the scheme depends on the hardness of the Ring Learning With Errors (\abr{RLWE}) problem. The \abr{RLWE} is a generalization of \abr{LWE} problem, but instead of using $\mathbb{Z}_q^n$ field, it is based on underlying algebra which is polynomials over a ring. \abr{RLWE} inherits useful properties of hardness from \abr{LWE}, but it is more space efficient. For an overview of \abr{LWE} see Appendix~\ref{app:lwe}.

\subsection{Description of processes inside CKKS}

The schematic process of how \ecryptname{CKKS} operates is depicted in Fig.\ref{fig:ckks_enc}. The general schema of \ecryptname{CKKS} consists of several steps.

\paragraph{First Step.} Firstly, a vector of values $m \in \mathbb{R}^{N/2}$ on which we want to perform certain computations is encoded into a plaintext polynomial $p(x)$. This is necessary because encryption, decryption, and other mechanisms work on polynomial commutative rings. In the case of encoding into polynomials from $\mathbb{Z}[X]/(X^N+1)$, the encoding process is reversible. In the original paper, \cite{cheon2017homomorphic} this process is carried out using canonical embedding. This embedding is carried in such a way that if evaluate the obtained polynomial from $\mathbb{Z}[X]/(X^N+1)$ at the roots of cyclotomic polynomial $X^N+1$ we will recover entries of the original message of $n \in \mathbb{R}^{N/2}$.

\paragraph{Second Step.} Next, the actual encryption part encrypted the plain polynomial with integer coefficients via a public key into polynomial $(c0, c1) \in (Z_q[x]/(X^N+1))^2$. The $q$ is a modulus number in the \ecryptname{CKKS} scheme. Here $q=\prod_{i=1}^{L}q_i$, where $q_i$ are prime numbers. So $q$ is not necessarily a prime number, but it is chosen to be a product of several prime numbers. To carry encryption the public key in the form of 
$Z_q[X]/(X^N+1))$ after specific transformation is used. It means that the size of the public key in bits is approximately equal to $N \cdot q$. To have \aesname{AES-128} security level for \ecryptname{CKKS} the key size is at least equal to $420$ KBytes.

\paragraph{Third Step.} Next, the algebraic operations are carried on encrypted messages $(c_0, c_1)$ by specific rules which we will not go deep into.

\paragraph{Fourth Step.} After carrying out the arithmetic operation, the result will be from the same space $(c0, c1) \in (Z_q(x)/(x^N+1))^2$. To carry decryption the private key in the form of $Z_q[X]/(X^N+1))$ after a specific transformation is used. Therefore it means that the size of the private key in bits is approximately equal to $N \cdot q$. To have \aesname{AES-128} security level the key size at least equal should be equal to $420$ KBytes.

\paragraph{Fifth Step.} Finally during decoding, a message from space $(Z_q(x)/(x^N+1))^2$ to plain text in $Z[x]/(x^N+1)$ and finally perform decoding into $\mathbb{R}^{N/2}$.

\newpage

\subsection{CKKS Configuration Equivalent to AES-128}

To provide privacy grantees similar to \aesname{AES-128} encryption: $N$ should satisfy this condition $N>16384$, and $q=\prod_{i=1}^{K}q_i$ should be at least $438$ bits long. This ensures that the security parameter $N$, $q$ is large enough to provide sufficient security. These details can be found in the reference implementation of Microsoft Research \libname{SEAL} Library
\footnote{\href{https://github.com/microsoft/SEAL/blob/master/native/src/seal/util/hestdparms.h}{https://github.com/microsoft/SEAL/blob/master/native/src/seal/util/hestdparms.h} - Microsoft Research SEAL Library.}. This is an underlying reason why \ecryptname{CKKS} has a more memory requirement compared to \aesname{AES-128}. The encoding consists of $2$ polynomials with $2 \max(d, N)$ coefficients, and each coefficient is not 32 (for FP32) or 64 (for FP64) bits long, but it's essentially $q=438$ bits long. In addition, when input/output vectors do not match $N$ \ecryptname{CKKS/HE} requires performing chunking of input and output - it requires additional operations to maintain the correctness and efficiency of the computation.

In all our experiments, we used the following \ecryptname{CKKS} configuration inside \libname{TenSeal} \citep{benaissa2021tenseal} library version 0.3.14. For this library to obtain \aesname{AES-128}, we have used the recommended configuration:

\begin{enumerate}
	\item Polynomial degree: $2^{14}=16\,384$
	\item Coefficient modulus: $q_1, q_2, q_3, q_4, q_5 = (60, 30, 30, 30, 60)$ bits, which corresponds to $q$ size of $210$ bits. This configuration is recommended, by \libname{TenSeal} for \aesname{AES-128} security level.
	\item Scale factor: $2^{30}$
	\item Scheme type: \ecryptname{CKKS}
\end{enumerate}

\tikzstyle{startstop} = [rectangle, rounded corners, minimum width=3cm, minimum height=1cm,text centered, draw=black, fill=red!30]
\tikzstyle{process} = [rectangle, minimum width=3cm, minimum height=1cm, text centered, draw=black, fill=orange!30]
\tikzstyle{arrow} = [thick,->;,>;=stealth]

\begin{figure}
	\centering
	\begin{tikzpicture}[node distance=2cm]
		\label{fig:ckks_enc}
		\node (input) [startstop] {Input message $m \in \mathbb{R}^{N/2}$ };
		\node (encode) [process, below of=input] {Encode message $m \in \mathbb{R}^{N/2}$ into $m' \in \mathbb{Z}[x]/(x^N+1)$ };
		\node (encrypt) [process, below of=encode] {Encrypt message $m' \in \mathbb{Z}[x]/(x^N+1)$ into $m'' \in \left(\mathbb{Z}_q[x]/(x^N+1)\right)^2$ };	
		\node (compute) [process, below of=encrypt] {Computation on encrypted messages $\left(m_1'', m_2'', \dots\right)$. CKKS allows to perform: addition, multiplication, and rotation.};
		\node (decrypt) [process, below of=compute] {Decrypt $m'' \in \left(\mathbb{Z}_q[x]/(x^N+1)\right)^2$ into $m' \in \mathbb{Z}[x]/(x^N+1)$};
		\node (decode) [process, below of=decrypt] {Decode message $m' \in \mathbb{Z}[x]/(x^N+1)$ into $m \in \mathbb{R}^{N/2}$ };	
		\node (output) [startstop, below of=decode] {Output message $m \in \mathbb{R}^{N/2}$ };	
		
		\draw [->] (input) -- (encode);
		\draw [->] (encode) -- (encrypt);
		\draw [->] (encrypt) -- (compute);
		\draw [->] (compute) -- (decrypt);
		\draw [->] (decrypt) -- (decode);
		\draw [->] (decode) -- (output);
	\end{tikzpicture}
	\caption{A high-level view of operations inside the \ecryptname{CKKS} schema.} \label{fig:ckks_enc}
\end{figure}
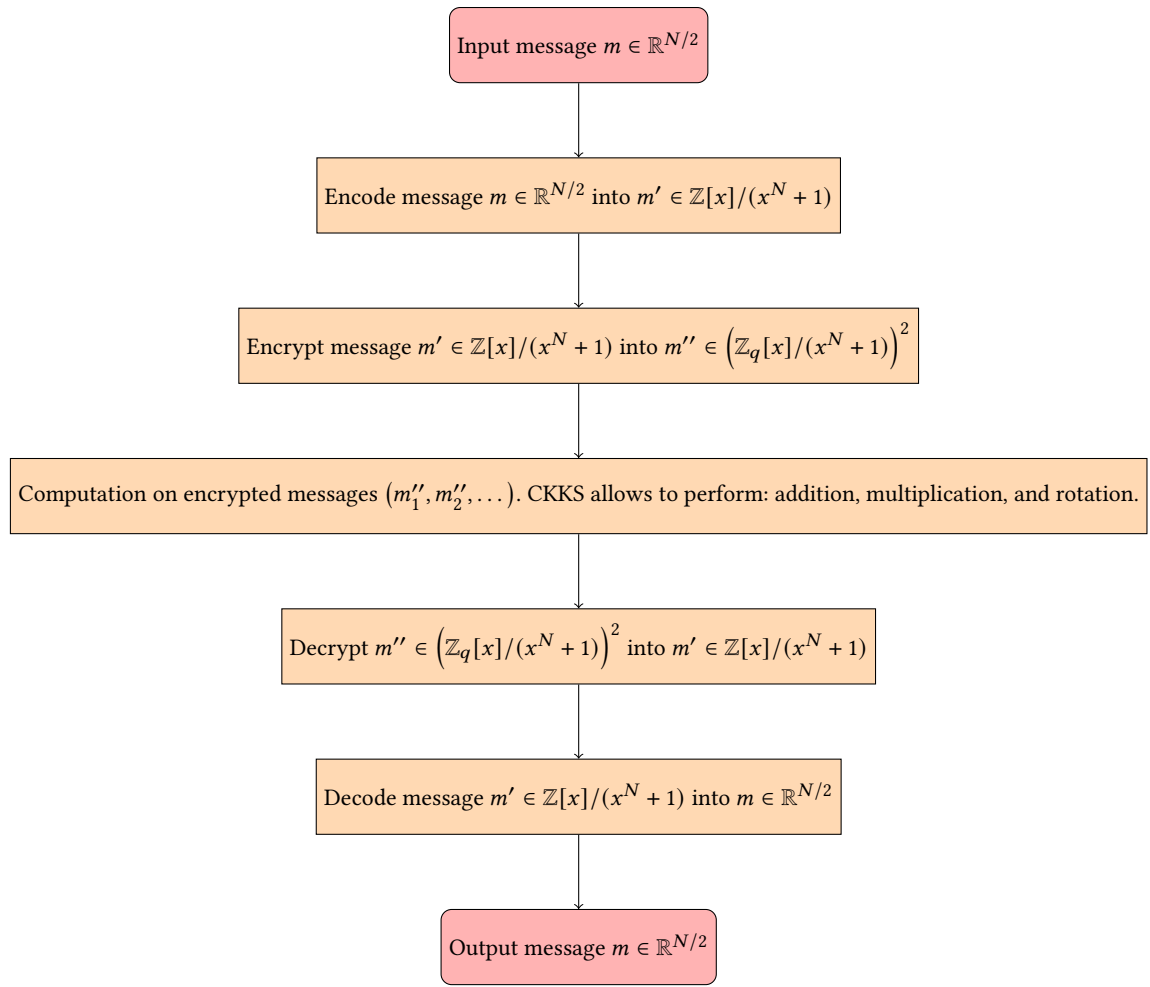

\clearpage

\section{Extra Experiments}
\label{app:extra_experiment}

\subsection{Exploring Problem Dimension}
\label{app:simulation_experiment}

\begin{figure*}
	\centering
	\captionsetup[sub]{font=scriptsize,labelfont={}}	
	\captionsetup[subfigure]{labelformat=empty}
	
	\begin{subfigure}[ht]{0.33\textwidth}
		\includegraphics[width=\textwidth]{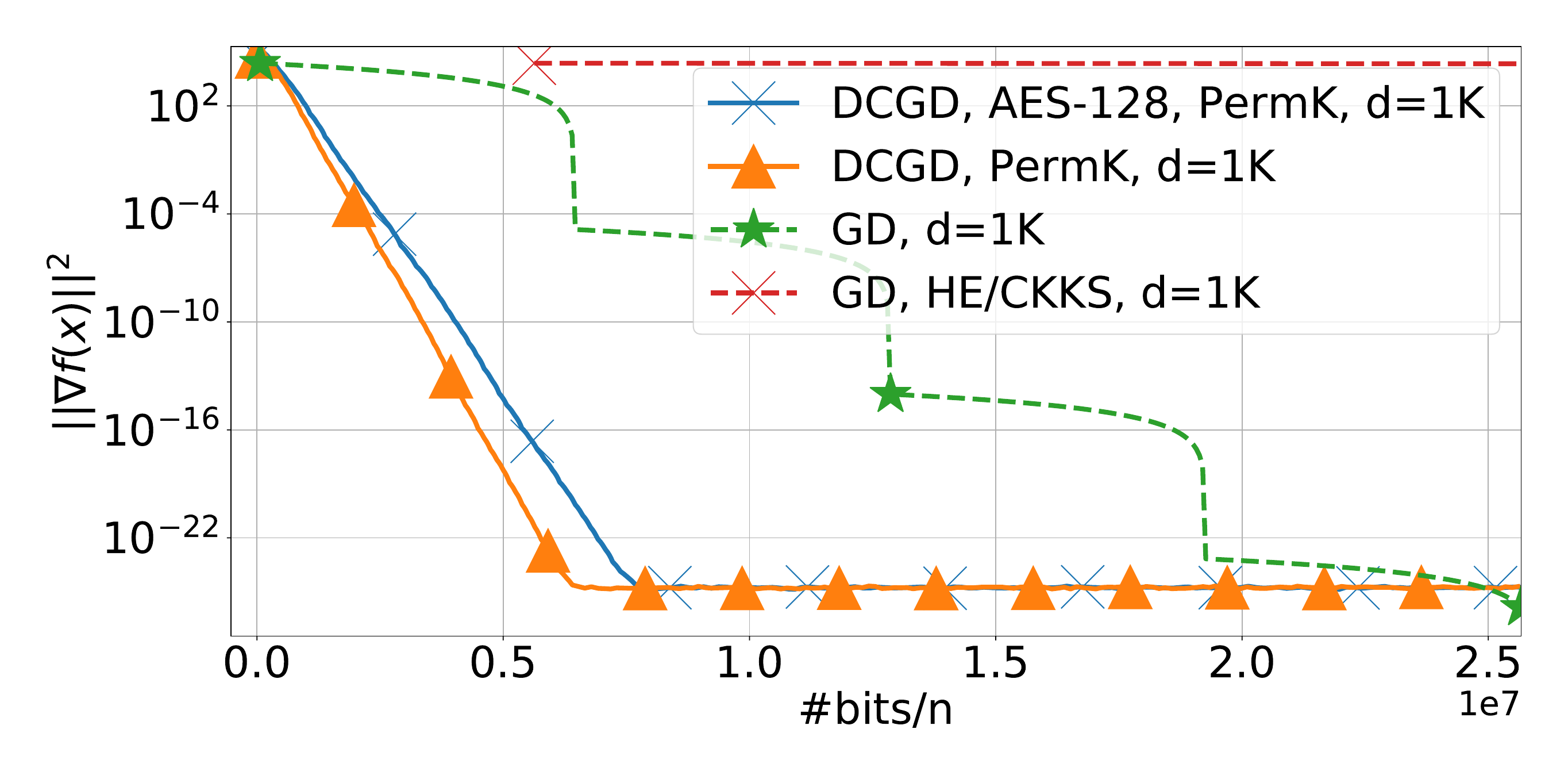} 
		\vspace{-1.5\baselineskip}
		\caption{{  }}
	\end{subfigure}
	\begin{subfigure}[ht]{0.33\textwidth}
		\includegraphics[width=\textwidth]{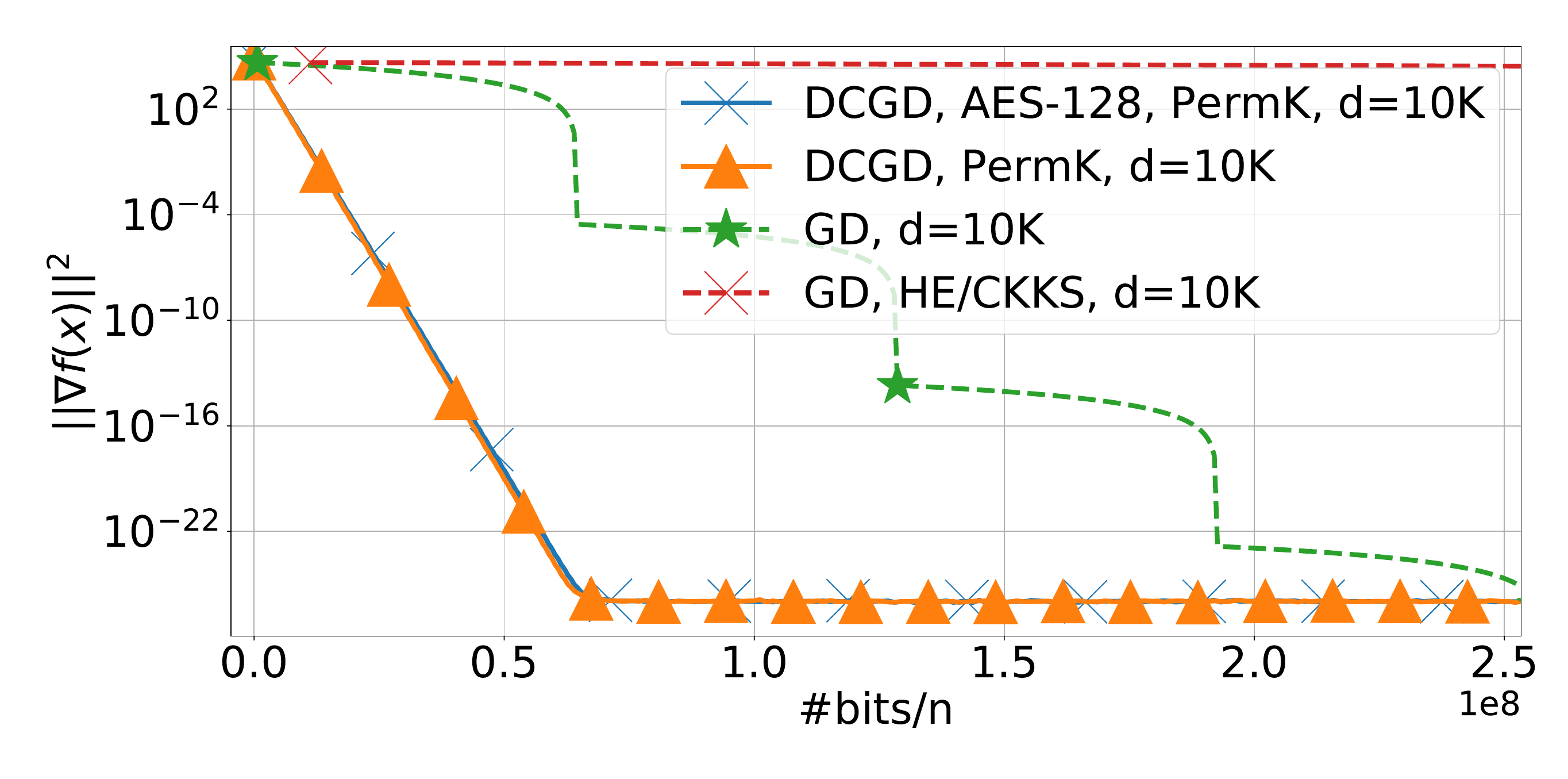} 
		\vspace{-1.5\baselineskip}
		\caption{{ }}
	\end{subfigure}
	\begin{subfigure}[ht]{0.33\textwidth}
		\includegraphics[width=\textwidth]{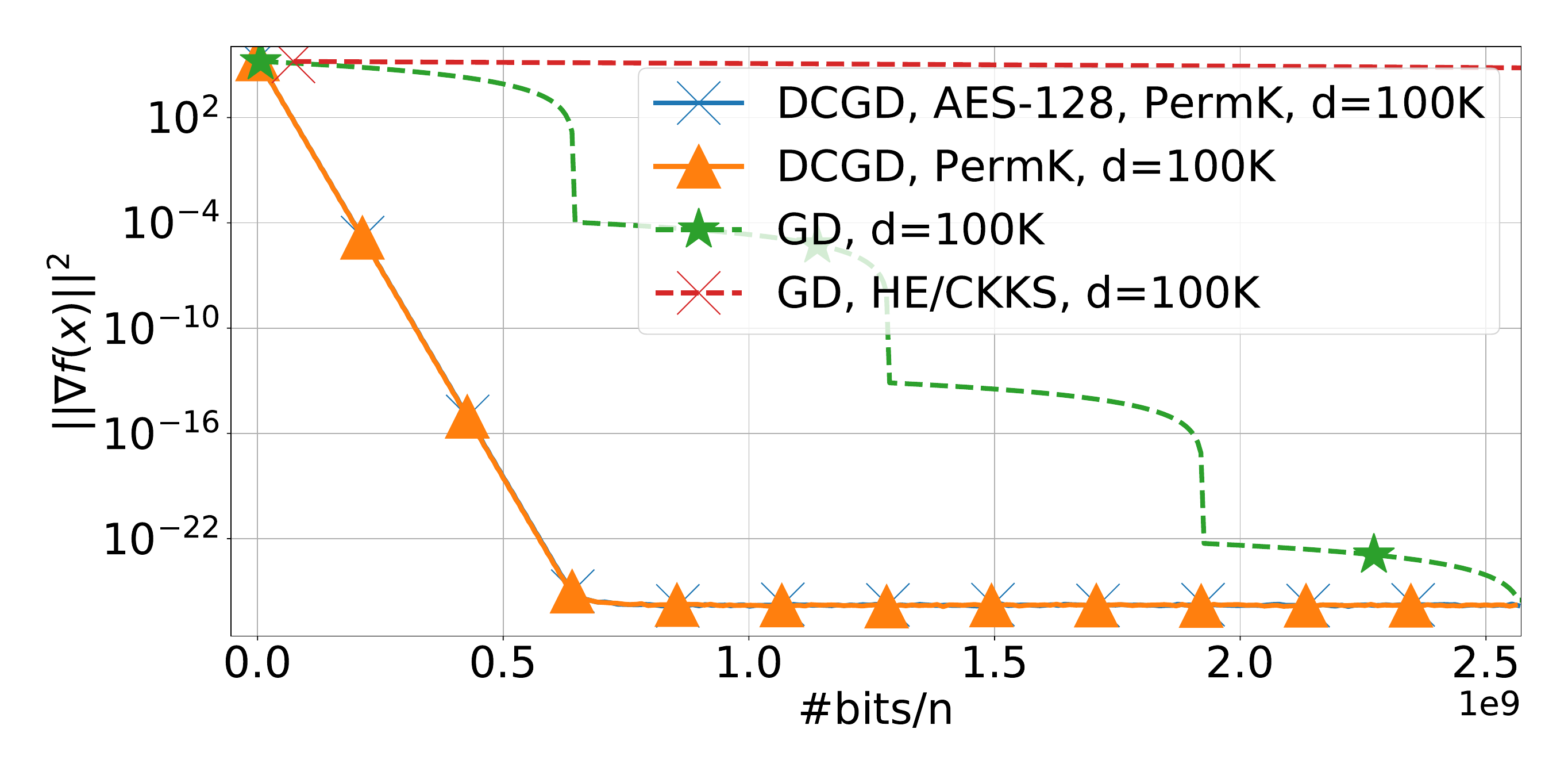}
		\vspace{-1.5\baselineskip}
		\caption{{  }}
	\end{subfigure}

	\begin{subfigure}[ht]{0.33\textwidth}
		\includegraphics[width=\textwidth]{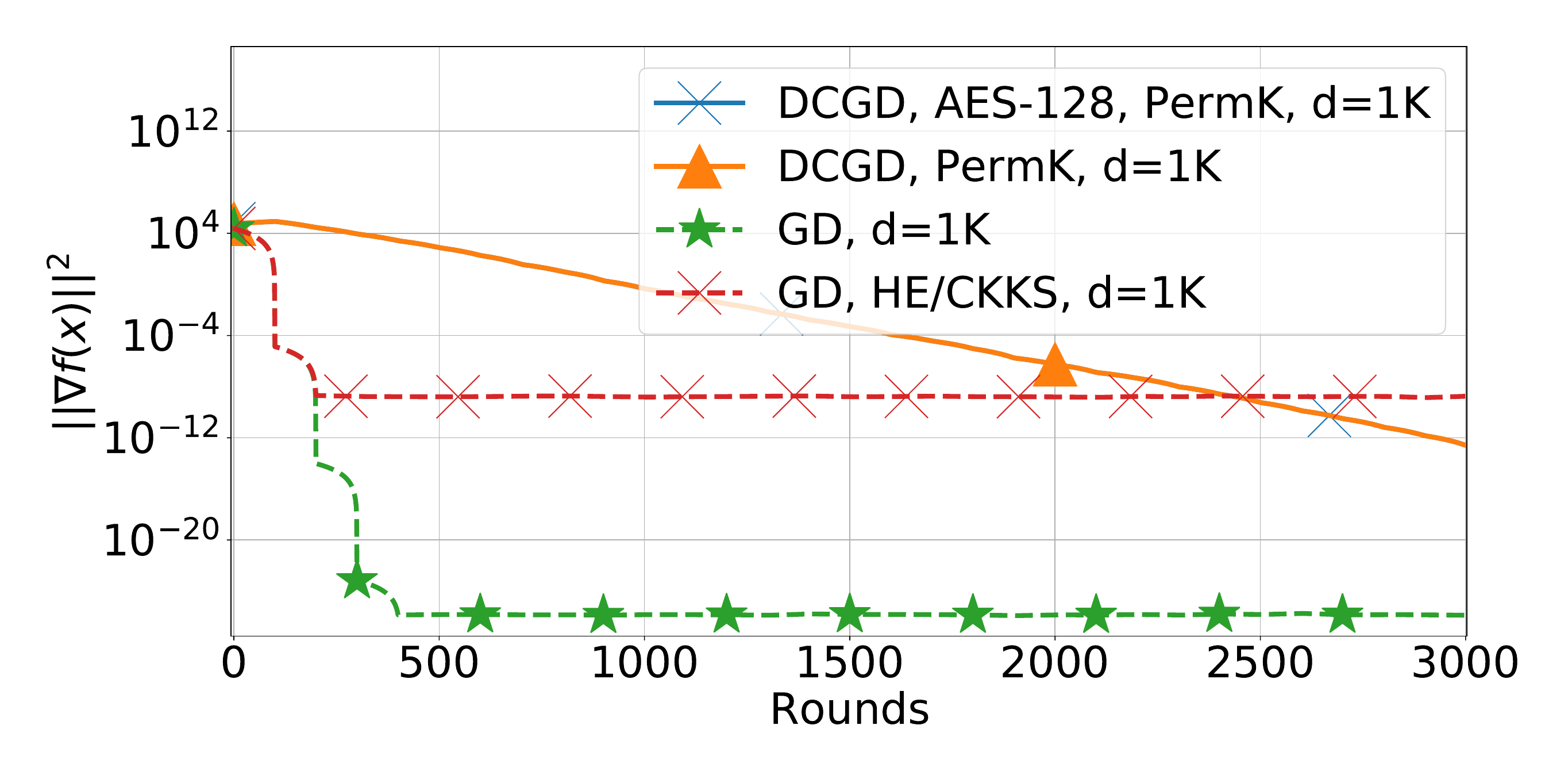} 
		\vspace{-1.5\baselineskip}
		\caption{{  }}
	\end{subfigure}	
	\begin{subfigure}[ht]{0.33\textwidth}
		\includegraphics[width=\textwidth]{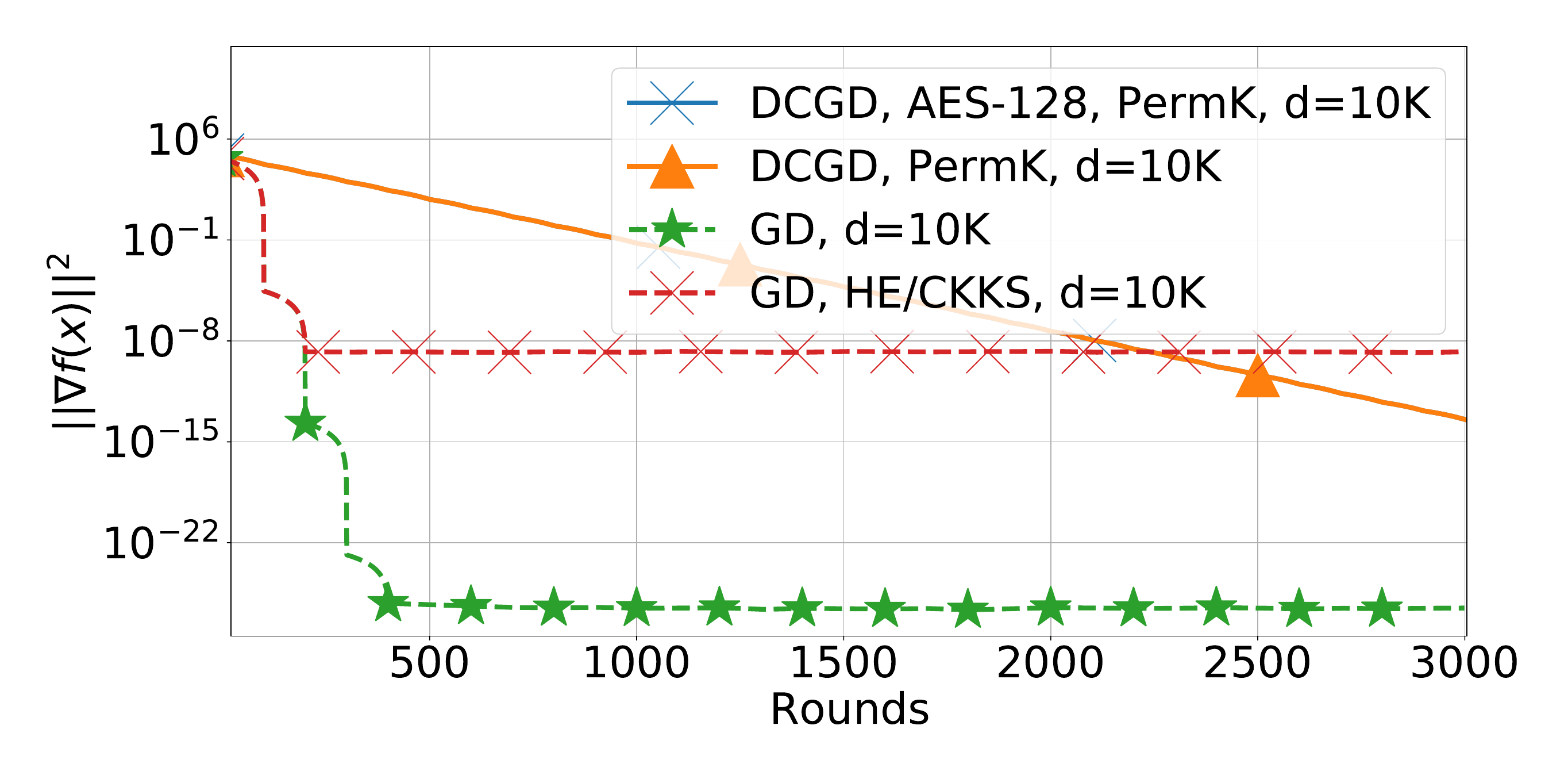} 
		\vspace{-1.5\baselineskip}
		\caption{{  }}
	\end{subfigure}
	\begin{subfigure}[ht]{0.33\textwidth}
		\includegraphics[width=\textwidth]{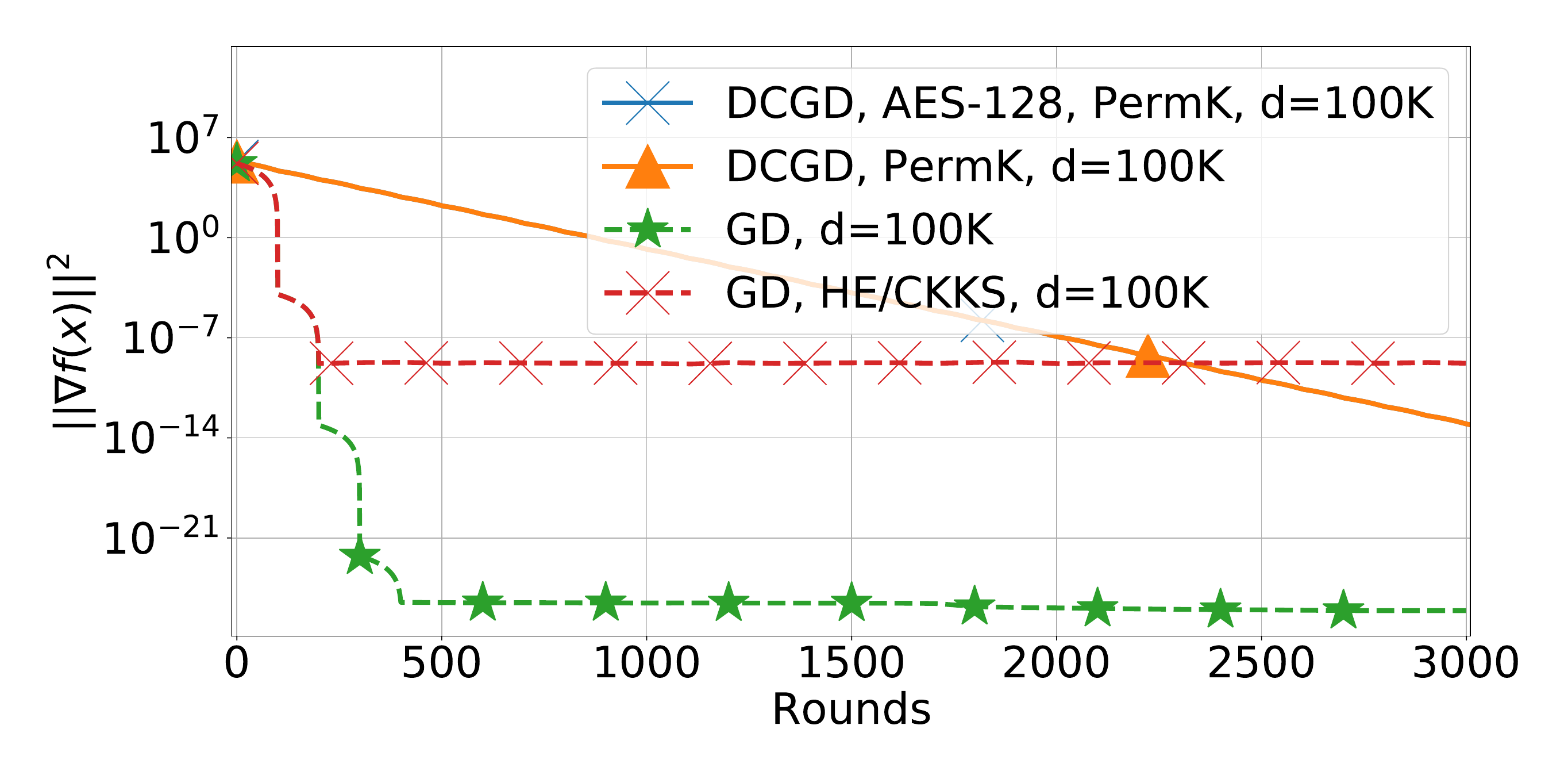} 
		\vspace{-1.5\baselineskip}
		\caption{{ }}
	\end{subfigure}

	\begin{subfigure}[ht]{0.33\textwidth}
		\includegraphics[width=\textwidth]{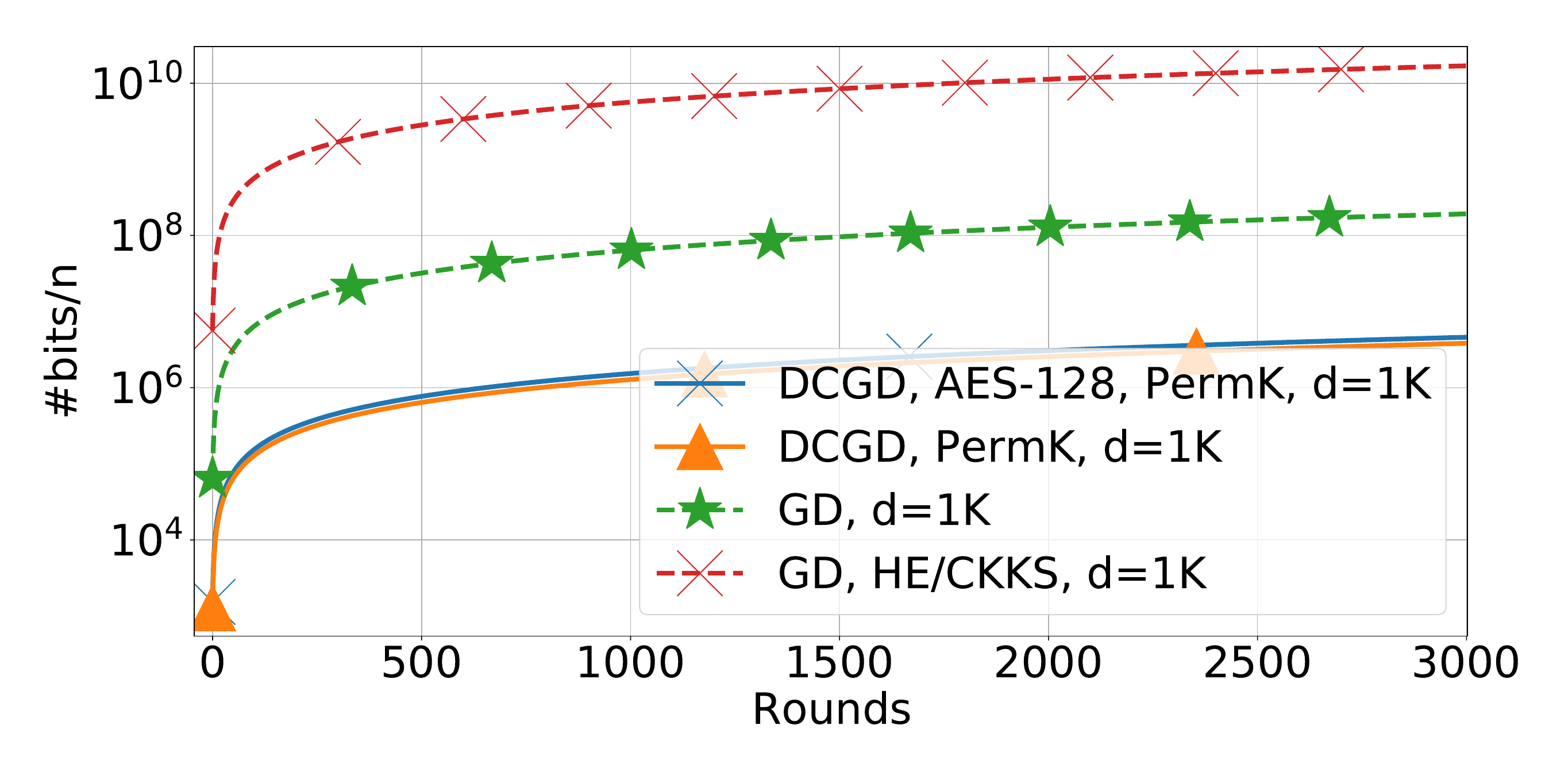} 
		\vspace{-1.5\baselineskip}
		\caption{{  }}
	\end{subfigure}
	\begin{subfigure}[ht]{0.33\textwidth}
		\includegraphics[width=\textwidth]{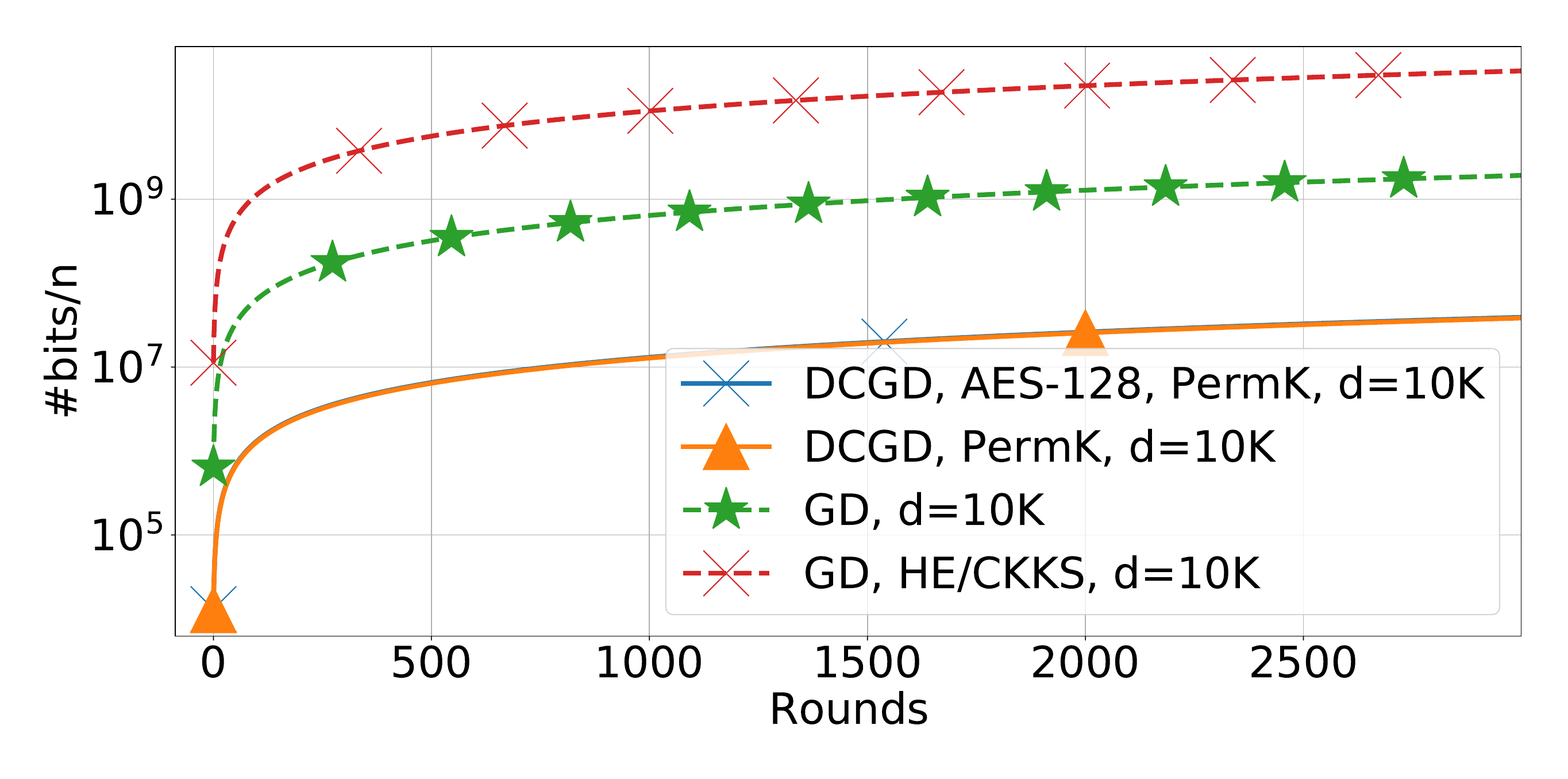} 
		\vspace{-1.5\baselineskip}
		\caption{{  }}
	\end{subfigure}
	\begin{subfigure}[ht]{0.33\textwidth}
		\includegraphics[width=\textwidth]{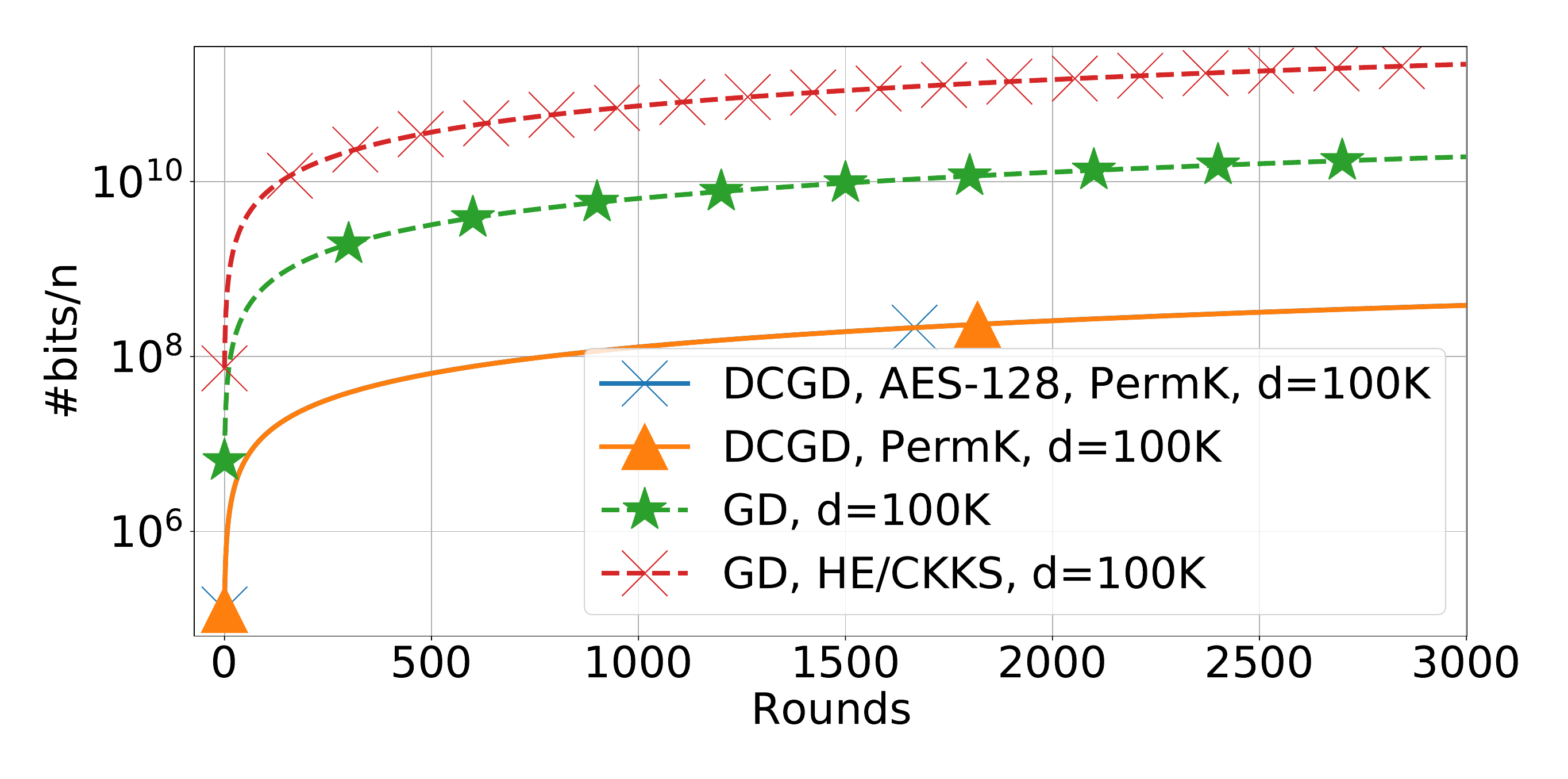} 
		\vspace{-1.5\baselineskip}
		\caption{{  }}
	\end{subfigure}

	\begin{subfigure}[ht]{0.33\textwidth}
		\includegraphics[width=\textwidth]{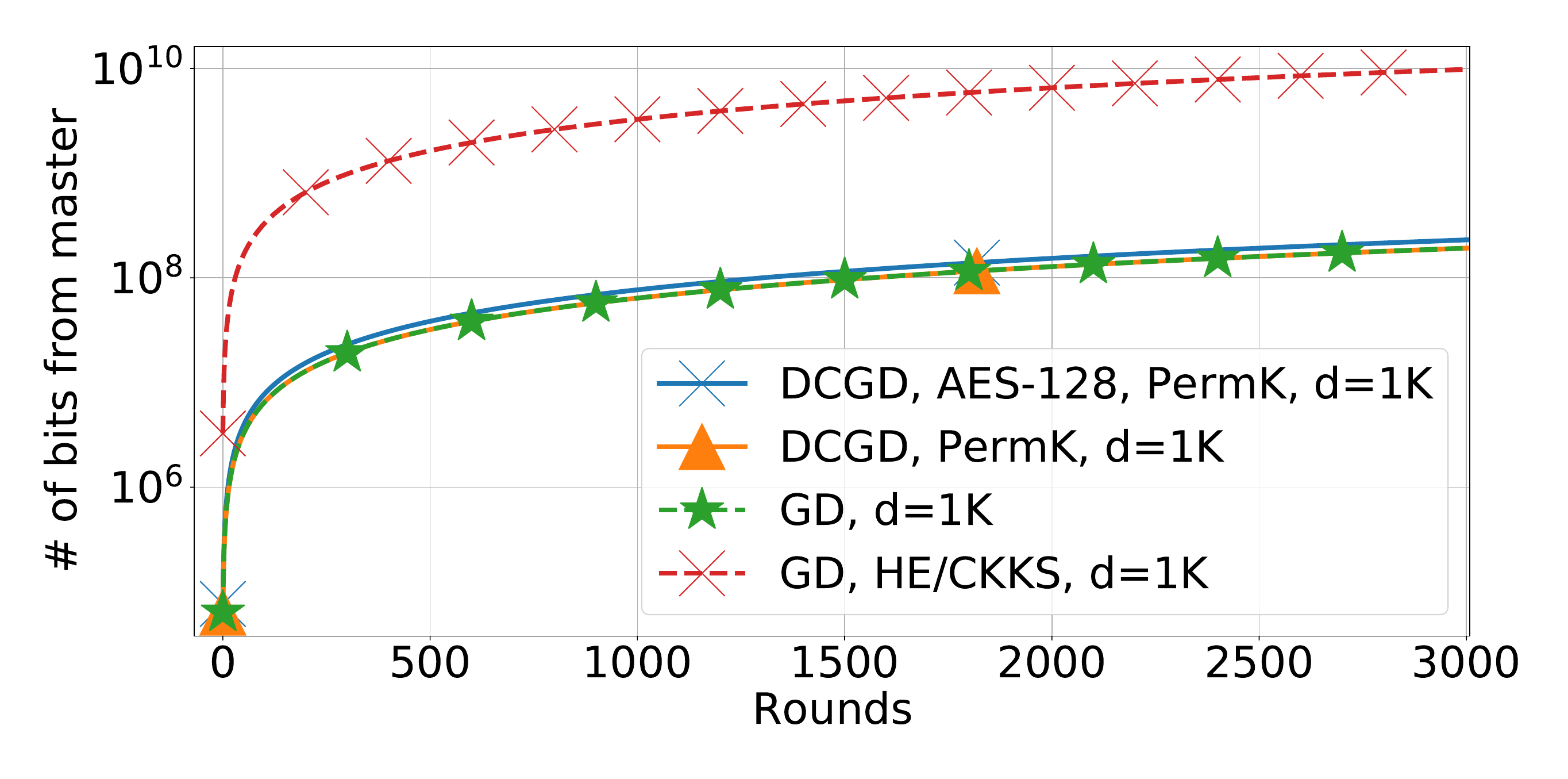} 
		\vspace{-1.5\baselineskip}
		\caption{{ (a) $d=1\,000$ }}
	\end{subfigure}
	\begin{subfigure}[ht]{0.33\textwidth}
		\includegraphics[width=\textwidth]{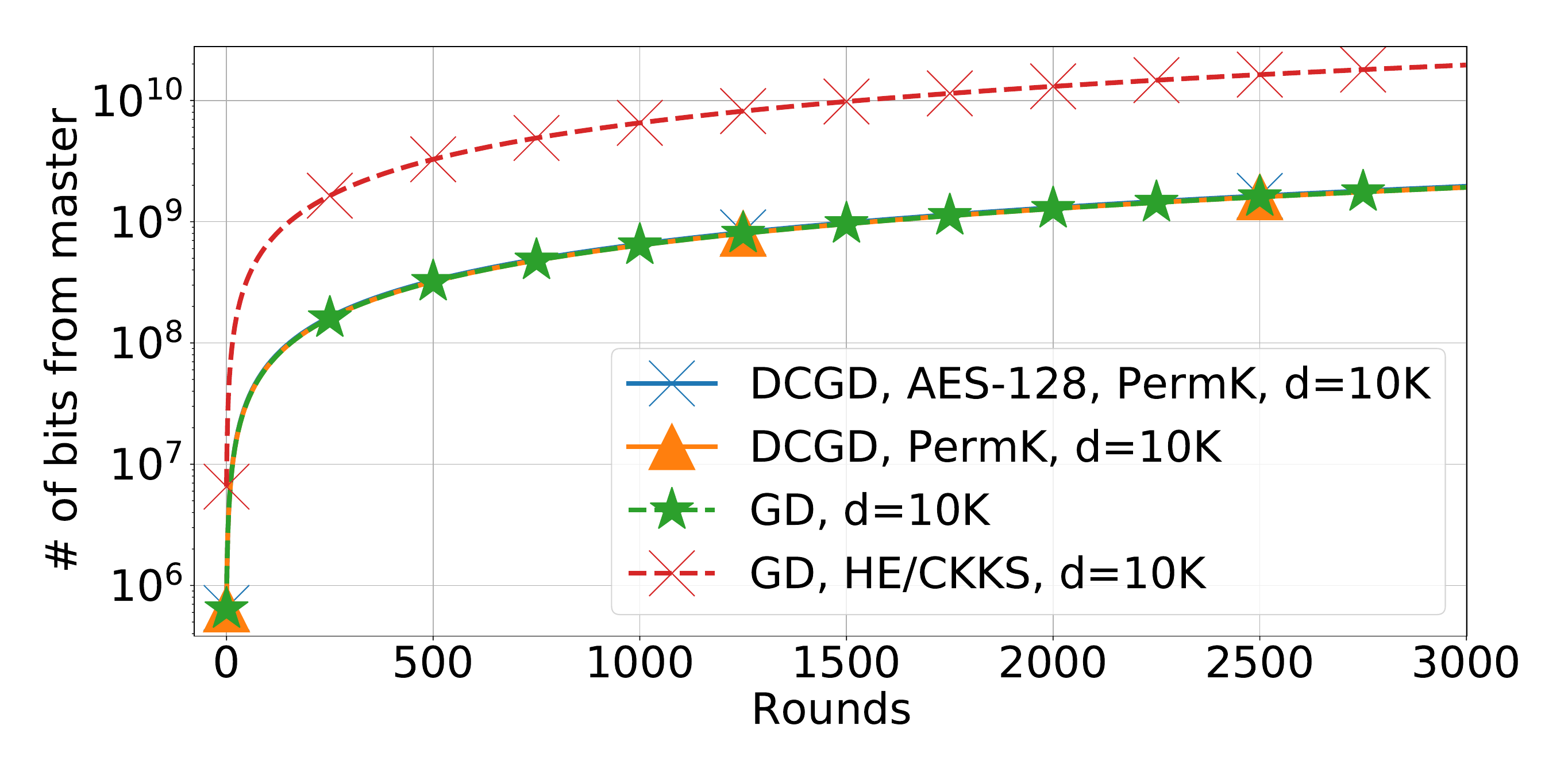} 
		\vspace{-1.5\baselineskip}
		\caption{{ (b) $d=10\,000$ }}
	\end{subfigure}
	\begin{subfigure}[ht]{0.33\textwidth}
		\includegraphics[width=\textwidth]{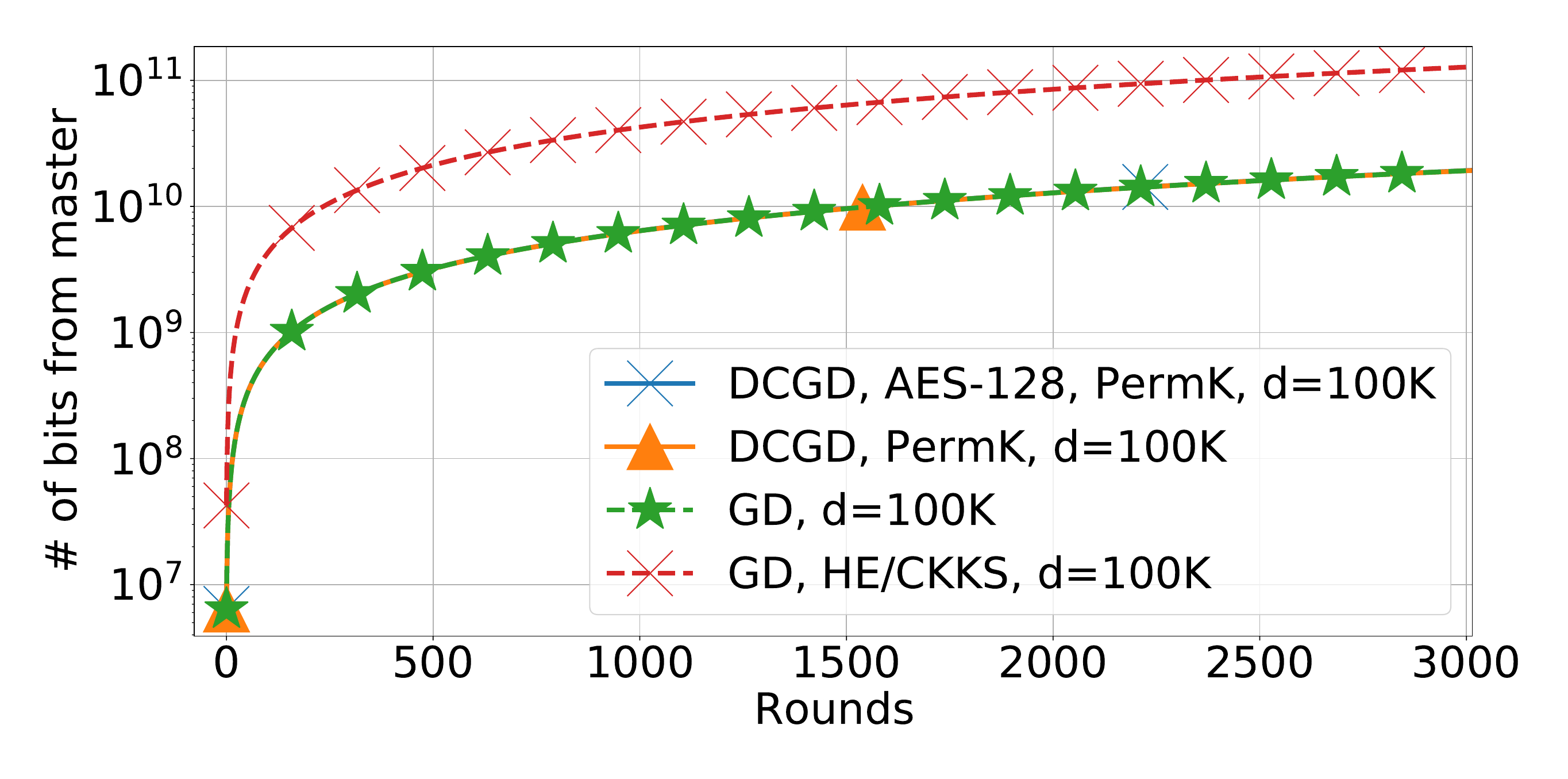} 
		\vspace{-1.5\baselineskip}
		\caption{{ (c) $d=100\,000$ }}
	\end{subfigure}
	
	\vspace{-0.5\baselineskip}
	\caption{{\modelname{Linear Regression} in interpolation, $n=50$, $n_i=12$, compute FP64. Tuned $\gamma=0.007$ for \algname{DCGD}. Theoretical $\gamma$ for \algname{GD}.}}
	\label{fig:exp_syn_7}
\end{figure*}

This experiment investigates the impact of problem dimension $d\in \{10^3,10^4,10^5\}$. Fig.~\ref{fig:exp_syn_7} shows that the \ecryptname{CKKS} overhead from encryption is $\times 10^3$ more both in master to client, and client to master communication direction compare to \aesname{DCGD/PermK/AES}. With $d=10^6$ the memory footprint for \ecryptname{CKSS} configured to guarantee the same guarantees as \aesname{AES-128} in the master to store $n=50$ encrypted gradients is $46$ GBytes, rendering storage of such information in the master challenging. The best convergence relative to the volume of information sent to the master is achieved with \algname{DCGD/PermK}. The behavior of \algname{DCGD/PermK} and \algnamewithaes{DCGD/PermK/AES} is indistinguishable for $d>10K$. Despite the ciphertext size being the same as the input when using \aesname{AES}, proper use of \aesname{AES} block ciphers for communication requires the addition of a Message Authentication Code (\abr{MAC}) for protection against malicious errors and a unique pseudo-random identifier (nonce). Each of these adds an overhead of $16$ bytes. It explains different behavior observed for \algname{DCGD/PermK} and \aesname{DCGD/PermK/AES} at $d=1K$ in Fig.~\ref{fig:exp_syn_7}. 

Given that, \algnamewithaes{DCGD/PermK/AES} emerges as a more viable alternative to \ecryptname{CKKS} in \abr{FL} context, in the setting when \abr{HE} previously has been applied.

\clearpage

\subsection{Compute and Communication Overlap for DCGD/PermK}
\label{app:simulation_experiment}

In this experiment, we simulated a compute and communication environment with realistic assumptions and evaluated the performance of \algname{DCGD/PermK}. We ignored overheads from the Operating System (OS) and its components (drivers, kernel, services). Firstly, we provide background on Network Communication and System Architecture to justify our modeling choices.

\subsubsection{Background in Network Communication}
\label{app:systems_background}



\paragraph{Communication Bandwidth During Master Client Communication}
We consider a scenario where clients are connected to the master via the Internet. These devices are called \textit{end systems} or \textit{hosts} in the context of a communication network. The communication between end systems is facilitated by intermediate \textit{routing devices}, such as routers and switches, connected by \textit{links}. Internet communication is based on \textit{packet switching}, which means that exchange messages $m$ are divided into smaller units called \textit{packets}. Then, they are delivered with the best effort through the communication network without reserving a path(circuit) from sender to receiver in advance. 

The \textit{bandwidth} of a link is the maximum rate at which data can be transferred over it and is measured in bits/second. This rate depends on various factors and generally can vary over time. It depends on:

\begin{enumerate}
	\item The distance between the linked devices.
	\item The type of physical medium for carrying the signal.
	\item The number of packets in the queue to transfer in routing devices.
	\item The probability that clients on the Internet will use the same link simultaneously.
	\item The probability that a packet will be dropped in a routing device.
	\item The number of clients attached to the link.
\end{enumerate}

We can distinguish between two types of bandwidth:
\begin{enumerate}
	\item \textit{Instantaneous bandwidth} - bandwidth at a specific moment of time.
	\item \textit{Average bandwidth} - average bandwidth over a complete message transfer.
\end{enumerate}

If the information flows from client to master via a directed path across $N$ links, then the effective bandwidth of the path is determined by the minimum bandwidth among all links:
\[B_{\mathrm{transfer}}=\min_{i\in [N]} B_{i}\].

The link with this minimum bandwidth is called the \textit{bottleneck link}.

\paragraph{Communication Delays During Master Client Communication}

The delay of a single link device is the actual time it takes for a bit of data to travel from one link device to the next. The delay consists of several components:

\begin{enumerate}
	\item  \textit{Processing delay} is the time required to examine the packet header and determine where to forward the packet in a link device with several output ports.
	\item  \textit{Transmission delay}, which is the time required for the router to push out the packet into the link, without considering the propagation of signal by itself.
	\item \textit{Propagation delay} is the time required for the signal to travel along the physical medium.
	\item  \textit{Queuing delay} is the time required for a packet to wait in the output buffer if the link output port is busy.
	\item  \textit{Loss delay} is the time required for a packet to be retransmitted if it is dropped due to buffer overflow or errors due to medium.
\end{enumerate}

Suppose a client and a server are connected by $N$ links sequentially, and the information flows through them in a directed path. In that case, the total delay for transferring a single package over this path is given by the sum of all delays along each link over the path:
\[\tau_{\mathrm{delay}}=\sum_{i=1}^{N} \tau_{\mathrm{delay},i}.\]

\subsubsection{Network Communication Model}

As we have seen, the total delay and bandwidth vary, even for the fixed topology of connecting the master and clients. According to \footnote{\href{https://www.speedtest.net/global-index}{Speedtest Global Index}} in the average download bandwidth across the globe is $B=41.54\,\mathrm{MBps}$ and communication latency is $RTT=28\,\mathrm{ms}$. The delay time for communication with messages $m \in \mathbb{R}^d$ where each component is represented by $bpp$ bits per component from master to client is modeled in the following way:
\[\tau_{delay} = \frac{RTT}{2} + \frac{d}{B} \cdot bpp.\]

Here, $B$ is the bandwidth (or throughput) of the communication channel between the master and the client. We assume it's constant. Next, $RTT$ is the time for a small packet to travel from client to server and back. It is only an approximation; for example, it does not model Packet Loss Delay. The worst-case maximum delay is very difficult even to estimate. We assume that $RTT/2$ and $B$ are not changing during training. However, in real communication networks, the path from \textit{client} to \textit{master} the exact communication path can change over time.

In reality, a Network Interface Controller (\abr{NIC}) implements network communication in a local computing device, and it is typically connected with a PCI-Express bus as an external input/output device to the whole local computation system. For modern PCI-Express buses such as PCI-E v5, the bandwidth, even for a single physical lane x1, is in the order of $4000.0$ MBps, and latency for this bus and latency is on the order of $35\cdot 10^{-3}$ ms. It means that in context when devices are communicated via the Internet, the effect of delays from PCI-Express is negligible, or at least it does not represent the bottleneck both in terms of bandwidth and latency. The same holds for involving communication time to transfer data from CPU to DRAM memory. Modern DDR5 memory has a bandwidth of $51 200$ MBps, and latency is measured in the order of nanoseconds. It means that what is represented as a bottleneck from a communication point of view (both in terms of latency and bandwidth) with the server is the connection to though \abr{NIC} installed in the client.

\subsubsection{CPU-based Computation in Clients} 

To evaluate the compressed gradient $C_i(\nabla f_i(x))$ in a client, different algorithms can be used, such as analytical, numerical, symbolic, and methods that leverage automatic differentiation. These algorithms need to run on some device that can execute them efficiently. We will focus on the execution aspects of modern Central Processing Units (CPUs), which are the most flexible devices from a programming perspective. CPUs have multiple cores that contain various components that work together to execute algorithms.

\subsubsection{Background in Modern Central Processing Unit}

\paragraph{Instruction Decode} In this stage, the CPU decodes the instructions and obtains information about the input, output, and operation type. Then, it splits the instructions into \textit{micro operations} and puts them into the Operation Issue Queue. This stage may introduce some complexities. One complexity is that some CPUs can decode multiple instructions simultaneously and issue them in parallel. This is called a multi-issue (or \textit{superscalar}) design, which aims to exploit instruction-level parallelism by executing independent instructions concurrently. Another complexity is that the CPU with \textit{out of order} issue capability can execute instructions whenever they are ready, regardless of the original order.

\paragraph{Operation Issue Queue} This is a hardware queue where decoded instructions, in the form of \textit{micro-operations}, are stored. The queue has a minimum length that can accommodate the longest sequence of \textit{micro-operations} for any instruction in the CPU's instruction set.

\paragraph{Control Unit (CU)} The CU operates at the level of \textit{micro-operations} and performs the following functions: (a) selecting the way to connect electrical components using multiplexers and demultiplexers; (b) turning on/off different electronic components; and (c) controlling the control lines of electronic components.

\paragraph{Multiplexers and Demultiplexers} During pipeline execution, intermediate inputs and results are stored in the latches of electrical components. To route signals within the CPU, multiplexers, and demultiplexers are utilized. Signals propagate through the component once all the components are connected, and data is applied to the input ports. The results are then produced when the enabled control signal reaches the electrical component.

\paragraph{Adder and Multiplier} The adders and multipliers are electrical circuits that perform addition or multiplication when turned on. The input for these devices is read from the intermediate buffer of the electronic component(latches). The typical input source (after intermediate routing with multiplexers and demultiplexers) is the Register File.

\paragraph{Register File} The Register File is a storage unit that holds all the registers in the CPU. It has multiple ports that allow parallel access to it. The Load and Store Units may have direct access to the Register File.

\paragraph{Load and Store Units (LS)} The Execution pipeline sends requests to the LS units for memory access. The LS units can access the Register File, the TLB for address translation, and the Memory Cache.	

\paragraph{Translation Lookaside Buffer (TLB)} To read code or data from memory in the user space or kernel space of the OS, the first step is to find the actual physical address of the specific memory location. This operation occurs for every instruction of a program. Without the TLB, the virtual addressing mechanism would require several accesses to different page tables, significantly increasing the time needed. The TLB is a cache that stores the mapping between virtual page numbers and physical frame numbers, speeding up the address translation process for memory access. The TLB relies on the locality of code and data in most algorithms.

\paragraph{CPU Memory Cache} The CPU Memory Cache is a fast storage unit that holds frequently accessed data and instructions. It is used to reduce the latency of accessing the DRAM memory. The Load and Store Units have access to the Cache. Modern high-end systems support three levels of Cache. The L1 cache is typically split between data and instructions, and it is the closest to the CPU. The L2 and L3 caches are larger and slower, and they can be shared by multiple cores. The cache implementation varies across different CPUs, depending on the trade-offs between speed, potential conflicts, hardware complexity, cache replacement policy, and power consumption. If the data is not available in the Cache, then the CPU Cache requests a block of memory from the Memory Controller (MU), which accesses the DRAM memory. When data from DRAM is stored in multiple CPU caches, it fundamentally means that data may be stored in several places. In this situation, another aspect becomes important: (a) \textit{cache consistency}, which essentially means that all copies of DRAM cache lines should be the same in all caches in the system;  (b) \textit{cache coherence} which essentially means that any read of memory returns the most recent update anywhere in the system.

\paragraph{DRAM Memory Controller (MC)} The DRAM Memory Controller is a device that manages access to the main DRAM memory Chips. It is used to fetch data from the Main Memory when it is not available in the CPU cache. The Memory Controller returns the data to the caches in blocks of a fixed size, called Cache Lines, and typically it is $64$ bytes. The Memory Controller is also responsible for running the memory bus transactions, which are the transfers of data between the MC and DRAM memory chips. The Memory Controller is typically implemented in the hardware as a device that is shared by multiple cores.

\paragraph{DRAM Memory Chips} Memory chips are devices that store data in binary form. They are usually specified by the number of bits stored and the number of bits accessed in one read or write operation. For example,  a common DRAM chip \textit{4Gbx1} means that it can store 4G bits and access one bit at a time. To protect data from corruption, extra logic may store bits for Error Correction Codes (ECC).

\paragraph{Input and Output Buses} Input and Output buses serve as pathways connecting external devices to the CPU. Examples of I/O buses include SATA, USB, and PCI-Express. For example, PCI-Express is commonly used to communicate with devices such as graphics cards and network cards. Communication with such external devices can be achieved using Direct Memory Access (DMA) or Programmed Input-Output (PIO). DMA enables devices to transfer data directly to or from memory without involving the CPU, while PIO requires the CPU to issue commands and wait for data. DMA is more efficient and faster than PIO but necessitates additional hardware support.


\subsubsection{Modeling of Computation}

We model clients' compute capability by assuming that they have a computation device similar to Intel-Xeon-E5-2666-v3 CPU \footnote{\href{https://www.intel.com/content/www/us/en/products/sku/81706/intel-xeon-processor-e52660-v3-25m-cache-2-60-ghz/specifications.html}{Intel Xeon Processor E5-2660 v3 Specification}}. We assume that the computation device of the client and the master is represented by a CPU with $10$ CPU cores, working at frequency $3.2\, GHz$, CPU support hyper-threading with executing $2$ computation works per core, we assume that Multiply - Add (MAD) operation is possible which effectively doubles compute throughput. There are $2$ functional units (FU) or compute ports per core for Floating Point arithmetic in such a device. We assume that add, subtract, and multiply operations for float numbers require a throughput of $1$ operation/clock and latency of $1$ clock per execution unit for FP32 arithmetic, which is realistic. With these assumptions, this device has a peak computation throughput of $238.41\, GFlops@FP32$. And we suppose all $n=4$ clients and masters are equipped with it.

In our model, we will assume that, on average, the train data is located in the L2 cache, and all memory operations in the client can be executed via accessing the L2 cache by utilizing one of $3$ Load/Store Units per Core. We assume access latency to read a cache line of size $64$ bytes requires $10$ clocks. Next, we assume that the Network Interface Controller(NIC) in the master and clients have data-direct I/O access to the L3 cache. Access latency to it is $40$ CPU cycles for $64$ byte cache line size. The effect of DRAM and caches can be ignored during inter-node communication but not during memory operation during gradient oracle computation.

\subsubsection{The Optimization Problem}

For modeling purposes, we consider solving a \modelname{Linear Regression} in the form:
\begin{eqnarray*}
	f(x) \eqdef \dfrac{1}{n} \sum_{i=1}^{n} f_i(x),\\
	f_i(x)=\dfrac{1}{n_i} \norm{A_i x - b_i}^2
\end{eqnarray*}

For this problem the value of  $\nabla f_i(x)=\dfrac{1}{n_i} A_i^\top(A_i x - b_i)$.  In implementing gradient oracles, we utilize dense matrix and matrix-vector operations. To add two vectors, we execute $d$ scalar additions and $2d$ memory access operations for reading. In modern computing, hardware loads are more expensive because writes essentially can be queued. The need time of inner product operation of two vectors of dimension $d$ is equal to $(d-1) \cdot \mathrm{add_{cost}} + d\cdot \mathrm{mult_{cost}} + 2d \cdot \mathrm{memaccess_{cost}}$. To estimate computing time for the Matrix-Vector and Matrix-Matrix Operations, we have assumed that their calculation is a sequence of inner products. In reality, not all data may fit into the Cache. If we go one step further with modeling, then Cache misses effects should modeled as well.

\subsubsection{Implementation Benefits of DCGD/PermK}

There are several flexibility aspects of DCGD/PermK that we will utilize in our experiment:

\begin{enumerate}
	\item The \compname{PermK} operator compressor behaviorism is independent of the input. This means that clients can a prior sample need coordinates for sparsification and increase the speed of $\nabla f_i(x)$ in clients.
	\item Next, we will exploit the fact that after the master receives the message, it can immediately broadcast it to all clients, regardless of whether the client has finished the work for the current round. This is possible in the context of using \algnamewithaes{DCGD/PermK/AES}, but it is impossible when using \ecryptname{CKKS}.
	\item If during the \abr{FL} process, there is a slow client with a slow CPU or with a big amount of samples, then \algname{DCGD/PermK} allow other clients to start several operations which are impossible for \algname{GD}: 
	\begin{itemize}
		\item Obtain (partial) results from the master by using a communication network for current round $\left[\nabla f(x^{k})\right]_{part}$
		\item Apply partial update for current model $x^k$ and obtain partially new model $x_{part}^{k+1}$
		\item Client can start perform partial computations for next iteration $\left[\nabla f(x_{part}^{k+1})\right]_{part}$
	\end{itemize}
\end{enumerate}

\subsubsection{Scheduling of Clients and Master Computation and Communication with Critical Path Method}

We modeled the \algname{GD} and \algname{DCGD} Algorithms with unrolling $r=4$ rounds of computation and communication over $n=4$ clients and $1$ master. We represented the optimization process as the directed graph of different elementary tasks.

Task dependencies are represented in the form of weighted, orientated, directed, acyclic graph:
$$G=(V, E, W), E=V \times V, W: \mathbb{E} \to \mathbb{R}_{+}.$$ 

The graph is constructed following the next rules:

\begin{enumerate}
	\item Tasks are represented as vertices in graph $G$.
	\item Task $s$ is connected to task $e$ with weight $0$, if and only if task $e$ can not start before task $s$.
	\item If the task $e$ obtains as input from task $s$, then the task $s$ is connected to task $e$ with the weight of duration of work (in seconds) which $s$ should perform to produce input for task $e$.
\end{enumerate}

The algorithm for the scheduling process involves the following steps:

\begin{enumerate}
	\item Introduce fake source and sink vertices.
	\item Add zero weight directed edge from a fake source vertex to all $v \in V$.
	\item From all original vertices $v \in V$ add zero weight edge to fake sink vertex.
	\item Add fake source and sink to set $V$.
	\item Compute topological order of $G$ starting from fake source $s$.
	\item Compute the longest path in the directed acyclic graph of tasks. It is achieved via temporally changing weights of $G=(V, E)$ to negative, computing the topological order of $G$, and relaxing all vertices in topological order.
	\item The longest path to all vertices $v \in V$ from fake source $s$ creates a schedule for executing operation $v$.
\end{enumerate} 

Essentially, this is a critical path method (CPM) to solve the parallel precedence-constrained scheduling problem. The running time of this algorithm is  $\mathcal{O}\left(V+E\right)$. 

\paragraph{Correctness Proof}

Let us examine the longest path $s {\leadsto} v$. The previously scheduled jobs before $v$ are vertices $x \in V$, such that $x {\leadsto} v$. The construction of the longest path implies that $w(s {\leadsto} v) \ge w(s {\leadsto} x)+w(x {\leadsto} v) \ge w(s {\leadsto} x), \forall s {\leadsto} x$.From this, we observe that the start times obtained by the longest paths are feasible because jobs $x$ that need to be executed before $v$ will be scheduled before $v$. Furthermore, the length of any path $s {\leadsto} v$ is a lower bound on the actual time to start $v$ because $v$ cannot begin earlier than previous tasks due to dependency constraints. This proves that the longest path from $s \in V$ to $v \in V$ determines the start time for task $v$.

Once task $v$ can begin execution in the timeline, it can potentially activate the execution of all tasks $(v, z) \in E$, where $z \in adj(v)$ and $W(v, z) > 0$.

\subsubsection{Critical Path Method Iterative Refinement}

Parallel precedence-constrained scheduling using the CPM method determines the execution schedule for a graph of jobs. However, there may be cases where the duration of a task (which is the input for the CPM Algorithm) is defined by the number of other tasks during specific time intervals. This creates a circular dependency between the input and output of the CPM Algorithm. In our scenarios, we encounter this situation due to the following reasons:

\begin{enumerate}
	\item If clients make a partial computation and the CPU is not busy with other works, parallelizable operations (such as Matrix-Vector multiply) can be parallelized across several CPU cores. Consequently, it increases clients' computational throughput for this partial update.
	\item If clients share the same bottleneck link to the master and some are still busy with compute $\nabla f_i(x^k)$, then other clients can transmit data at a fast bandwidth because the bottleneck link is shared across a smaller number of clients. This observation leads to a situation in which effective bandwidth can be increased.
\end{enumerate}

\begin{figure*}[t]
	\centering
	\captionsetup[sub]{font=scriptsize,labelfont={}}	
	\captionsetup[subfigure]{labelformat=empty}
	
	\begin{subfigure}[ht]{0.48\textwidth}
		\includegraphics[width=\textwidth]{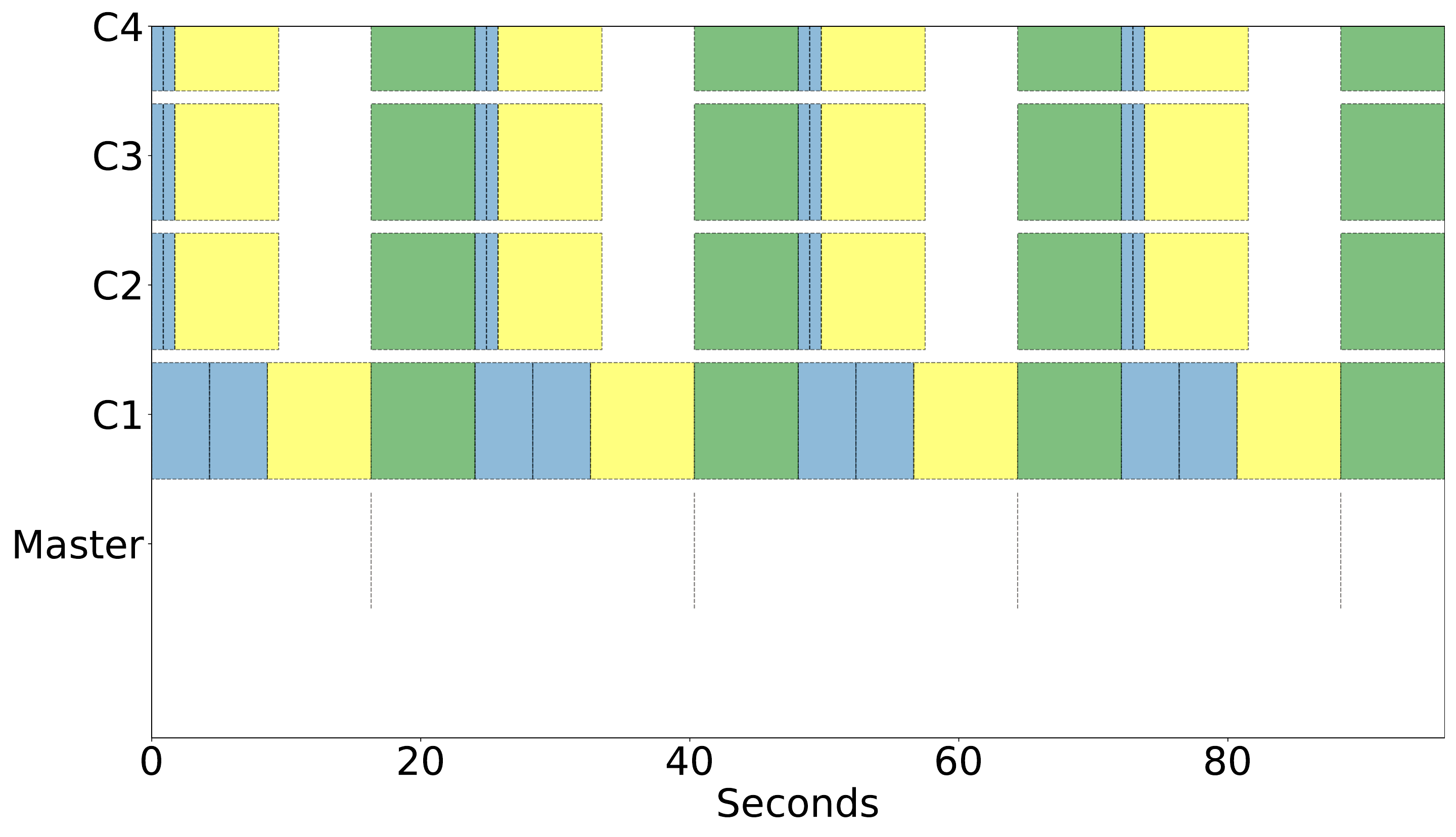} \caption{{(a) \algname{GD}. Modeling: $96.11$s.}}
	\end{subfigure}
	\begin{subfigure}[ht]{0.49\textwidth}
		\includegraphics[width=\textwidth]{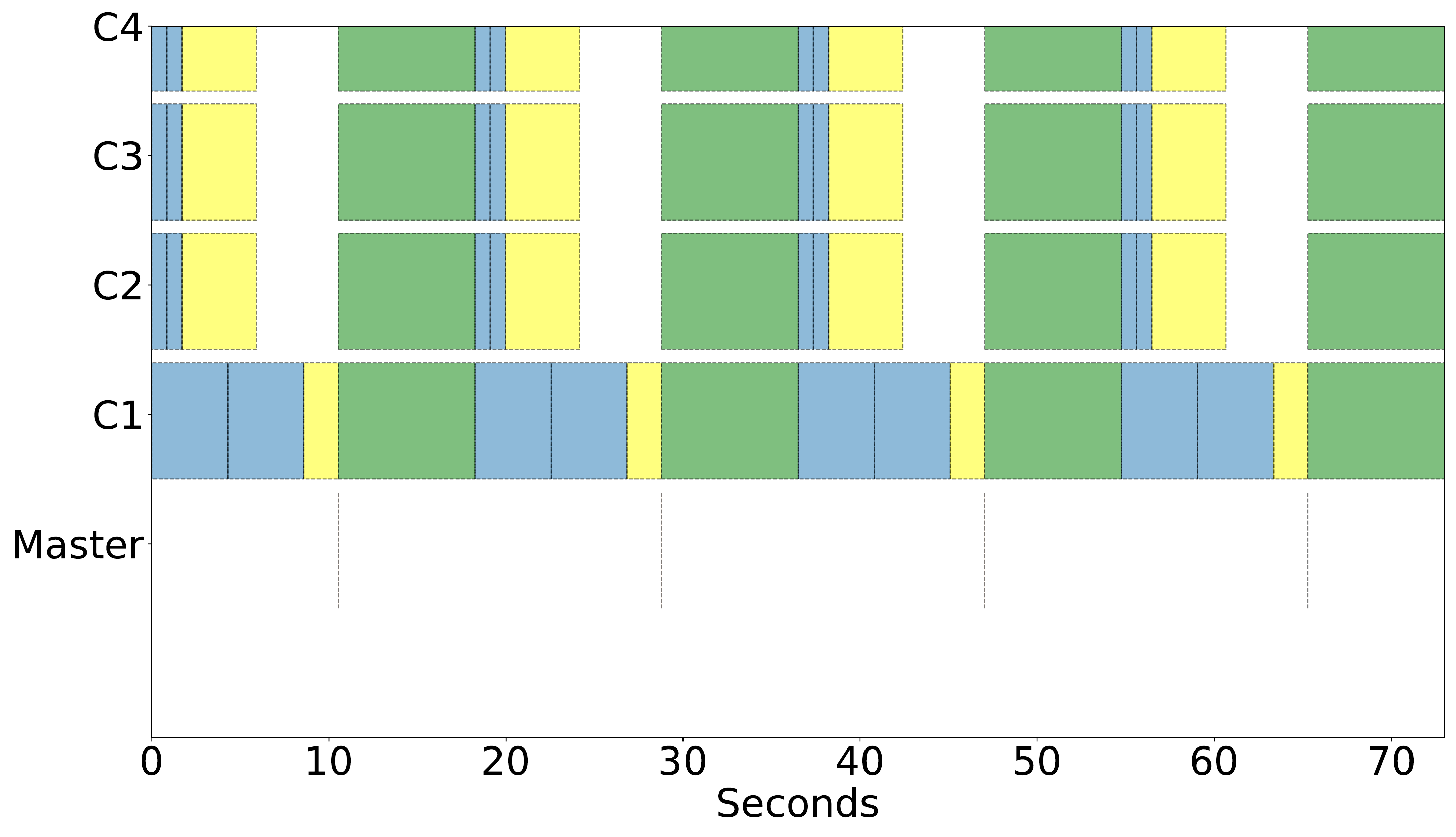} \caption{{(b) Refined \algname{GD}. Modeling: $73.00$s.}}
	\end{subfigure}
	
	\begin{subfigure}[ht]{0.48\textwidth}
		\includegraphics[width=\textwidth]{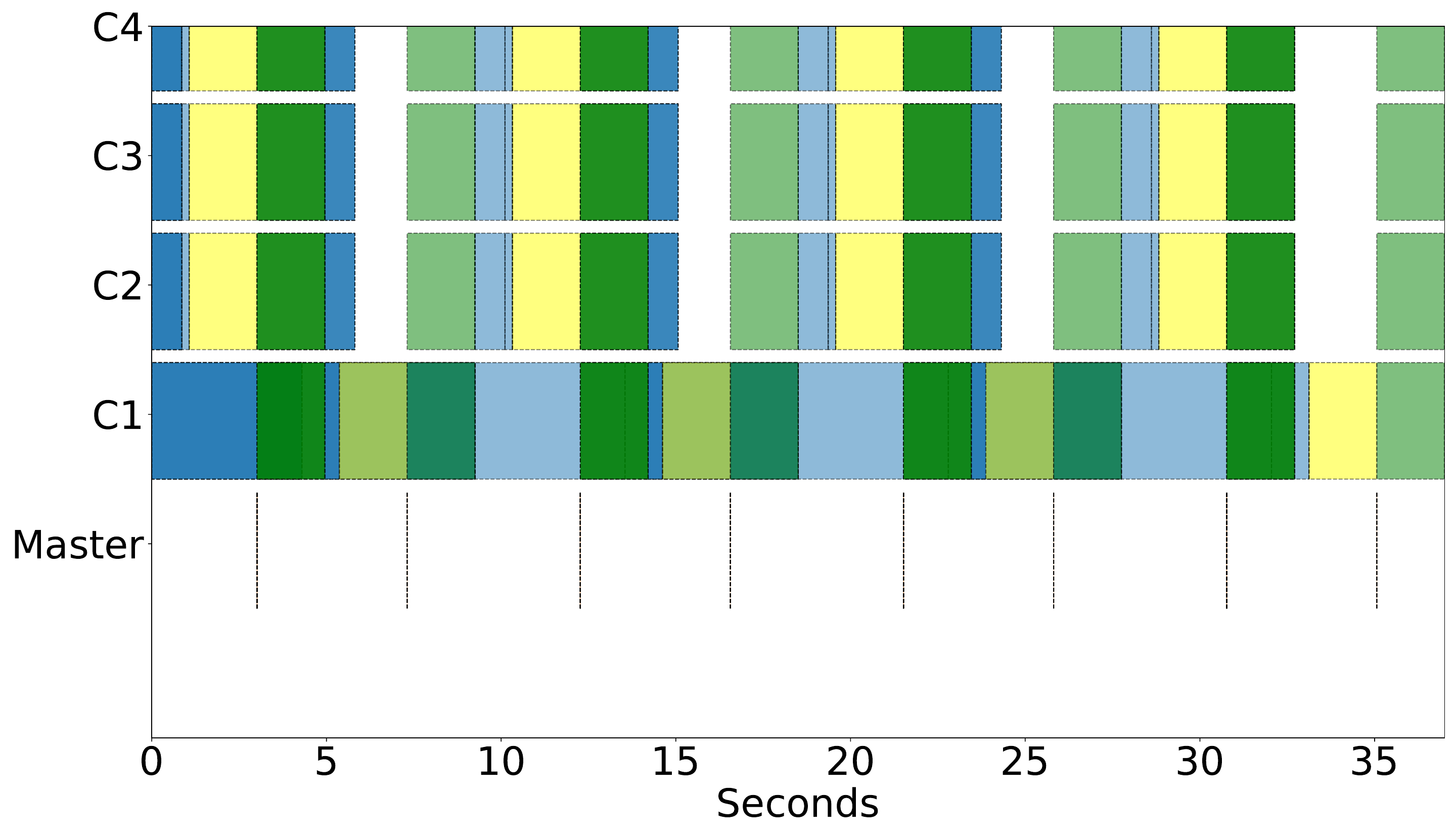} \caption{{(c) \algnamewithaes{DCGD/PermK/AES} Modeling: $37.003$ s.}}
	\end{subfigure}
	\begin{subfigure}[ht]{0.49\textwidth}
		\includegraphics[width=\textwidth]{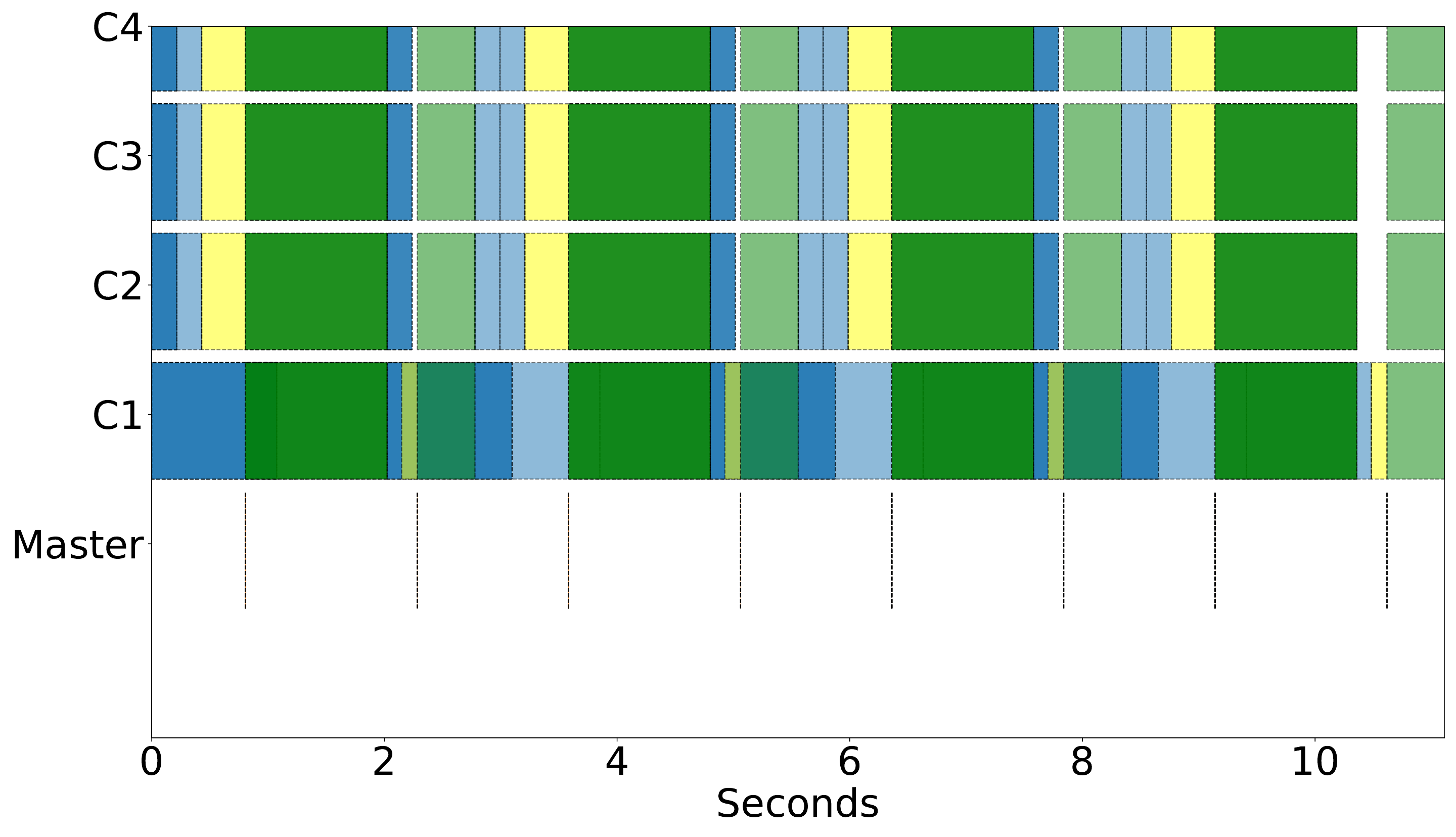} \caption{{(d) Refined \algnamewithaes{DCGD/PermK/AES}. Modeling: $11.113$s.}}
	\end{subfigure}
	
	\caption{
		Event-based modeling for training \modelname{Linear Regression} across $n=4$ clients, $d=10\cdot 10^6$, $n_1=55000, n_2=n_3=n_4=11000$ during $4$ rounds. The uplink and downlink bandwidth is $41.54$ MBps,  latency $28$ms, and computational thought of modeled CPUs is $238.41$ GFLOPS. Legend:\\  
		\phantom{x} \colorbox{blue}{\phantom{x}} - Computation and local memory access in the client (Client can use all available CPU cores), \\
		\phantom{x} \colorbox{yellow}{\phantom{x}} - Communication from client to the master (Clients share the same bottleneck link),\\ \phantom{x} \colorbox{green}{\phantom{x}} - Communication from the master to client (Clients share the same bottleneck link).
	}
	\label{fig:exp_syn_8}
\end{figure*}

\subsubsection{Results of Scheduling and Refinement for {GD} and {DCGD/PermK/AES}}

Scheduling results are presented in Fig.~\ref{fig:exp_syn_8}. In our experiment, all clients have the same compute power equal to $238.41$ GFLOPS. Communication bandwidth and latency from all clients to the master and vice versa is the same, namely $B_i=41.54$ Mbps, and latency (or delay) for transmission is $28\cdot10^{-3}$ seconds. The assumption for computation is that each memory operation is carried with an L2 cache. In case there is a need to perform read and write, we treat it as two memory operations. We assume all clients during communication to master share the same bottleneck communication link. Also, we assume that the target CPU supports \aesname{AES}. The detailed generated execution plans can not be represented in this paper due to their size. However, they can be found in the source code attached to the paper. 

As illustrated in Fig.~\ref{fig:exp_syn_8}, overlapping communication, computation, and the refined scheduling due to time-varying into shared communication bus and time-varying load into computing devices in clients leads to different speedups in execution plans (Fig.~\ref{fig:exp_syn_8}, (d), (b)) for \algname{GD} and \algname{DCGD/PermK}. The total computation speedup per round when compared to refined \algname{GD} is $6.578$. We hope this simulation will prove valuable for those seeking to adopt \algnamewithaes{DCGD/PermK/AES} to align with hardware requirements closely.




\clearpage

\section{Flexibility in Training Deep Learning Models}

\label{app:flexibility_for_dl_training}

\subsection{Introduction to Gradient Oracle Computation with Backpropagation}
In Appendix \ref{app:simulation_experiment} we have explored that \algname{DCGD/PermK} exhibits practical flexibility for training linear models in detail. However, described communication and computation overlap can be important not only for linear models but also for \abr{DL} models. 

The evaluation of the gradient of the score function in modern computational frameworks such as \libname{TensorFlow} \citep{abadi2016tensorflow} or \libname{PyTorch} \citep{paszke2019pytorch} for complex computational graph is automatized via leveraging algorithms for the numeric evaluation of derivatives. In most cases, this is achieved using \textit{Automatic Differentiation(AD) in Reverse Accumulation mode} \citep{linnainmaa1970representation} named as a \textit{Backpropagation algorithm} \citep{rumelhart1986learning} in \abr{ML} literature. A composed Loss function for Deep Learning models typically has the following structure: 
\begin{eqnarray}
	\label{eq:loss_for_dl}
	f_i(x) = \sum_{j=1}^{n_i} \mathcal{L}(b_{real,j}, \textcolor{red}{g_1}([x]_{Q_1}, a_j, \textcolor{red}{g_2}([x]_{Q_2}, a_j, \dots))) + R(x)\\
	Q_1 \cup Q_2 \dots = [d], Q_i \cap Q_j \dots = 0, \forall i,j. \notag
\end{eqnarray}

The score function $f_i(x)$ is represented by a computation graph and has a nested structure where prediction is driven by a nested composition of functions $\textcolor{red}{g_i}$ and the predicted label is scored with a true label for training input-output pair $(a_{j}, b_{real,j})$ with a score loss function $\mathcal{L}$. Each function $\textcolor{red}{g_i}$ has the following source of inputs: output from the prevision function (or layer) $g_{i-1}$, and input (trainable) scalar parameters $x_p$ where $p \in {Q_1, Q_2, \dots}$, and $Q_i \subseteq [d]$.

Assume that $f_i(x)$ differentiable in points where derivatives according to the chain rule are evaluated. In this case, the needed partial derivatives $[\nabla f_i(x)]_k = \frac{\partial f_i}{\partial x_k}$ can be computed with chaining Jacobians calculated with intermediate variables. The \textit{Backpropagation algorithm} allows for fixed input samples $D_i={(a_1,b_1), \dots, (a_{n_i},b_{n_i})}$ stored in client number $i$ and for fixed computational graph compute the required gradient $\nabla f_i(x)$ in two \textit{passes} or \textit{phases}:
\begin{enumerate}
	\item \textit{Forward Pass}. Compute all intermediate variables namely values of $g_i()$ for every training sample, for the whole compute graph. All these variables are stored in memory for the next pass.
	\item \textit{Backward Pass}. Compute the intermediate Jacobians and produce final partial derivatives of a full gradient. 
\end{enumerate}

\subsection{Examples of Parallelization inside  Backpropagation}
\label{app:parallforbackprop}

In fact, there is no need to store all intermediate Jacobians simultaneously. They are stored implicitly in special variables typically denoted as $\delta$ in implementation of \textit{Backpropagation Algorithm}. However, there is a need to store all intermediate outputs explicitly (named as activation in \abr{ML} literature) after \textit{Forward Pass}. Computationally \textit{Backpropagation} is often a preferred strategy when the entire gradient needs to be computed. The compute scheduling and parallelism strategies for effective computation of $\nabla f_i(x)$ is an active line of research by itself in Deep Learning System literature right now \cite{jia2019beyond},
\cite{krizhevsky2014one}. The standard strategies for performing parallelism inside computation of $\nabla f_i(x)$ involve the following:

\begin{enumerate}
	\item \textbf{Data Parallelism.} Assign train samples $(a_j, b_j)$ to different computation devices in client $i$.
	\item \textbf{Model Parallelism.} Assign different functions $g_i$ to different computation devices available in client $i$.
	\item \textbf{Attributes Parallelism.} Split the image of functions $g_{i-1}$ into parts and process it by parts if function $g_{i}$ allows to do it.
	\item \textbf{Parameters Parallelism.} Partition trainable variables for function $g_i(x)$, i.e. $[x]_{Q_i}$ into smaller chunks. After partitioning $[x]_{Q_i}$ if $g_i(x)$ can be computed in parallel then compute it in parallel in different devices.
\end{enumerate}

\subsection{Research Opportunities for Parallelization in Backpropagation from DCGD/PermK}

\label{app:flexibility_for_dl_training_from_permk}

Assume that clients know current iterate $x^{k}$, however, if there is even one straggler $s$ which still did not send ${g_s}^k$  to the Master, then no clients can proceed in training. Master has to wait for gradient estimator ${g_s}^k$ from
straggler $s$, and what can be done for \algname{DCGD/RandK} is at least challenging in these circumstances. However, \algname{DCGD/PermK} exhibits useful properties that can partially helpful in dealing with this situation which we previously discussed in Appendix \ref{app:simulation_experiment}:

\begin{itemize}
	\item Clients can obtain (partial) results from the master for current round $\left[\nabla f(x^{k})\right]_{part}$
	\item Clients can apply partial update for current model $x^k$ and obtain partially new model $x_{part}^{k+1}$
\end{itemize}

It depends on the specific situation, but during the waiting of a  straggler, the forward pass (\textit{forward pass} in practice takes at most $50\%$ of $\nabla f_i(x)$ computation) potentially can be started using the available $x_{part}^{k+1}$. Assume that we have statistical information about stragglers, then potentially this information can be provided to Deep Learning schedulers and parallelization strategies described in the previous section Appendix~\ref{app:parallforbackprop} to optimize $\nabla f_i(x^{k+1})$. This opens new opportunities to refine parallelization strategies for training in general and \textbf{Parameter Parallelism} in particular.

\clearpage
\section{Flexibility of DCGD/PermK for Different Communication Topologies}
\label{app:comm_networks}

In High-Performance Computation and Network Communication, the \textit{Physical Network Topologies} describes the physical arrangement of the devices and routing devices for organizing communication. Next, we will explain what benefits and flexibility \algname{DCGD/PermK} can bring if this algorithm runs in the mentioned popular physical network topologies.

\subsection{Potential Benefits in Point-to-point Topology}
In this type of topology, some clients are directly connected with a single link pairwise. It is possible to instantiate \algname{DCGD/PermK} so that clients do not coordinate during runtime for aggregation computation. It is possible because there is no need to perform reduction in the sense of averaging $\nabla f_i(x)$. If some pair clients are connected with the fast link, this pair of clients can utilize this channel and deliver $\left[\nabla f(x_{part}^{k+1})\right]_{part}$ for his neighbor with a fast point-to-point link without obtaining it from a master.

\subsection{Potential Benefits in Bus Topology}
A single cable or bus connects all the nodes in this type of topology. In this type of topology, if some clients have slow computing, it removes them from competition for a shared bus and decreases communication contention. In a bus topology, the benefit of \algname{DCGD/PermK} is that messages from each client can be effectively broadcast. In fact, \algname{DCGD/PermK} can be implemented without a centralized server, and a shared bus is enough for organized communication. Employing \aesname{AES} based encryption guarantees protection from eavesdropping on the shared bus.

\subsection{Potential Benefits in Star Topology}
In this type, all the nodes are connected to a central master device with individual links. In the case of using \algname{DCGD/PermK}, the master in this topology only plays the role of communication hub. Interestingly, while obtaining parts of the gradients from clients, computation is unnecessary, and synchronization is far more relaxed, as we have observed in Appendix~\ref{app:simulation_experiment}. While using \algname{DCGD/PermK}, the server can potentially utilize communication links with Full Duplex mode. Examples of Full Duplex communication buses include PCI-Express, InfiniBand\footnote{\url{https://network.nvidia.com/pdf/whitepapers/HPC_Clustering_130.pdf}}, and some forms of Ethernet (see \citep{spurgeon2000ethernet} Table 4.1 in Chapter 4). In our setting, we consider the situation when clients are connected to the Internet, and as it has been described in Appendix~\ref{app:simulation_experiment}, PCI-Express is not a bottleneck. However, in data centers, PCI-Express can also represent a bottleneck \citep{li2019priority}. With \algname{DCGD/PermK}, during some period, a single link can be used simultaneously to obtain information from clients and deliver information to the clients. This is not the case for \algname{DCGD/RandK} or \algname{GD} because they require explicit synchronization. But it is the case for \algname{DCGD/PermK}. This doubles the maximum bandwidth during master client communication and decreases latency by factor two.

\subsection{Potential Benefits in Ring Topology}
This type of topology connects all the nodes circularly. The ring topology is sometimes preferable because it reduces the number of \abr{NIC} and cables required for connecting multiple devices. Specifically, each device only needs one \abr{NIC} to participate in all-reduce. Also, ring topology avoids congestion and collisions at the central hub as messages are distributed evenly along the ring. The time delay during performing aggregation for \algname{GD} or \algname{DCGD/RandK} in a ring topology represents a sum of delays along the circle and is equal to $\tau_{\mathrm{ring\,delay}} = \sum_{i=1}^{n} \frac{RTT_i}{2} + \frac{d}{B_i} \cdot bpp$. Fundamentally, it is because during employing \algname{GD} or \algname{DCGD/RandK}, clients can not start any computation for the next iteration until they do not compute the whole gradient. For \algname{DCGD/PermK} clients, after obtaining $\left[\nabla f(x_{part}^{k})\right]_{part}$ from the neighbor, can start computation for the next iteration.

\subsection{Potential Benefits in Mesh Topology}
This topology connects every node to every other node with dedicated links. It has high bandwidth and fault tolerance but requires a lot of cables and ports and is complex to install. In this form of topology, the \algname{DCGD/PermK} is the most natural choice because underlying communication topology naturally maps to communication pattern in \algname{DCGD/PermK}, which can observed if view Algorithm~\ref{alg:dcgd_permk_aes} as a series of broadcast operations to reconstitute the global direction for optimization step.

\subsection{Potential Benefits in Tree Topology}
This topology combines multiple star topologies into a hierarchical structure with a root node. In this topology, the broadcasting can be implemented very effectively. One possible vision of our algorithm is that it is implemented without a central server; instead, each client broadcasts information to other clients in Line 6 of Algorithm \ref{alg:dcgd_permk_aes}.

\clearpage

\section{Acknowledgements}

The work of Peter Richt\'{a}rik and Konstantin Burlachenko was supported by the KAUST Baseline Research Scheme (KAUST BRF) and also supported by the SDAIA-KAUST Center of Excellence in Data Science and Artificial Intelligence (SDAIA-KAUST AI).

We acknowledge two members of Peter Richt\'{a}rik's Optimization and Machine Learning Laboratory, Alexander Tyurin and Egor Shulgin, for useful and insightful discussions before the start of this project.

\clearpage

\section{Reproducibility}

To ensure reproducibility, we use the following 
\libname{FL\_PyTorch} simulator \cite{burlachenko2021fl_pytorch} features: random seeds were fixed for data reshuffling, and random seeds were fixed for the runtime pseudo-random generators involved in variants of \algname{DCGD} for randomized compressors. 

The source code of our experiments is a part of our submission to \textit{"4th International Workshop on Distributed Machine Learning, co-located with CoNEXT 2023"}. If you are interested in the source code for experiments, please either find this publication and supplementary materials on a dedicated conference website\footnote{\href{https://distributedml.org/}{https://distributedml.org/}} \footnote{\href{https://conferences2.sigcomm.org/co-next/2023/}{https://conferences2.sigcomm.org/co-next/2023/}} or contact the authors.

\clearpage

\section{Limitations and Future Research}
\label{app:limitation_and_future_research}

One limitation of our work is the assumption that clients trust each other. However, it is also presented in \abr{HE}. The second limitation is that our work studies the case $d>n$. Our paper did not provide a rigorous theoretical analysis and focused on practical aspects of the proposed framework. Our method achieves strong privacy guarantees without compromising efficiency or accuracy via leveraging existing Cryptography protocols. 

For future research, in addition to developing rigorous theory and developing strategy for $d>n$, we believe that our work provides a bridge to utilizing other fields of science that work only on the level of bits. In addition to Cryptography another class of methods that operate on a bitwise representation of information is lossless data compression. This line can be investigated within \algname{DCGD/PermK} framework in future research. Next, as we have described in Appendix \ref{app:flexibility_for_dl_training} our work potentially opens an extra degree of freedom for research in Systems and Compilers for Deep Learning that investigates different parallelism scheduling strategies for optimize gradient oracles $\nabla f_i(x)$. Our work opens the opportunity to refine \textit{{Parameter Parallelism}} schedulers (see Appendix \ref{app:flexibility_for_dl_training_from_permk}).

\clearpage

\end{document}

%% file: new_commands.tex
\newcommand{\norm}[1]{\left\| #1 \right\|}

\newcommand{\cC}{\mathcal{C}}

\newcommand{\bA}{\mathbf{A}}

\newcommand{\del}[1]{}

\newcommand{\eqdef}{\stackrel{\text{def}}{=}}

\newcommand{\R}{\mathbb{R}}

\newcommand{\E}{\mathbf{E}}
\newcommand{\Exp}[1]{\mathbb{E}\left[#1\right]}

\newcommand{\squeeze}{\textstyle}

\newcommand{\RD}{\mathbb{R}^d}
\newcommand{\RS}{\mathbb{R}}

\newcommand{\ZS}{\mathbb{Z}}

\def\lfk{\left\lfloor}
\def\rfk{\right\rfloor}

\definecolor{linenumbercolor}{rgb}{0.98, 0.81, 0.69}

\definecolor{classcolor}{RGB}{0,0,255}

\definecolor{apicolor}{RGB}{255,0,0}

\definecolor{linenumbercolor}{rgb}{0.1, 0.1, 0.1}

\definecolor{modelcolor}{RGB}{108,57,0}
\definecolor{datacolor}{RGB}{108,57,0}
\definecolor{abrcolor}{RGB}{108,57,0}
\definecolor{algcolor}{RGB}{108,57,0}
\definecolor{aescolor}{RGB}{0,0,255}
\definecolor{algnamewithaescolor}{RGB}{0,0,255}
\definecolor{encryptcolor}{RGB}{108,57,0}
\definecolor{attackcolor}{RGB}{108,57,0}
\definecolor{compcolor}{RGB}{108,57,0}
\definecolor{libcolor}{RGB}{108,57,0}

\newcommand{\modelname}[1]{{\color{modelcolor}\sf \small {#1}}} 
\newcommand{\dataname}[1]{{\color{datacolor}\sf \small {#1}}}  
\newcommand{\abr}[1]{{\color{abrcolor}\sf \small {#1}}}       
\newcommand{\algname}[1]{{\color{algcolor}\sf \small {#1}}}   
\newcommand{\aesname}[1]{{\color{aescolor}\sf \small {#1}}}   
\newcommand{\algnamewithaes}[1]{{\color{algnamewithaescolor}\sf \small {#1}}} 
\newcommand{\ecryptname}[1]{{\color{encryptcolor}\sf \small {#1}}} 
\newcommand{\attackname}[1]{{\color{attackcolor}\sf \small {#1}}} 
\newcommand{\compname}[1]{{\color{compcolor}\sf \small {#1}}}   
\newcommand{\libname}[1]{{\sf \color{libcolor} \small {#1}}}    

\newcommand{\myred}[1]{{\color{red} #1}}

\newcommand{\myblue}[1]{{\color{blue} #1}}

%% file: theorem_environments.tex
\newtheorem{theorem}{Theorem}

\theoremstyle{plain}

\theoremstyle{definition}

\newtheorem{definition}[theorem]{Definition}

%% file: notes.tex
\usepackage[colorinlistoftodos,bordercolor=orange,backgroundcolor=orange!20,linecolor=orange,textsize=scriptsize]{todonotes}

\newcommand{\abdulmajeed}[1]{\todo[inline]{\textbf{Abdulmajeed: } #1}}
\newcommand{\fahad}[1]{\todo[inline]{\textbf{Fahad: } #1}}
 
\newcommand{\kostya}[1]{\todo[inline]{\textbf{Kostya: } #1}}